\documentclass[iop,apj,numberedappendix,twocolappendix]{emulateapj}
\usepackage{geometry} 
\usepackage{natbib}
\usepackage{graphics}
\usepackage{graphicx}
\usepackage{amssymb}
\usepackage{amsmath}
\usepackage{epstopdf}
\usepackage{epsfig}
\usepackage[usenames,dvipsnames]{color}
\usepackage[normalem]{ulem}

\newcommand{\srd}{\textsuperscript{rd}\ }

\newcommand{\beqn}{\begin{equation}}
\newcommand{\eeqn}{\end{equation}}

\newcommand{\ie}{\emph{i.e.},\ }

\newcommand{\eg}{\emph{e.g.},\ }

\newcommand{\mpc}{M_\sun\ \mathrm{pc}^{-2}}

\newcommand{\vlsr}{$v_{_\mathrm{LSR}}$}

\newcommand{\dml}{$d_{_\mathrm{ML}}$}
\newcommand{\dbar}{$\overline{d}$}
\newcommand{\dsun}{$d_{_\sun}$}
\newcommand{\rgal}{$R_\mathrm{gal}$}
\newcommand{\spitzer}{\emph{Spitzer}}
\newcommand{\herschel}{\emph{Herschel}}
\newcommand{\mml}{$M_{_\mathrm{ML}}$}

\newcommand{\cc}{cm$^{-3}$}
\newcommand{\msun}{$M_{_\sun}$}

\newcommand{\HI}{\ion{H}{1}}
\newcommand{\HII}{\ion{H}{2}}
\newcommand{\htwo}{H$_2$}
\newcommand{\hcop}{HCO$^{^+}$}

\newcommand{\nhhh}{NH$_3$}
\newcommand{\thco}{$^{13}$CO}
\newcommand{\twco}{$^{12}$CO}

\shorttitle{BGPS XIII. Physical Properties}
\shortauthors{Ellsworth-Bowers {\it et al.}}

\slugcomment{Accepted Version: April 6, 2015}

\begin{document}
\bibliographystyle{apj}

\title{The Bolocam Galactic Plane Survey. XIII. Physical Properties and Mass Functions of Dense Molecular Cloud Structures}

\author{Timothy P. Ellsworth-Bowers\altaffilmark{1,2}, Jason Glenn\altaffilmark{1}, Allyssa Riley\altaffilmark{1}, Erik Rosolowsky\altaffilmark{3}, Adam Ginsburg\altaffilmark{4}, Neal J. Evans II\altaffilmark{5}, John Bally\altaffilmark{1}, Cara Battersby\altaffilmark{6}, Yancy L. Shirley\altaffilmark{7,8}, and Manuel Merello\altaffilmark{9}}

\altaffiltext{1}{CASA, University of Colorado, UCB 389, Boulder, CO 80309, USA}
\altaffiltext{2}{email: \texttt{timothy.ellsworthbowers@colorado.edu}}
\altaffiltext{3}{Department of Physics, 4-183 CCIS, University of Alberta, Edmonton, AB T6G 2E1, Canada}
\altaffiltext{4}{European Southern Observatory,
Karl-Schwarzschild-Stra\ss e 2, 85748,
Garching bei M\"{u}nchen, Germany}
\altaffiltext{5}{Department of Astronomy, University of Texas, 2515 Speedway, Stop C1400, Austin, TX 78712, USA}
\altaffiltext{6}{Harvard-Smithsonian Center for Astrophysics, 60 Garden Street, Cambridge, MA 02138 USA}
\altaffiltext{7}{Steward Observatory, University of Arizona, 933 North Cherry Avenue, Tucson, AZ 85721, USA}
\altaffiltext{8}{Adjunct Astronomer at the National Radio Astronomy Observatory}
\altaffiltext{9}{Istituto di Astrofisica e Planetologia Spaziali-INAF, Via Fosso del Cavaliere 100, I-00133 Roma, Italy}

\begin{abstract}

We use the distance probability density function (DPDF) formalism of \citet{EllsworthBowers:2013,EB14a} to derive physical properties for the collection of 1,710 Bolocam Galactic Plane Survey (BGPS) version 2 sources with well-constrained distance estimates.  To account for Malmquist bias, we estimate that the present sample of BGPS sources is 90\% complete above $400\,M_{_\sun}$ and 50\% complete above $70\,M_{_\sun}$.  The mass distributions for the entire sample and astrophysically motivated subsets are generally fitted well by a lognormal function, with approximately power-law distributions at high mass.  Power-law behavior emerges more clearly when the sample population is narrowed in heliocentric distance (power-law index $\alpha = 2.0\pm0.1$ for sources nearer than 6.5\,kpc and $\alpha = 1.9\pm0.1$ for objects between 2\,kpc and 10\,kpc).  The high-mass power-law indices are generally $1.85 \leq \alpha \leq 2.05$ for various subsamples of sources, intermediate between that of giant molecular clouds and the stellar initial mass function.  The fit to the entire sample yields a high-mass power-law $\hat{\alpha} = 1.94_{-0.10}^{+0.34}$.  Physical properties of BGPS sources are consistent with large molecular cloud clumps or small molecular clouds, but the fractal nature of the dense interstellar medium makes difficult the mapping of observational categories to the dominant physical processes driving the observed structure.  The face-on map of the Galactic disk's mass surface density based on BGPS dense molecular cloud structures reveals the high-mass star-forming regions W43, W49, and W51 as prominent mass concentrations in the first quadrant.  Furthermore, we present a 0.25-kpc resolution map of the dense gas mass fraction across the Galactic disk that peaks around 5\%.
\end{abstract}



\section{INTRODUCTION}\label{ch4:intro}

As the possible precursors to stellar clusters, OB associations, or smaller stellar groups, molecular cloud clumps and cores have become a primary focus for understanding the process of high-mass star formation \citep[\eg][and references therein]{McKee:2007}.  The recent advent of large-scale continuum surveys of the Galactic plane at (sub-)millimeter wavelengths (BGPS, \citealp{Aguirre:2011}, \citealp{Ginsburg:2013}; ATLASGAL, \citealp{Schuller:2009}; Hi-GAL, \citealp{Molinari:2010a}) have detected tens of thousands of these objects in thermal dust emission.  A major payoff of blind surveys in this portion of the electromagnetic spectrum is the derivation of physical properties of regions hosting high-mass star formation.  A detailed census of these dense molecular cloud structures can help constrain star-formation and galactic-evolution theories \citep[\eg][]{Kennicutt:2012}.  Studies of the physical properties \citep[\eg][]{Peretto:2010b,Giannetti:2013} and mass distributions \citep[\eg][]{Netterfield:2009a,Olmi:2013,Gomez:2014} of these objects have begun, but despite the richness of the current data sets, a coherent picture has not yet emerged for the evolution of the dense interstellar medium and the origin of the stellar initial mass and stellar cluster mass functions.

While giant molecular clouds (GMCs) have been studied for several decades using the lowest rotational transitions of the isotopologues of CO and other simple molecules \citep[\eg][and references therein]{Scoville:1975,Cohen:1980,Dame:1987,Dame:2001}, and studies of the apparently uniform stellar and cluster initial mass functions \citep[\eg][]{Bastian:2010} go back to \citet{Salpeter:1955}, the observational technology (detectors and angular resolution) has only recently developed to study the dense substructures intermediate between these two extremes.  And while studies of nearby molecular cloud complexes have recently yielded estimates of the core mass function \citep[\eg][]{Motte:2007,Enoch:2007,Swift:2010} for the possible progenitors of single stellar systems, it has only been with the recent large-scale blind surveys that the study of the \emph{clump} mass function has been possible.  

Theoretical modeling of molecular cloud structure evolution is beginning to place constraints on the clump mass function \citep[\eg][]{Donkov:2012,Veltchev:2013}, and two primary functional forms are discussed in the literature: the power-law and lognormal distributions.  Both arise from physical processes in molecular clouds; supersonic turbulence within molecular clouds produces a lognormal density distribution \citep[\eg][although \citealp{Tassis:2010} find other means to produce a lognormal]{Padoan:1997}, while gravitational collapse of dense structures tends to produce a power law \citep[\eg][]{Padoan:2002}.  The distinctions between these have implications for competing theories of high-mass star formation \citep[][]{Elmegreen:1985}.  Furthermore, the observed mass function for dense molecular cloud structures should resemble a combination of these forms due to the complex interaction of physical processes \citep[\eg][]{Offner:2013}; for a discussion of the details, see \citet{Hopkins:2013d}.

In a broader context, the Galactic distribution of dense molecular gas and star formation has implications for using the Milky Way as ground truth for studies of extragalactic star formation and galaxy formation.  Cosmological simulations of galaxy formation and evolution have made remarkable progress in the last two decades reproducing some of the observed properties of galaxies \citep[\eg][]{Katz:1996,Keres:2009,Stinson:2013}, but numerical simulations have largely been unable to \emph{a priori} produce the strong winds and instantaneous inefficiency of star formation observed in galaxies \citep{Hopkins:2014e}.  By incorporating feedback from stars via `sub-grid' physics recipes \citep[\eg][]{Kim:2014}, numerical galaxy evolutions models can now reliably produce spiral galaxies.  Very recently, \citet[][]{Hopkins:2014e} introduced simulations that \emph{do} include realistic feedback to produce galaxies commensurate with locally observed examples.  The continued refinement of these models must rely upon the constraints of derived physical properties of dense molecular cloud structures in the Milky Way.  Additionally, Galactic structures can provide a comparison for scaling relationships for star formation rates with respect to more active galaxies \citep{Kamenetzky:2014}.

As the first completed dust-continuum millimeter-wavelength survey of the northern Galactic plane, the Bolocam Galactic Plane Survey (BGPS) has provided a foundation for studying dense molecular cloud structures in a variety of Galactic environments.  The majority of objects detected in the BGPS correspond to molecular cloud clumps \citep[][hereafter D11]{Dunham:2011c}, each of which will likely form a cluster of stars.\footnote{Molecular cloud clumps are typically $10^2 - 10^3$\,\msun\ with sizes $0.3-3$\,pc \citep[][]{Bergin:2007}.  Following D11, we generally adopt the \citeauthor[][]{Bergin:2007} classifications for dense molecular cloud structures.}  Derivation of physical properties of these objects requires robust distance estimates; \citet{EllsworthBowers:2013,EB14a} developed a distance probability density function (DPDF) methodology for utilizing all available information towards this goal.  By combining kinematic distance information with prior DPDFs based on ancillary data (such as \spitzer/GLIMPSE mid-infrared images and trigonometric parallax measurements) in a Bayesian framework, \citet[][hereafter EB15]{EB14a} presented a catalog of 1,710 BGPS objects with well-constrained distance estimates.

In this work, we build on that distance catalog to derive the physical properties of and fit mass functions to this vital population of sources.  The probabilistic description of the distance to each object in the survey catalog may be used to directly propagate uncertainties into the derived physical properties via Monte Carlo methods.  We present here a catalog of physical properties for this set of sources, and fit mass functions to the entire sample, as well as astrophysically motivated subsets.  Furthermore, we explore the extension of fundamental molecular cloud scaling relationships \citep[][]{Larson:1981} to dense substructures and map the distribution of dense molecular gas throughout the Galactic disk.


\section{DATA}\label{ch4:data}

The Bolocam Galactic Plane Survey version 2 \citep[BGPS V2;][hereafter G13]{Aguirre:2011,Ginsburg:2013}, is a $\lambda = 1.1$~mm continuum survey covering 192~deg$^2$ at 33\arcsec\ resolution.  It is one of the first large-scale blind surveys of the Galactic plane in this region of the spectrum, covering $-10\degr \leq \ell \leq 90\degr$ with $|b| \leq 0\fdg5$ and selected larger cross-cuts to $|b| \leq 1\fdg5$, plus selected regions in the outer Galaxy.  For a map of BGPS V2 coverage and details about observation methods and the data reduction pipeline, see G13.  From the BGPS V2 images, 8,594 millimeter dust-continuum sources were identified using a custom extraction pipeline.  BGPS V2 pipeline products, including image mosaics and the catalog, are publicly available.\footnote{Available through IPAC at\\ \texttt{http://irsa.ipac.caltech.edu/data/BOLOCAM\_GPS}}

The BGPS data pipeline removes atmospheric signal using a principle component analysis technique that discards common-mode time-stream signals correlated among bolometers in the focal plane array.  The pipeline attempts to iteratively identify astrophysical signal and prevent that signal from being discarded with the atmospheric signal.  This sky-subtraction behaves like a high-pass filter that may be characterized by an angular transfer function passing scales from approximately 33\arcsec\ to 6\arcmin\ (see G13 for a full discussion).  The effective angular size range of detected BGPS sources therefore corresponds to anything from molecular cloud cores up to entire GMCs depending on heliocentric distance (D11).

\citet{EB14a} presented a distance catalog for BGPS V2 sources using a Bayesian DPDF framework.  The DPDFs describe the relative probability of finding an object at points along the line of sight and are constructed from multiple sources of distance-related information via 
\beqn\label{eqn:dpdf}
\mathrm{DPDF}(d_{_\sun}) = \mathcal{L}(v_\mathrm{LSR},l,b;d_{_\sun})~\prod_i P_i(l,b;d_{_\sun})~,
\eeqn
where $\mathcal{L}(v_\mathrm{LSR},l,b;d_{_\sun})$ is the kinematic distance likelihood function based on a line-of-sight velocity (\vlsr), and the $P_i(l,b;d_{_\sun})$ are prior DPDFs based on ancillary data.  Given circular orbits about the Galactic center, there are two points along a line of sight looking interior to the Solar circle that correspond to a single \vlsr, the kinematic distance ambiguity (KDA).  The $\mathcal{L}(v_\mathrm{LSR},l,b;d_{_\sun})$ for such objects is double-peaked, representing the equal probability of finding the object at either kinematic distance; prior DPDFs are used principally to resolve the KDA, placing an object at either the near or far distance (see EB15 for further discussion).  The resulting catalog contains 1,710 sources with well-constrained distance estimates (distance uncertainties generally $\lesssim 0.5$\,kpc), hereafter referred to as the Distance Catalog.  The posterior DPDFs\footnote{Available with the BGPS release at IPAC.} are normalized to unit integral probability ($\int_0^\infty \mathrm{DPDF}~\mathrm{d}d_{_\sun} = 1$) to facilitate marginalization over distance, and allow for the Monte-Carlo propagation of uncertainty in an object's distance to the derived physical properties by randomly sampling the DPDFs many times.


\section{METHOD}\label{ch4:method}

\subsection{Computing Physical Properties Using the DPDF}\label{meth:phys}

It is straightforward to compute physical properties of dense molecular cloud structures using single-value distance estimators.  The complete information contained in an object's posterior DPDF, however, represents the combination of knowledge and ignorance about its distance based on a wide collection of data encapsulated in the suite of prior DPDFs applied.  Use of this full information, therefore, requires careful consideration.  Below, we discuss mass calculation for BGPS objects in some detail (\S\ref{meth:mass}), and then apply those methods to the computation of other physical properties (\S\ref{meth:prop}).

\subsubsection{Mass Derivation}\label{meth:mass}

A simple estimate of a dense molecular cloud structure's mass may be computed from optically thin millimeter dust continuum data via
\beqn\label{eqn:mass}
M = \frac{\mathsf{r}~S_{\nu,\mathrm{int}}~d_{_\sun}^2}{\kappa_\nu~B_\nu(T_d)}~,
\eeqn
where $S_{\nu,\mathrm{int}}$ is the source-integrated flux density, \dsun\ is the estimated heliocentric distance, $\kappa_\nu$ is the opacity per mass of dust, $\mathsf{r} \equiv (m_{_\mathrm{g}}/m_{_\mathrm{d}})$ is the gas-to-dust mass ratio, and $B_\nu(T_d)$ is the Planck function evaluated at dust temperature $T_d$.  For the specific case of the BGPS,
\beqn\label{eqn:bgps}
M = 13.1\,M_{_\sun}~\left(\frac{e^{13.0~\mathrm{K}/T_d}-1}{e^{13.0~\mathrm{K}/20.0~\mathrm{K}}-1} \right) \left(\frac{S_{_{1.1}}}{\mathrm{Jy}} \right) \left(\frac{d_{_\sun}}{\mathrm{kpc}} \right)^2~,
\eeqn
where $S_{_{1.1}}$ is the $\lambda = 1.1$~mm source-integrated flux density, $\kappa_\nu = 1.14$~cm$^2$~g$^{-1}$ of dust \citep[][Table 1, Column 5]{Ossenkopf:1994},\footnote{This table of $\kappa_\nu$ represents dust grains with ice mantles, coagulating in cold, dense molecular regions at $n = 10^5$\,\cc\ for $10^6$~yr.  This is appropriate for molecular cloud \emph{cores}, and may be an upper limit for less-dense \emph{clumps} \citep[\eg][]{Martin:2012}.  The computed masses may, therefore, be biased low.  See a recent study of dust opacity for dense molecular gas by \citet[][]{Juvela:2015}.} $\mathsf{r}$ = 100 \citep[][]{Hildebrand:1983}, and the equation has been normalized to $T_d = 20$\,K.
Standard error-propagation methods may be used to determine the mass uncertainty, but the asymmetric nature of many DPDF error bars (EB15) complicates the issue.  The uncertainty in the distance may be marginalized over to produce the expectation value of the mass via
\beqn\label{eqn:mbar}
\langle M \rangle = \frac{\mathsf{r}~S_{\nu,\mathrm{int}}}{\kappa_\nu~B_\nu(T_d)} \int_0^\infty \mathrm{DPDF}~d_{_\sun}^2~\mathrm{d}d_{_\sun}~,
\eeqn
where $\langle M \rangle$ is proportional to the second moment of the DPDF.  The bimodal nature of many posterior DPDFs in the inner Galaxy, however, causes this to be a biased estimator of the mass, in much the same way \dbar\ (the first-moment distance) can be a biased estimator of the distance.  For instance, an object with a well-constrained distance estimate may have up to $\approx20\%$ of the integrated DPDF in the non-favored kinematic distance peak; the difference between \dbar\ and the maximum-likelihood distance (\dml) can be greater than 1\,kpc, or $\geq2\times$ the uncertainty in \dml\ (see \citealp{EllsworthBowers:2013} for a complete discussion).

This bias, coupled with the measured source flux-density and assumed dust temperature uncertainties, suggests using a Monte Carlo approach for estimating an object's mass and associated uncertainty.  To determine the maximum-likelihood mass for a given catalog object, Equation~(\ref{eqn:mass}) is computed many times with each realization drawing a distance, flux density, and temperature from the suitable probability density functions.  The resulting (properly normalized) mass probability density function (MPDF) may then be used to estimate the maximum-likelihood mass (\mml) and the 68.3\% error bars in the same fashion as done with DPDFs in EB15.

Being the probability of an object having a mass between $M$ and $M + \delta M$, the MPDF of a given object is created from the histogram of masses produced by many ($\sim10^6$) Monte Carlo realizations of Equation~(\ref{eqn:mass}).  Given the fractional nature of the uncertainties in the derived mass, we chose to use logarithmic binning for computation of MPDFs and the probability density functions for other physical properties.  Note that the computation of DPDFs (EB15) is done with respect to a fixed linear distance scale and is not subject to the need to create histograms from Monte Carlo calculations.

\subsubsection{Physical Radius and Number Density}\label{meth:prop}

In addition to the masses of dense molecular cloud structures, the catalog of heliocentric distances enables the computation of physical sizes and number densities  of these objects.  Not only can these physical properties help to disentangle the various populations of objects (\ie cloud, clump, core) detected by the BGPS and others, but they can also test the extension of various empirical relationships for GMCs down to more dense substructures.

The physical radius is computed from
\beqn\label{eqn:prad}
R = 0.29~\mathrm{pc}~\left(\frac{\theta_R}{\mathrm{arcmin}}\right)\left(\frac{d_{_\sun}}{\mathrm{kpc}}\right)~,
\eeqn
where $\theta_R$ is the deconvolved angular radius of a detected object computed from geometric mean of the deconvolved major ($\sigma_\mathrm{maj}$) and minor ($\sigma_\mathrm{min}$) axes of the flux density distribution,
\beqn\label{eqn:theta_r}
\theta_R = \eta \left[ \left(\sigma_\mathrm{maj}^2 - \sigma_\mathrm{bm}^2 \right)  \left(\sigma_\mathrm{min}^2 - \sigma_\mathrm{bm}^2 \right) \right]^{1/4}~
\eeqn
\citep{Rosolowsky:2010}.  For the BGPS, the rms size of the beam is given by $\sigma_\mathrm{bm} = \theta_{_\mathrm{FWHM}} / \sqrt{8 \ln 2} = 14\arcsec$, and $\eta = 2.4$ is a factor relating the rms size of the emission distribution to the angular radius of the object.\footnote{The appropriate value of $\eta$ to be used depends on the true emission distribution of the object and its size relative to the beam.  \citet{Rosolowsky:2010} computed $\eta$ for a large range of models, varying the emissivity distribution, angular size relative to the beam, and signal-to-noise ratio.  The chosen value for the BGPS catalog is the median value from these simulations, but the variations span more than a factor of two.  Use of a specific source model in conjunction with the BGPS catalog would require the appropriate value of $\eta$ and rescaled catalog values for $\theta_R$}  Objects whose measured minor axis dispersion is smaller than $\sigma_\mathrm{bm}$ do not have a real solution to Equation~(\ref{eqn:theta_r}); actual physical objects should not fall into this category, but effects of the cataloging process lead to 29\% of the full BGPS V2.1 catalog and 20\% of the Distance Catalog having a complex computed $\theta_R$.  These objects are excluded from the analyses of physical radius and number density in \S\ref{ch4:results}.

As a means for classifying dense molecular cloud structures, the mean number density offers possible insight into the physical processes, such as structure formation via turbulence or gravitational collapse, that may be at play in a given catalog object.  The number density computed with assumed cylindrical geometry (\ie a circle on the sky with radius $R$, and depth $R$),\footnote{This parameterization of the volume results in $V = \pi R^3$; assuming spherical geometry increases the volume by a factor of 4/3, and assuming a cylinder of depth $2R$ increases it by a factor of 2.  As shown in \S\ref{res:cat}, the number density is a poorly constrained quantity for these objects, and source-to-source geometry differences play a significant role.} $n = M / \pi \mu m_H R^3$, may be parameterized in terms of mass and radius or observable quantities as
\begin{eqnarray}\label{eqn:nden}
n &=& 4.60\times10^2~\mathrm{cm}^{-3}~\left(\frac{M}{100\,M_{_\sun}}\right) \left(\frac{R}{\mathrm{pc}}\right)^{-3} \nonumber \\
 &=& 2.47\times10^3~\mathrm{cm}^{-3}~\left(\frac{e^{13.0~\mathrm{K}/T_d}-1}{e^{13.0~\mathrm{K}/20.0~\mathrm{K}}-1} \right) \times \nonumber \\
 & & \left(\frac{S_{1.1}}{\mathrm{Jy}}\right) \left(\frac{\theta_R}{\mathrm{arcmin}}\right)^{-3} \left(\frac{d_{_\sun}}{\mathrm{kpc}}\right)^{-1}~.
\end{eqnarray}
The first parameterization relates the physical quantities in units appropriate to dense molecular cloud structures, while the second is specific to the observational quantities of the BGPS and the assumptions used in Equation~(\ref{eqn:bgps}).  Note that the number density is not as strong a function of heliocentric distance as is the mass; incorrect KDA resolution will have a smaller effect on the derived value.

\subsection{Constructing Mass Distributions and Estimating Completeness}\label{meth:mfn}

The mass function describing the relative numbers of dense molecular cloud structures is of particular interest for connecting star-formation theory to observation.  Functional forms are fitted to the distribution of masses and we explore some of the possible mass distribution constructions.  Additionally, since the range of masses over which conclusions may be drawn is restricted by completeness considerations, we present an analysis estimating the mass completeness levels for a heterogenous data set such as the BGPS.

\subsubsection{Mass Distributions}\label{meth:mdistr}

\begin{figure*}[!t]
  \centering
  \includegraphics[width=2.2in]{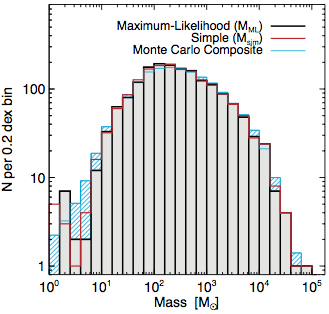}
  \includegraphics[width=2.1in]{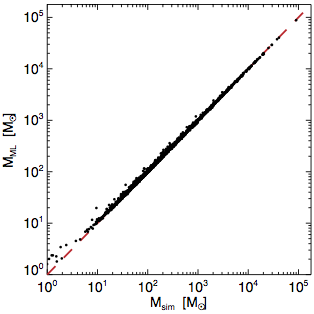}
  \includegraphics[width=2.1in]{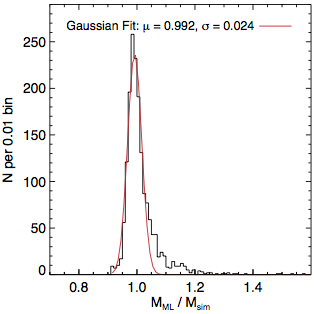}
  \caption{Comparison of mass computation methods.  \emph{Left}: The mass distribution computed by applying Equation~(\ref{eqn:mass}) to each object, using the appropriate single distance estimate from the posterior DPDF (as discussed in \citet{EllsworthBowers:2013}, `Simple $M_\mathrm{sim}$', red) plotted atop the mass distribution using the maximum-likelihood masses from the MPDFs (`Maximum-Likelihood \mml', black).  Also shown is the Monte Carlo composite mass distribution, as discussed in \S\ref{res:mass_dist} (hashed cyan).  \emph{Middle}: Comparison of \mml with $M_\mathrm{sim}$ for each object in the Distance Catalog.  The red dashed line marks the 1:1 relationship.  \emph{Right}: Histogram of the ratio \mml$/ M_\mathrm{sim}$ for each object in the Distance Catalog, with Gaussian fit overlaid in red.}
  \label{fig:md_ex}
\end{figure*}

In its simplest form, the mass distribution of dense molecular cloud structures may be compiled from the direct application of Equation~(\ref{eqn:mass}) to each object with a well-constrained distance estimate, where \dsun\ is either the maximum-likelihood distance or the first-moment distance, as discussed in \citet{EllsworthBowers:2013}.  This form of the distribution for the Distance Catalog sources is shown as the red histogram in Figure~\ref{fig:md_ex} (\emph{left}) and represents a na\"ive, but effective, realization of the mass distribution.

While the simple mass distribution utilizes the best distance estimate from the DPDF formalism, it discards much of the information contained in the posterior DPDFs.  Given the asymmetric nature of many DPDFs, the most straightforward way to capitalize on this pool of information is through Monte Carlo realizations utilizing the MPDFs.  Plotted in black in Figure~\ref{fig:md_ex} (\emph{left}) is the mass distribution computed using the \mml\ from the Monte-Carlo generated MPDFs.  While there are noticeable deviations in the wings of the distributions, a two-sample KS test between these two distributions reveals that they are indistinguishable to better than 99.9\%.  Despite the negligible differences between the distributions, the power of the MPDF-based mass distributions lies in the ability to create many Monte Carlo realizations of the mass distribution, randomly drawing a mass from each object's MPDF for each trial.

If the aggregate mass distributions between the two methods are indistinguishable, what of the masses of individual objects?  The comparison of the mass derived from a simple application of Equation~(\ref{eqn:mass}) using the single-value distance estimate from EB15 ($M_\mathrm{sim}$) with the maximum-likelihood value from the MPDF (\mml) is shown in the middle panel of Figure~\ref{fig:md_ex}, where the dashed line marks the 1:1 relationship.  The histogram of the ratio of these masses (\mml$/ M_\mathrm{sim}$) is plotted in the right panel along with a Gaussian fit.  The mean of the distribution is consistent with unity, and has a standard deviation of $\approx 2\%$.  Differences between the masses are more significant for low-mass objects ($M \lesssim 100\,M_{_\sun}$), with \mml becoming larger than the simple mass by up to 60\% in the most extreme case.

Going beyond constructing a mass distribution for the Distance Catalog using a single mass for each object, a more complete Monte Carlo approach is shown as the hashed cyan histogram in Figure~\ref{fig:md_ex} (\emph{left}).  Rather than taking the maximum-likelihood value from the MPDF of each object (as does the black histogram in that panel), each object's MPDF was randomly sampled 20,000 times to produce a mass distribution containing 34.2 million entries.  The resulting histogram was then normalized to represent the $N=1710$ objects of the Distance Catalog.  The resulting distribution is much smoother at both mass extrema because of the lack of small-number statistics.  Interestingly, the mode (peak) of the distribution is shifted to slightly higher mass, likely due to the $d_{_\sun}^2$ dependence in Equation~(\ref{eqn:mass}).  This Monte Carlo composite distribution is discussed further in \S\ref{res:mass_dist}.

\subsubsection{Mass Completeness Function}\label{meth:complete}

\begin{figure}[!t]
  \centering
  \includegraphics[width=3.1in]{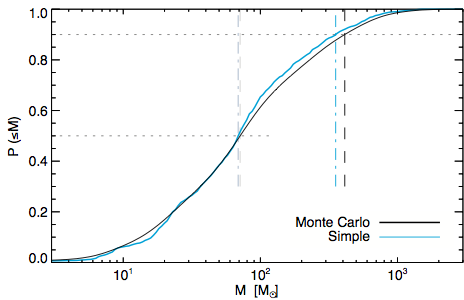}
  \caption{Computed mass completeness levels.  Cumulative distribution of the mass corresponding to $5\times$~rms flux-density noise and the source distance from the posterior DPDF.  The cyan curve represents the `Simple' distribution, using the appropriate single distance estimator from the DPDF \citep[see][]{EllsworthBowers:2013}.  The black curve is the aggregate of $10^3$ Monte Carlo realizations, with each realization randomly drawing a distance from each object's posterior DPDF and dust temperature from the lognormal distribution discussed in \S\ref{res:catalog}.  Dotted horizontal lines identify 50\% and 90\% completeness.  Light vertical dashed and dot-dashed lines identify the 50\% completeness levels for both distributions, and the dark vertical lines mark the 90\% completeness levels.}
  \label{fig:complete}
\end{figure}

(Sub-)Millimeter continuum surveys of thermal dust emission are flux-limited, so Malmquist bias must be addressed when analyzing the mass distribution of detected objects.  It is relatively straightforward to identify flux-density completeness levels, but since higher mass (and therefore brighter) objects are detectable through a larger Galactic volume than low-mass objects, it is imperative to define the mass criteria at which a survey is complete to some level.  While small surveys of isolated regions at known distances \citep[\eg][]{Ridge:2006a,Enoch:2006,Rosolowsky:2008} can directly estimate strict mass completeness levels, heterogenous surveys of the entire Galactic plane must be content to have a more complex criterion.

We cast the issue of completeness levels in terms of a mass completeness \emph{function}, which describes the fraction of objects in the survey area at any given mass that \emph{should} be detected.  Mass completeness begins with flux-density completeness, usually defined in terms of the rms noise in survey maps.  For instance, the BGPS V1 catalog is 99\% complete at $5\times$ the rms noise flux density for any given field \citep[][]{Rosolowsky:2010}, a value that is unchanged in version~2 (G13).  While surveys that do not have uniform noise properties across the Galactic plane pose an additional complication to the process, variable backgrounds throughout the Galactic plane (especially at shorter wavelengths) will also affect the completeness calculation.  The flux-density completeness level must, therefore, be estimated on a field-by-field\footnote{The BGPS has a variable rms noise from field to field (see G13, their Figures~1 and~2).} (or, better, source-by-source) basis.  Estimates of rms noise in survey images, however, usually do not take the confusion limit into account, so these completeness levels should be viewed as lower limits.

To begin, we compute the local flux-density completeness level for each catalog object as $5\times$ the mean of the BGPS noise maps within the object's catalog boundary \citep[see][]{Rosolowsky:2010}.  The minimum complete flux density for each object implies that any source at this location in the survey map brighter than this value will be detected $\geq99\%$ of the time.  Approximately 13\% of the objects in the Distance Catalog have flux densities below their own completeness value, meaning that they are $<5\sigma$ detections.  These objects represent a tip of the iceberg; similar objects exist in those same fields but were not recovered.

Combining the completeness flux density with the object's distance information allows an estimate of the ``minimum complete mass'' for each object, meaning that any object at the specified ($\ell,b,$\dsun) with mass higher than this value will be detected $\geq 99\%$ of the time.  Repeating this process on a source-by-source basis allows for the effective marginalization over the variable noise properties across the Galactic plane.  The mass completeness function is therefore the cumulative distribution of the minimum complete mass for all objects in the sample.

Figure~\ref{fig:complete} illustrates the mass completeness functions computed using the two techniques described in Figure~\ref{fig:md_ex}.  The completeness function for the single distance estimate from EB15 for each object is plotted in cyan, and an aggregation of $10^3$ Monte Carlo realizations pulling distances from each object's DPDF is shown in black.  The curves are similar, but a two-sample KS test reveals they are not drawn from the same population at greater than 99.5\% significance.  Shown as horizontal dotted lines are the 50\% and 90\% completeness levels, and the vertical lines mark the intersections of the curves with these levels.  The 50\% completeness level is $\approx 70\,M_{_\sun}$ for both curves, and the 90\% completeness level is $\approx 400\,M_{_\sun}$.  The importance of these values is in limiting conclusions about the mass distributions (or other derived physical properties) to larger masses; extending conclusions to smaller-mass objects is not supported by the available data.

\subsection{Fitting a Functional Form to the Mass Distribution}\label{meth:func}

To compare the observed mass distribution for dense molecular cloud structures with theory and other observations, we fit both power law and lognormal functions to the data.  The formulae and procedures for this fitting are discussed below, and are based on maximum-likelihood methods to avoid the pitfalls of working with binned data.

\subsubsection{Power Law}\label{meth:pl}

The use of power-law functions for describing the mass distributions of stars and their precursors goes back to the original studies of the stellar initial mass function by \citet{Salpeter:1955}.  The mass function is defined as the number of objects either per logarithmic mass interval,
\beqn
\xi(\log M) = \frac{dN}{d \log M} \propto M^{-x}~,
\eeqn
or per linear mass interval, 
\beqn\label{eqn:mfn}
\xi(M) = \frac{dN}{dM} = \frac{\xi(\log M)}{M~\ln 10} = \left(\frac{1}{M \ln 10} \right)\frac{dN}{d \log M}~,
\eeqn
where $\xi(M) \propto M^{-\alpha}$, and $x = \alpha - 1$ \citep{Chabrier:2003}.  Within this framework, the \citeauthor{Salpeter:1955} power-law index is  $x = 1.35$, $\alpha = 2.35$.  For a detailed description of this parameterization and its consequences, see \citet{Swift:2010}, \citet{Olmi:2013}, and references therein.  The mathematical nature of a power law requires a finite lower limit on the mass range over which it is a valid descriptor of the data in addition to the observational constraint that the mass distribution turns over at $M \lesssim 1\,M_{_\sun}$ \citep[\eg][]{Kroupa:2001}.

To fit a power-law function to the observed mass distribution, we use the maximum-likelihood method described by \citet{Clauset:2009}\footnote{\texttt{http://tuvalu.santafe.edu/\textasciitilde aaronc/powerlaws/}} to estimate the power-law index and range over which it is valid.  Briefly, assuming $\alpha > 1$ and $M_\mathrm{max} \gg M_\mathrm{min}$, the probability (or \emph{likelihood}) that the observed masses ($M_i$) are drawn from a power-law distribution is proportional to
\beqn\label{eqn:pl_like}
p_\mathrm{pl}\,(M\,|\,\alpha) = \prod_{i=1}^n\ \frac{\alpha - 1}{M_\mathrm{min}}\, \left(\frac{M_i}{M_\mathrm{min}} \right)^{-\alpha}~,
\eeqn
where $\alpha$ is defined as in Equation~(\ref{eqn:mfn}).  The maximum-likelihood value is computed by taking the derivative of the logarithm of the likelihood ($\mathcal{L}$)\footnote{The log-likelihood is used for mathematical expediency, as the product in Equation~(\ref{eqn:pl_like}) becomes a sum.  While the numerical value of the function is changed by taking the logarithm, the location of the maximum is not.} with respect to $\alpha$ and finding the root (\ie $\partial \mathcal{L} / \partial{\alpha} = 0$).   The power-law index estimator is
\beqn\label{eqn:alpha}
\hat{\alpha} = 1 + n \left[ \sum_{i=1}^n~\ln\frac{M_i}{M_\mathrm{min}}\right]^{-1}~.
\eeqn
The minimum mass ($M_\mathrm{min}$) for which a power law is a good descriptor of the mass distribution is computed from the data.  For each value in the mass distribution, Equation~(\ref{eqn:alpha}) is computed using that mass as $M_\mathrm{min}$.  The cumulative distribution of $M \geq M_\mathrm{min}$ is compared with the analytic power law with index $\hat{\alpha}$, and the maximum difference between them is computed (\ie the K-S test $D$-statistic); the value of $M_\mathrm{min}$ that minimizes $D$ is returned by the algorithm (see \citealp{Clauset:2009} for a complete description of the algorithm).  The results presented here were computed using a \texttt{python} translation of the \texttt{PLFIT} routine from \citeauthor{Clauset:2009}\footnote{\texttt{https://github.com/keflavich/plfit}}

\subsubsection{Lognormal}\label{meth:ln}

Predicated on the properties of supersonic turbulence and the turnover in the IMF at $M \lesssim 1\,M_{_\sun}$, the lognormal form of the mass function is also commonly fit to the mass distributions of dense molecular cloud structures \citep[\eg][]{Swift:2010,Olmi:2013}.  Analogous to the power-law case, we use a maximum-likelihood method for computing the best fit to the mass distribution.  The likelihood of the observed masses being drawn from a lognormal distribution is proportional to
\begin{eqnarray}\label{eqn:ln_like}
p_\mathrm{ln}\,(M\,|\,\mu,\sigma) &=& \prod_{i=1}^n\ \frac{C_\mathrm{ln}}{M} \exp \left[ - \frac{(\ln M - \mu)^2}{2\sigma^2} \right] \nonumber \\
&=& \prod_{i=1}^n\ \frac{C_\mathrm{ln}}{M} \exp \left[-x^2\right]~,
\end{eqnarray}
where the normalization factor
\beqn\label{eqn:ln_norm}
C_\mathrm{ln} = \sqrt{\frac{2}{\pi \sigma^2}}\ \left[\mathrm{erfc}(x_\mathrm{min}) - \mathrm{erfc}(x_\mathrm{max})   \right]^{-1}~,
\eeqn
and erfc$(x)$ is the complementary error function evaluated at $x$.  The parameters [$\mu,\sigma$] are the mean and width of the lognormal Gaussian, respectively, and [$x_\mathrm{min},x_\mathrm{max}$] are related to the minimum and maximum masses, respectively, over which the lognormal fit is valid.  The maximum-likelihood estimators $[\hat{\mu},\hat{\sigma}]$ are found by numerically solving $\nabla_{\mu,\sigma}~\mathcal{L} = 0$ (\ie numerically maximizing $\mathcal{L}$), where 
\begin{eqnarray}\label{eqn:ln_loglike}
\mathcal{L} &=& \ln\; p_\mathrm{ln}\,(M\,|\,\mu,\sigma) \nonumber \\
&=& n\,\ln\,C_\mathrm{ln}\;-\;\sum_{i=1}^n\, \ln\,M_i\;-\;\sum_{i=1}^n\, x_i^2~.
\end{eqnarray}
As with the power-law case, the limits $[M_\mathrm{min},M_\mathrm{max}]$ over which the lognormal fit best describe the data must be estimated from the data themselves.  We employ the same K-S test $D$-statistic analysis as described above, but use a numerical minimizer to explore the 2-dimensional parameter space including $M_\mathrm{max}$.  Following the algorithmic structure of \texttt{PLFIT} (\S\ref{meth:pl}), we wrote a routine in \texttt{python} to maximize Equation~(\ref{eqn:ln_loglike}) using Powell's Method inside a numerical minimization of the K-S test $D$-statistic using a downhill simplex optimization.\footnote{Both optimization steps utilized routines from the \texttt{scipy} library (\texttt{http://www.scipy.org}).}


\section{RESULTS}\label{ch4:results}

\subsection{Physical Properties of BGPS Sources}\label{res:cat}

\subsubsection{Catalog}\label{res:catalog}


\begin{deluxetable*}{ccccccccc}
  \tablecolumns{9}
  \tablewidth{0pt}
  \tabletypesize{\small}
  \tablecaption{Physical Properties of BGPS Distance Catalog Sources\label{table:physics}}
  \tablehead{
    \multicolumn{5}{c}{BGPS V2.1 Catalog Properties} & & \multicolumn{3}{c}{Derived from the PDFs\tablenotemark{a}}\\
    \cline{1-5} \cline{7-9}
    \colhead{Catalog} & \colhead{$\ell$} & \colhead{$b$} & \colhead{$S_\mathrm{int}$\tablenotemark{b}} & \colhead{$\theta_R$\tablenotemark{c}} & 
    \colhead{\dsun\tablenotemark{d}} & \colhead{$M$} & \colhead{$R$} & \colhead{$n$}\\
    \colhead{Number} & \colhead{(\degr)} & \colhead{(\degr)} & \colhead{(Jy)} & 
    \colhead{(\arcsec)} & \colhead{(kpc)} & \colhead{($\log\,M_{_\sun}$)} & \colhead{(pc)} & \colhead{(log\ \cc)}
  }
  \startdata
2235 & 7.993 & $-0.268$ & $6.23(0.49)$ & 74.3 & $11.94_{-0.42}^{+0.58}$ & $4.07_{-0.11}^{+0.12}$ & $4.31_{-0.16}^{+0.19}$ & $2.81_{-0.10}^{+0.12}$ \\
2254 & 8.187 & \phs$0.482$ & $0.61(0.15)$ & 35.9 & \phn$1.64_{-1.06}^{+0.84}$ & $1.47_{-0.71}^{+0.43}$ & $0.36_{-0.19}^{+0.11}$ & $3.58_{-0.28}^{+0.39}$ \\
2256 & 8.207 & \phs$0.190$ & $1.44(0.22)$ & 56.7 & \phn$2.46_{-1.16}^{+0.84}$ & $2.09_{-0.50}^{+0.33}$ & $0.73_{-0.28}^{+0.19}$ & $3.24_{-0.23}^{+0.28}$ \\
2261 & 8.249 & \phs$0.180$ & $2.29(0.29)$ & 77.6 & \phn$3.16_{-0.86}^{+0.64}$ & $2.48_{-0.30}^{+0.22}$ & $1.23_{-0.29}^{+0.19}$ & $2.93_{-0.17}^{+0.19}$ \\
2262 & 8.263 & \phs$0.168$ & $2.82(0.32)$ & 79.8 & \phn$2.68_{-0.96}^{+0.74}$ & $2.43_{-0.38}^{+0.27}$ & $1.09_{-0.37}^{+0.24}$ & $3.05_{-0.19}^{+0.22}$ \\
2265 & 8.281 & \phs$0.164$ & $1.00(0.15)$ & 23.8 & \phn$3.06_{-0.88}^{+0.66}$ & $2.09_{-0.32}^{+0.23}$ & $0.36_{-0.09}^{+0.06}$ & $4.13_{-0.18}^{+0.19}$ \\
2295 & 8.545 & $-0.342$ & $1.00(0.20)$ & 35.5 & \phn$4.30_{-0.78}^{+0.60}$ & $2.36_{-0.18}^{+0.16}$ & $0.75_{-0.09}^{+0.06}$ & $3.44_{-0.22}^{+0.21}$ \\
2296 & 8.551 & $-0.296$ & $0.25(0.10)$ & \nodata & $12.20_{-0.40}^{+0.48}$ & $2.72_{-0.21}^{+0.19}$ & \nodata & \nodata \\
2297 & 8.579 & $-0.344$ & $0.86(0.18)$ & 37.3 & \phn$4.04_{-0.56}^{+0.46}$ & $2.25_{-0.21}^{+0.18}$ & $0.73_{-0.10}^{+0.08}$ & $3.35_{-0.16}^{+0.17}$ \\
2300 & 8.669 & $-0.406$ & $0.44(0.11)$ & \nodata & $11.80_{-0.36}^{+0.44}$ & $2.92_{-0.16}^{+0.16}$ & \nodata & \nodata  \enddata
  \tablenotetext{a}{See \S\ref{meth:phys}.}
  \tablenotetext{b}{The source-integrated $\lambda = 1.1$~mm flux density, uncertainties in parentheses.}
  \tablenotetext{c}{The deconvolved radius of the catalog object, computed using Equation~(\ref{eqn:theta_r}).}
  \tablenotetext{d}{The appropriate distance estimate from EB15; \dbar\ for objects near the tangent point and \dml\ otherwise.}
  \tablecomments{This table is available in its entirety in a machine-readable form.}
\end{deluxetable*}

We present a physical properties catalog for the 1,710 sources in the Distance Catalog.  The mass, physical radius, and number density for each object were computed from the Monte-Carlo probability density functions as described in \S\ref{meth:phys}, where a dust temperature was drawn for each realization from a lognormal distribution with a mean of 20\,K and full-width at half-maximum of 8\,K \citep[][]{Battersby:2011}.  This choice of temperature distribution was recently affirmed by the analysis of analogous dense molecular cloud structures from the ATLASGAL survey by \citet{Wienen:2015}.  We note that the study of \citet{Merello:2015} combining $\lambda = 1.1$~mm and $\lambda = 350$\,\micron\ data finds a lower mean temperature for BGPS sources ($T_d = 13.3$\,K for dust spectral index $\beta = 2.0$, and $T_d = 16.3$\,K for $\beta = 1.7$).  Using a mean $T_d=20$\,K would lead to an underestimate of the mass of each object by a factor of 1.3 to 1.8 for the $\beta = 1.7$ and $\beta = 2.0$, respectively, of \citet{Merello:2015}.  While the computed mass and number density values are affected by this uncertainty in temperature, Monte Carlo simulations of BGPS source mass distributions by \citet{Schlingman:2011} showed the power-law slope of the mass function (\S\ref{res:plwhole}) is not very sensitive to the choice of underlying dust temperature distribution.

The pdfs for mass, physical radius, and number density were constructed in log space to capture the spread in values over orders of magnitude, and error bars were computed to enclose 68.3\% of the total probability, where the endpoints occur at equal probability (see the discussion of DPDF error bars in EB15).  The error bars approximate the Gaussian $\pm1 \sigma$ region, but account for the asymmetric nature of the pdfs.  The catalog of physical properties is shown in Table~\ref{table:physics}, along with relevant information from the BGPS V2.1 catalog (G13) and heliocentric distance (EB15).

\begin{figure}[!t]
  \centering
  \includegraphics[width=3.1in]{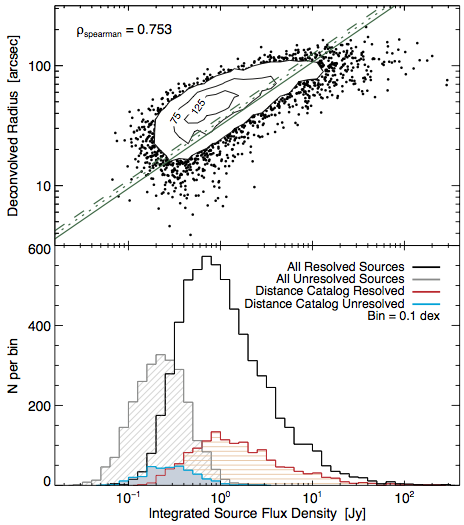}
  \caption[Brief analysis of resolved versus unresolved BGPS V2 sources.]{Brief analysis of resolved versus unresolved BGPS V2 sources.  \emph{Top}: Scatter plot of deconvolved radius versus integrated $\lambda = 1.1$~mm flux density.  Evenly spaced contours represent the density of points per 0.1~dex~$\times$ 0.1~dex bin.  Only resolved sources are shown.  Plotted in green are Equation~(\ref{eqn:flux_rad}) evaluated at \dsun~= 1\,kpc (solid), 5\,kpc (dotted), and 20\,kpc (dashed) (see \S\ref{disc:larson} for discussion).  \emph{Bottom}: Histograms of integrated $\lambda = 1.1$~mm flux density.  Black represents all resolved sources (\ie the projection of the points in the top panel), and the gray diagonally hashed histogram represents the unresolved sources.  Red horizontally hashed and blue solid histograms represent the resolved and unresolved populations, respectively, of the Distance Catalog (EB15).}
  \label{fig:flux_rad}
\end{figure}

Of particular note in the catalog are the 20\% of sources in the BGPS Distance Catalog whose angular extents on the sky do not yield real solutions to Equation~(\ref{eqn:theta_r}) for $\theta_R$.  A brief analysis of these objects in the context of the flux-density distribution for all catalog sources reveals a strong trend whereby objects with non-physical $\theta_R$ are preferentially dim (Figure~\ref{fig:flux_rad}).  The gray and blue histograms represent the ``unresolved'' sources in the full V2.1 catalog and the Distance Catalog, respectively, while the black and red histograms represent the resolved sources in the same groups.

The obvious correlation in the data points in the top panel (Spearman rank correlation coefficient $\rho = 0.753$) is a direct consequence of a universal scaling relationship in the dense interstellar medium first identified by \citet{Larson:1981} whereby the mass of a molecular cloud structure is roughly proportional to the square of its radius.  Section~\ref{disc:larson} explores this relationship in more detail, culminating in Equation~(\ref{eqn:flux_rad}), which is plotted as green lines in Figure~\ref{fig:flux_rad} for three different heliocentric distances.

The \citeauthor{Larson:1981} relationships in the figure extend to low flux-density, but since the cataloging process truncates source boundaries where the flux density meets the local rms value \citep{Rosolowsky:2010}, source masks do not extend to the theoretical zero-flux-density isophot.  The resulting catalog source may have an extent in the map smaller than the beam size, leading to a complex solution to Equation~(\ref{eqn:theta_r}).  To further aggravate the situation, the beam deconvolution step assumes that sources are spherical (or at least have circular projections on the sky), which is invalid for many of the fainter filamentary structures throughout the Galactic plane \citep[\eg][]{Molinari:2010b}.  Future cataloging routines may be constructed that divide emission between filamentary and compact sources, and may decrease the number of BGPS sources with invalid deconvolved angular extent.  For the present, however, unresolved sources are excluded from the following analyses that rely on physical radius or number density, leaving a sample of $N=1369$ objects with a tabulated physical size.

\subsubsection{Ensemble Physical Property Distributions}\label{res:ensemble}

\begin{figure}[!t]
  \centering
  \includegraphics[width=3.1in]{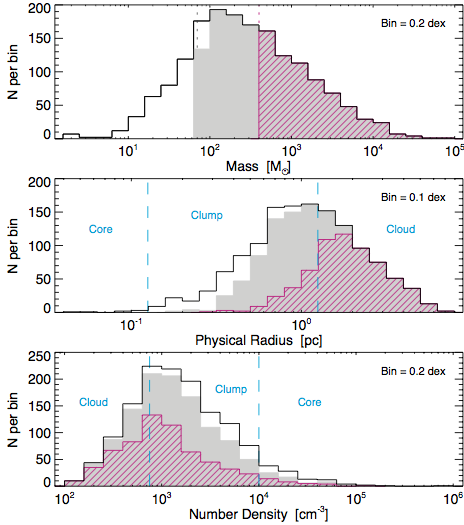}
  \caption{Physical property histograms for the Distance Catalog, computed using the maximum-likelihood values from the Monte Carlo method described in \S\ref{meth:phys}.  The gray filled and magenta hashed histograms represent sources above the 50\% and 90\% mass completeness levels, respectively (see \S\ref{meth:complete}).  \emph{Top}: Source mass distribution.  Vertical lines mark the locations of the 50\% (gray) and 90\% (magenta) completeness levels.  \emph{Middle}: Physical radius distribution.  The vertical dashed lines identify the boundaries between cloud / clump / core as used by \citet{Dunham:2011c}.  \emph{Bottom}: Number density distribution; vertical lines as in the middle panel.}
  \label{fig:phys_cat}
\end{figure}

The ensemble physical property distributions from Table~\ref{table:physics} are shown in Figure~\ref{fig:phys_cat} to illustrate the range of BGPS objects in the Distance Catalog.  This subsample is generally representative of the entire BGPS catalog, except that detection in a molecular transition line that traces dense gas (\eg \hcop(3-2)) is strongly correlated with continuum flux density; dim sources are largely absent from the Distance Catalog (see EB15 for a complete discussion).  In all three panels, sources above the 50\% (70\,\msun) and 90\% (400\,\msun) mass completeness levels are indicated by gray solid and magenta hashed histograms, respectively.

The mass distribution (top panel) peaks around 100\,\msun, but the completeness levels indicated by the filled histograms imply that the location of the distribution peak is observationally biased.  The location of the 50\% completeness level demonstrates that  the predicted turnover at low mass is not constrained by the current observations.  Physical radius is illustrated in the middle panel of Figure~\ref{fig:phys_cat}, with vertical dashed lines marking plausible boundaries between molecular clouds (large $R$), clumps (intermediate $R$), and individual cores.  We place approximate radius boundary markers at 1.25\,pc between cloud and clump and 0.125\,pc between clump and core, the same boundaries used by D11 which were based on the general guidelines from \citet[][]{Bergin:2007}.  These divisions, while rooted in physical distinctions based on virial ratio and capacity for Jeans fragmentation, are flexible and the boundaries are easily blurred by the continuum of structure in molecular cloud complexes (see \S\ref{disc:larson}), and by variations in Galactic environment.  Note, however, the near complete lack of Distance Catalog objects with radii $\lesssim 0.1$\,pc).  From the relationship in Figure~\ref{fig:flux_rad}, smaller sources have lower flux density and are therefore less likely to be included in the Distance Catalog (EB15).

Definitions notwithstanding, the typical size scale of BGPS-identified dense molecular cloud structures is clearly in the vicinity of 1\,pc and may encompass multiple true but unresolved molecular cloud clumps.  The mass completeness subsets illustrate the effects of Malmquist bias (coupled with the \citeauthor{Larson:1981} relationship connecting mass and physical radius) whereby the 90\% complete sample is nearly all cloud-scale ($R \gtrsim 1$\,pc) objects, and the BGPS does not detect any large ($R\geq2$\,pc), lower-mass objects ($M \leq 400\,M_{_\sun}$).  This places a rough constraint on the minimum number density ($n \gtrsim 230$\,\cc, Equation~\ref{eqn:nden}) detectable by the BGPS for large-scale objects.

The bottom panel of Figure~\ref{fig:phys_cat} depicts the number density distribution of resolved sources, which spreads across a wide dynamic range.  Approximate number density boundaries at $n = 750$\,\cc\ between cloud and clump, and $n = 10^4$\,\cc\ between clump and core (D11) are indicated by vertical dashed lines.  Whereas there is nearly complete overlap between high-mass and large-radius objects (`clouds'; middle panel), there exist a handful of low-mass ($M \leq 70$\,\msun) objects with low computed number density (`clouds'; bottom panel).  These objects, along with the spread of higher-mass objects ($M \geq 400$\,\msun) across classifications in the bottom panel insinuate that number density computed via Equation (\ref{eqn:nden}) for BGPS sources is not a reliable indicator of physical classification.  It is possible that non-uniform geometry of sources is causing the spread of the completeness samples over a large range, or that the algorithm by which the cataloging routine decomposes emission into discrete sources is doing so in a manner inconsistent with the underlying physical structure in molecular clouds (\eg  by assuming spherical clouds when computing $N$(\htwo)).

\subsubsection{Physical Properties as a Function of Heliocentric and Galactocentric Position}

\begin{figure}[!t]
  \centering
  \includegraphics[width=3.1in]{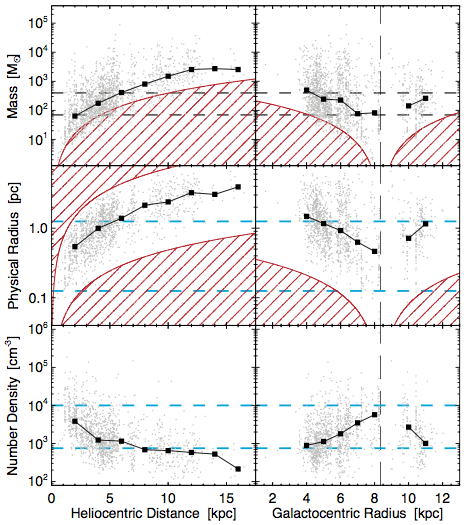}
  \caption[Physical properties as a function of heliocentric distance and Galactocentric radius.]{Physical properties as a function of heliocentric distance and Galactocentric radius.  Grey dots mark individual objects from the BGPS Distance Catalog, and black squares identify the median values in 2-kpc (\dsun) or 1-kpc (\rgal) bins.  \emph{Top Row}: Computed mass for each object.  The hashed region represents the mean $\lesssim5\sigma$ flux-density completeness level as a function of heliocentric distance.  For the right panel, this region is computed for \rgal\ along the $\ell = 0\degr, 180\degr$ line.  Horizontal dashed lines mark the 50\% and 90\% completeness levels (as in Figure~\ref{fig:phys_cat}).  \emph{Middle Row}: Physical radius.  The lower hashed region corresponds to $\theta_R = 10\arcsec$, while the upper marks $\theta_R = 150\arcsec$.  For the right panel, the lower region is again computed for \rgal\ along $\ell = 0\degr, 180\degr$.  Horizontal dashed lines mark the approximate cloud / clump / core boundaries (as in Figure~\ref{fig:phys_cat}).  \emph{Bottom Row}: Number density.  Horizontal dashed lines as above.}
  \label{fig:phys_relate}
\end{figure}

A major challenge in characterizing the ensemble  physical property distributions of detected molecular cloud structures is separating observational and data-processing systematic effects from underlying physical relationships, as systematic effects on the measurement of physical properties can lead to significant biases in the derived quantities.  Understanding the physics of dense molecular cloud structures requires well-constrained distance estimates.  It is instructive, therefore, to analyze the functional dependences.

The left panels of Figure~\ref{fig:phys_relate} illustrate the dependence on heliocentric distance of the mass, physical radius, and number density for sources in the Distance Catalog.  Sources are plotted as gray dots, with the median value per 2-kpc bin marked by black squares.  The systematic dependence on heliocentric distance can be clearly seen in the median values as $M\sim d_{_\sun}^2$, $R\sim d_{_\sun}$, and $n\sim d_{_\sun}^{-1}$, following Equations~(\ref{eqn:mass}), (\ref{eqn:prad}), and (\ref{eqn:nden}), respectively.  For the distribution of mass (top left panel), the hashed region identifies the mass associated with the mean $\leq5\sigma$ flux density completeness level ($S_\mathrm{complete} = 0.30$~Jy, Section~\ref{meth:complete}).  Especially given the variable rms noise from field to field in the BGPS, sources may be detected in this region but completeness is not assured.  The horizontal dashed lines mark the 50\% and 90\% completeness levels (as in Figure~\ref{fig:phys_cat}).  The large collection of sources around \dsun~$= 5$\,kpc has a median mass $\approx 300\,M_{_\sun}$, whereas the median grows to $\approx 2000\,M_{_\sun}$ for objects $\gtrsim 10$\,kpc.  

The physical radius plot (middle left panel) shows similar effects, with the hashed regions marking $\theta_R \leq 10\arcsec$ (typical minimum angular radius) and $\theta_R \geq 150\arcsec$ (approximate maximum recoverable radius; see G13).  The horizontal dashed lines represent the $R = 0.125$\,pc and $R = 1.25$\,pc divisions between physical object categories discussed above.  Finally, the number density distribution (bottom left panel) displays the expected trend, except that the median values remain somewhat constant between 500~cm$^{-3} \leq n \leq 1000$\,\cc\ over a large range of heliocentric distance.  Since this flattening is not observed for physical radius (which also depends linearly on \dsun), this range of $n$ may represent a typical value for BGPS dense molecular cloud structures, subject to the geometric and cataloging caveats discussed above.

The right panels of Figure~\ref{fig:phys_relate} illustrate the dependence of these physical properties on Galactocentric radius.  The vertical dashed line identifies our adopted $R_0 = 8.34$\,kpc \citep{Reid:2014} for reference.  Black squares mark the median values in 1-kpc bins, and roughly show the systematic effects of heliocentric distance (\ie objects far from \rgal~= $R_0$ are at large \dsun).  Gaps in the black function represent bins with $N \leq 20$~sources, where the median value is skewed by outliers.  The gap in sources (gray dots) at \rgal~= $8.5-9.5$\,kpc is pointed out in EB15 as arising from a unique combination of BGPS coverage region and a kinematic avoidance zone (where kinematic distances are unreliable in the face of non-circular motions about the Galactic center), and is not likely a true feature of the underlying Galactic dense molecular gas distribution.  Many of the dense molecular cloud structures observed in the Molecular Ring / Scutum-Centarus Arm feature at \rgal~$= 4-5$\,kpc, \dsun~$= 3-6$\,kpc, correspond the physical properties typical of molecular cloud clumps.  

The systematic effects of heliocentric distance will obscure some of the underlying physical trends in properties as a function of Galactocentric radius.  For instance, the spike in median number density near \rgal~$= R_0$ is due to the population of nearby (\dsun~$\lesssim 2$\,kpc) core-scale objects.  The observed decrease in mass for dense molecular cloud structures from \rgal~= 4\,kpc to \rgal~= 7\,kpc is consistent with the trend observed for GMCs by \citet{RomanDuval:2010}, but is likely primarily an observational bias.

\subsection{Mass Function Fits}\label{res:mfns}

\subsubsection{Mass Distribution Trials}\label{res:mass_dist}

The full information available from the MPDFs (and ultimately the DPDFs from EB15) allow for the utilization of a large number of Monte Carlo trials to fit functional forms to the dense molecular cloud structure mass distribution.  For each trial, a single value was drawn from the MPDF of each object (\S\ref{meth:mass}) to create an independent realization of the mass distribution of the Distance Catalog.  Both a power law (\S\ref{meth:pl}) and lognormal (\S\ref{meth:ln}) function were fit to each realization, and the fit parameters recorded.  This process was repeated $N = 20,000$ times, a value chosen to balance appropriately sampling the parameter spaces for the functional fits with computation time.  The collected fit parameters from the Monte Carlo process were then plotted as two-dimensional histograms to identify joint confidence intervals.  Below, we discuss the results of the power-law (\S\ref{res:plwhole}) and lognormal (\S\ref{res:lnwhole}) fits to the entire data set.  Attempts to subdivide the Distance Catalog into astrophysically meaningful subsets are described in Appendix~\ref{res:types}, but most fits resemble those of the entire sample. Maximum-likelihood fit parameter values for all fits are listed in Table~\ref{table:mfn_fit_values}.

\begin{figure*}[!t]
  \centering
  \includegraphics[width=6.5in]{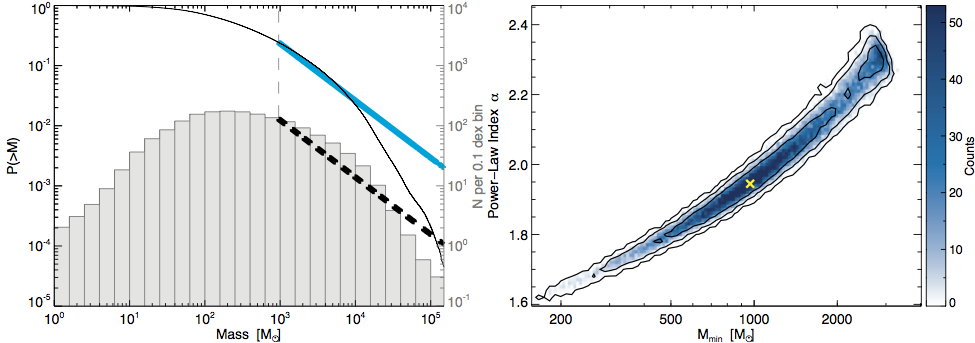}
  \caption[Power-law fit to the aggregate mass distribution.]{Power-law fit to the aggregate mass distribution.  \emph{Left}:  The complementary cumulative distribution function (CCDF, solid black curve) and differential mass distribution (gray histogram) for the Distance Catalog.  This distribution is the aggregate of 20,000 Monte Carlo realizations of the mass function, normalized to a sum of $N=1710$ to create an average mass distribution that marginalizes over the uncertainties in the mass of each object.  The solid cyan and dashed black curves represent the maximum-likelihood power-law index (from right panel), and the vertical gray dashed line marks the maximum-likelihood minimum mass for which a power law describes the data.  \emph{Right}: The two-dimensional histogram of the power-law fit parameters for each of the 20,000 Monte Carlo trials, whereby a power law was fit to a single realization of the mass distribution.  Contours enclose 68.3\% (highest), 95.4\%, and 99.7\% of the points in the plane.  The yellow $\mathsf{X}$ marks the largest concentration of points, or the joint maximum-likelihood values of $\hat{\alpha}$ and $M_\mathrm{min}$.}
  \label{fig:pl_whole}
\end{figure*}

\subsubsection{Power-Law Fit}\label{res:plwhole}

The collected power-law fits to the complete mass distribution are shown in Figure~\ref{fig:pl_whole}.  The left panel illustrates the mass distribution both as the complementary cumulative distribution function (CCDF, solid black)\footnote{The CCDF is simply 1 - CDF, where CDF is the typical cumulative distribution function.  The CCDF measures the probability of finding an object with mass \emph{greater} than a given value, whereas the CDF describes the probability of finding an object with mass \emph{less} than a given value.  The CCDF is used to illustrate the distribution in a manner robust against fluctuations caused by finite sample size \citep[][]{Clauset:2009}.} and as the differential mass distribution (gray histogram).  Both distributions represent the aggregate of 20,000 Monte Carlo realizations of a single mass distribution, normalized so the sum equals 1,710 objects.  To smooth the effects of small sample size and utilize the full amount of information encoded in the collected set of MPDFs, the aggregate is shown rather than any single one of the Monte Carlo trials.  The power-law fits (derived from the data in the right panel) are shown for the CCDF (cyan line) and the differential mass distribution (black dashed line).

The power-law fitting algorithm (\texttt{PLFIT}) returns two parameters, the power-law index $\hat{\alpha}$ and the minimum mass over which the data may be described by a power law ($M_\mathrm{min}$).  The two-dimensional histogram of the 20,000 parameter pairs from the Monte Carlo trials are shown in the right panel of Figure~\ref{fig:pl_whole}.  Color scale intensity indicates the number of points in each pixel of the histogram, and contours enclose 68.3\% (highest contour), 95.4\%, and 99.7\% of the parameter pairs, meant to illustrate joint confidence regions.  The high degree of correlation between $\hat{\alpha}$ and $M_\mathrm{min}$ is indicative of the continuously steepening nature of the mass distribution shown in the left panel (where a power law is a straight line).  The yellow $\mathsf{X}$ in the right panel marks the peak of the two-dimensional histogram, or the joint maximum-likelihood parameter values ($\hat{\alpha} = 1.95,\ M_\mathrm{min} = 966\,M_{_\sun}$).  This minimum mass is above the 90\% completeness level described in \S\ref{meth:complete} ($\approx 400\,M_{_\sun}$), meaning the power-law fit is not affected by Malmquist bias, although other selection effects may be at play.  Note the secondary peak of points near $\hat{\alpha} = 2.3,\ M_\mathrm{min} = 3000\,M_{_\sun}$ in the right panel; there appears to be a sharp steepening of the mass distribution at the high-mass end (see \S\ref{disc:steep} for discussion).  The distribution truncates around $M_\mathrm{min} = 3500\,M_{_\sun}$ due to a requirement in the fitting algorithm for a minimum number of data points required for a robust fit \citep[see][]{Clauset:2009}.  This truncation is a general feature of the astrophysical subset fits discussed in Appendix~\ref{res:types}.

\subsubsection{Lognormal Fit}\label{res:lnwhole}

\begin{figure*}[!t]
  \centering
  \includegraphics[width=6.5in]{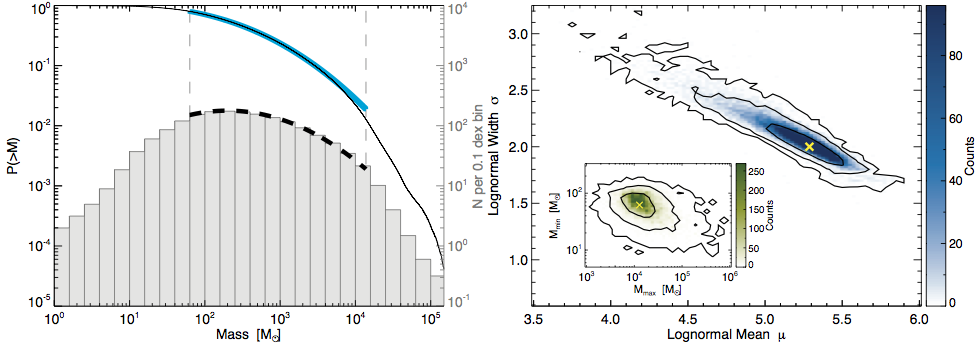}
  \caption{Lognormal fit to the aggregate mass distribution.  \emph{Left}:  The CCDF (solid black curve) and differential mass distribution (gray histogram) are identical to Figure~\ref{fig:pl_whole}.  The blue and black dashed curves represent the maximum-likelihood lognormal fit (from the right panel), and the vertical gray dashed lines mark the maximum-likelihood minimum and maximum mass for which a lognormal describes the data.  \emph{Right}: The two-dimensional histogram of the lognormal fit parameters [$\mu,\sigma$] for each of the 20,000 Monte Carlo realizations of the mass distribution.  Contours enclose 68.3\% (highest), 95.4\%, and 99.7\% of the points in the plane.  The yellow $\mathsf{X}$ marks the largest concentration of points, or the joint maximum-likelihood values.  The inset shows the parameter space distribution for the pair [$M_\mathrm{min},M_\mathrm{max}$].}
  \label{fig:ln_whole}
\end{figure*}

The continuously steepening nature of the mass CCDF makes fitting or interpreting a power law function for the entire data set difficult at best.  Furthermore, distance inhomogeneities in large-scale surveys such as the BGPS tend to drive the composite observed mass distribution toward a lognormal (see \S\ref{res:inevit}).  We applied our lognormal fitting routine to the $N=20,000$ Monte Carlo realizations of the mass distribution, and the resulting fit parameter distributions are shown in Figure~\ref{fig:ln_whole}.  The CCDF and differential mass distributions in the left panel are identical to Figure~\ref{fig:pl_whole} (\emph{left}).  The functional fits are drawn from the right panel as was done with the power law, where the cyan line fits the CCDF and the dashed black line fits the differential mass distribution.

Whereas the power-law fit returns two parameters, the lognormal fitting optimizes the functional fit over four parameters: the lognormal characteristic [$\mu,\sigma$], and also both ends of the mass range [$M_\mathrm{min},M_\mathrm{max}$] over which the lognormal fit is valid.  For computational expediency, we divided the fitting into a nested structure following the recipe from \citet[][Appendix B]{Olmi:2013}.  The lognormal characteristic is fit for a given pair of mass range values, then the mass range is adjusted to find the best match between the model CCDF and the CCDF constructed from the source masses.  Shown in the right panel of Figure~\ref{fig:ln_whole} are the distributions of lognormal best-fit parameters paired in this fashion.  The more fundamental is the [$\mu,\sigma$] plane, shown in the background, since these define the position and shape of the function.  As with the power-law fit, the two-dimensional histogram of the 20,000 Monte-Carlo trial fit parameters is shown, with contours representing the same enclosed fractions of points.  Both parameters are tightly constrained, with $\hat{\mu} = 5.30_{-0.20}^{+0.15}$ ($160\,M_{_\sun} \leq M_\mu \leq 230\,M_{_\sun}$) and $\hat{\sigma} = 2.00_{-0.08}^{+0.16}$, where the uncertainties represent the marginalized 68.3\% confidence intervals.  The joint maximum-likelihood values (marginalized over [$M_\mathrm{min},M_\mathrm{max}$], yellow $\mathsf{X}$) occur at $\mu = 5.30,\ \sigma = 2.01$, and are reflected in the solid cyan and dashed black curves in the left panel.  Since the peak occurs at $M_\mu = e^{5.30} =$ 200\,\msun, below the 400\,\msun\ 90\% completeness level, it may be biased and should be viewed as an upper limit.

Inset in the right panel is the two-dimensional histogram of the [$M_\mathrm{min},M_\mathrm{max}$] plane, showing the relationship between the bounding parameters.  There is scant correlation, except that the spread along the diagonal from top-left to bottom-right represents expansion and contraction about a central value.  The optimum values of these parameters (marginalized over [$\mu,\sigma$]) are indicated by the yellow $\mathsf{X}$ in the inset ($M_\mathrm{min} = 70\,M_{_\sun},\ M_\mathrm{max} = 12000\,M_{_\sun}$), and are marked in the left panel by vertical gray dashed lines.  This minimum mass extends down to the 50\% mass completeness level ($\approx 70\,M_{_\sun}$), and its location may not be entirely reliable, as the peak of the fitted lognormal function always falls below the 90\% completeness level ($\mu < 6.0 = \ln\,400\,M_{_\sun}$).

Comparison of the lognormal fit in Figure~\ref{fig:ln_whole} with the power-law fit in Figure~\ref{fig:pl_whole} yields some interesting notions.  As expected, the minimum mass for the power-law fit is well above the peak of the lognormal distribution.  In fact, the 99.7\% contour in Figure~\ref{fig:pl_whole} (\emph{right}) terminates at about this value.  Furthermore, the curvature of the mass distribution is well-matched to a lognormal function over nearly 2.5 orders of magnitude.  The continual curve of the mass distribution may be an effect of observational and cataloging biases against the highest mass sources (see \S\ref{disc:steep}).


\begin{deluxetable*}{ccccccccccc}
  \tablecolumns{11}
  \tablewidth{0pt}
  \tabletypesize{\footnotesize}
  \tablecaption{Mass Distribution Functional Fit Parameter Values\label{table:mfn_fit_values}}
  \tablehead{
    \colhead{} & \multicolumn{2}{c}{} & \colhead{} & \multicolumn{2}{c}{Power-Law Fit} & \colhead{} & \multicolumn{4}{c}{Lognormal Fit} \\
    \cline{5-6} \cline{8-11}
    \colhead{Cut} & \multicolumn{2}{c}{Subset\tablenotemark{a}} & \colhead{} & \colhead{$\hat{\alpha}$} & \colhead{$M_\mathrm{min}$} & \colhead{} & \colhead{$\hat{\mu}$} & \colhead{$\hat{\sigma}$} & \colhead{$M_\mathrm{min}$\tablenotemark{b}} & \colhead{$M_\mathrm{max}$\tablenotemark{b}} \\
    \cline{2-3}
    \colhead{Type} & \colhead{Name} & \colhead{$N$} & \colhead{} & \colhead{} & \colhead{($\log\,M_{_\sun}$)} & \colhead{} & \colhead{($\ln\,M_{_\sun}$)} & \colhead{($\ln\,M_{_\sun}$)}  & \colhead{($\log\,M_{_\sun}$)} & \colhead{($\log\,M_{_\sun}$)}
  }
  \startdata
     & Distance Catalog & 1710 & & $1.94_{-0.10}^{+0.34}$ & $2.98_{-0.06}^{+0.46}$ & & $5.30_{-0.20}^{+0.15}$ & $2.00_{-0.08}^{+0.16}$ & $1.84_{-0.16}^{+0.08}$ & $4.08_{-0.12}^{+0.24}$ \\
  \\[-7pt]
  \hline
  \\[-7pt]
\dsun & Nearby & 1328 & & $2.01_{-0.09}^{+0.11}$ & $2.79_{-0.19}^{+0.20}$ & & $4.85_{-0.08}^{+0.15}$ & $1.72_{-0.02}^{+0.08}$ & $1.84_{-0.12}^{+0.12}$ & $4.14_{-0.24}^{+0.18}$ \\
 & Distant & 382 & & $1.93_{-0.05}^{+0.19}$ & $3.33_{-0.29}^{+0.01}$ & & $7.10_{-0.05}^{+0.05}$ & $1.48_{-0.08}^{+0.08}$ & $1.84_{-0.08}^{+0.12}$ & $4.14_{-0.18}^{+0.24}$ \\
  \\[-7pt]
  \hline
  \\[-7pt]
\rgal & Sct-Cen Arm & 866 & & $1.98_{-0.11}^{+0.09}$ & $2.96_{-0.16}^{+0.18}$ & & $5.38_{-0.25}^{+0.18}$ & $1.90_{-0.20}^{+0.12}$ & $1.84_{-0.08}^{+0.12}$ & $3.96_{-0.18}^{+0.30}$ \\
 & Sgr / Local Arms & 610 & & $2.02_{-0.28}^{+0.02}$ & $3.19_{-0.42}^{+0.04}$ & & $4.90_{-1.08}^{+0.88}$ & $2.82_{-0.64}^{+0.52}$ & $1.84_{-0.08}^{+0.12}$ & $3.96_{-0.18}^{+0.18}$ \\
 & Sgr Arms & 330 & & $1.97_{-0.24}^{+0.01}$ & $3.17_{-0.35}^{+0.03}$ & & $4.90_{-1.08}^{+0.88}$ & $2.82_{-0.64}^{+0.52}$ & $1.84_{-0.08}^{+0.12}$ & $3.96_{-0.18}^{+0.18}$ \\
 & Per / Outer Arms & 234 & & $1.94_{-0.10}^{+0.03}$ & $2.56_{-0.13}^{+0.04}$ & & $5.40_{-0.15}^{+0.25}$ & $1.70_{-0.22}^{+0.06}$ & $1.84_{-0.16}^{+0.12}$ & $3.96_{-0.18}^{+0.30}$ \\
  \\[-7pt]
  \hline
  \\[-7pt]
$R$ & Clump / Core & 807 & & $1.90_{-0.06}^{+0.26}$ & $3.24_{-0.44}^{+0.01}$ & & $5.32_{-0.32}^{+0.25}$ & $2.12_{-0.18}^{+0.26}$ & $1.84_{-0.16}^{+0.12}$ & $4.14_{-0.24}^{+0.18}$ \\
 & Cloud & 562 & & $1.86_{-0.24}^{+0.05}$ & $2.90_{-0.59}^{+0.18}$ & & $5.25_{-0.55}^{+0.30}$ & $2.22_{-0.24}^{+0.34}$ & $1.84_{-0.16}^{+0.12}$ & $4.14_{-0.18}^{+0.36}$ \\
  \\[-7pt]
  \hline
  \\[-7pt]
$n$ & Clump / Core & 993 & & $1.87_{-0.10}^{+0.17}$ & $2.91_{-0.10}^{+0.41}$ & & $5.35_{-0.50}^{+0.15}$ & $2.22_{-0.20}^{+0.26}$ & $1.84_{-0.12}^{+0.12}$ & $4.14_{-0.24}^{+0.12}$ \\
 & Cloud & 376 & & $1.76_{-0.06}^{+0.18}$ & $3.03_{-0.48}^{+0.02}$ & & $5.35_{-0.55}^{+0.27}$ & $2.12_{-0.38}^{+0.26}$ & $1.84_{-0.16}^{+0.12}$ & $4.14_{-0.24}^{+0.24}$ \\
  \\[-7pt]
  \hline
  \\[-7pt]
$\Sigma$ & Mean $\leq 120\,M_{_\sun}$\,pc$^{-2}$ & 1444 & & $2.38_{-0.36}^{+0.03}$ & $3.04_{-0.38}^{+0.03}$ & & $5.10_{-0.05}^{+0.12}$ & $1.68_{-0.10}^{+0.04}$ & $1.84_{-0.20}^{+0.12}$ & $4.00_{-0.24}^{+0.30}$ \\
 & Mean $\geq 120\,M_{_\sun}$\,pc$^{-2}$ & 358 & & $2.06_{-0.38}^{+0.02}$ & $3.44_{-0.45}^{+0.04}$ & & $7.20_{-0.05}^{+0.10}$ & $1.72_{-0.08}^{+0.18}$ & $1.84_{-0.08}^{+0.16}$ & $4.18_{-0.18}^{+0.30}$ \\
 & Peak $\leq 120\,M_{_\sun}$\,pc$^{-2}$ & 392 & & $2.15_{-0.25}^{+0.01}$ & $2.14_{-0.34}^{+0.01}$ & & $3.62_{-0.20}^{+0.85}$ & $1.70_{-0.40}^{+0.02}$ & $1.84_{-0.16}^{+0.12}$ & $4.00_{-0.06}^{+0.30}$ \\
 & Peak $\geq 120\,M_{_\sun}$\,pc$^{-2}$ & 1410 & & $1.93_{-0.08}^{+0.36}$ & $3.47_{-0.50}^{+0.01}$ & & $5.85_{-0.12}^{+0.07}$ & $1.80_{-0.08}^{+0.10}$ & $1.84_{-0.12}^{+0.12}$ & $4.18_{-0.24}^{+0.18}$ \\
  \\[-7pt]
  \hline
  \\[-7pt]
 \vlsr\ Type & Dense Gas & 1372 & & $1.98_{-0.08}^{+0.31}$ & $3.07_{-0.07}^{+0.36}$ & & $5.32_{-0.15}^{+0.15}$ & $1.96_{-0.08}^{+0.14}$ & $1.84_{-0.12}^{+0.12}$ & $4.14_{-0.18}^{+0.24}$ \\
 & \thco & 422 & & $1.97_{-0.08}^{+0.05}$ & $2.59_{-0.17}^{+0.07}$ & & $4.55_{-0.28}^{+0.22}$ & $1.72_{-0.04}^{+0.14}$ & $1.84_{-0.20}^{+0.12}$ & $3.96_{-0.06}^{+0.36}$ \\
  \\[-7pt]
  \hline
  \\[-7pt]
``Protocluster'' & Blind ($\ell \leq 90\degr$) & 1513 & & $1.94_{-0.09}^{+0.34}$ & $2.97_{-0.02}^{+0.47}$ & & $5.38_{-0.22}^{+0.15}$ & $2.08_{-0.14}^{+0.16}$ & $1.84_{-0.12}^{+0.12}$ & $4.18_{-0.24}^{+0.12}$ \\
 & $R \geq 0.2$\,pc & 1331 & & $1.92_{-0.12}^{+0.11}$ & $2.99_{-0.20}^{+0.19}$ & & $5.30_{-0.30}^{+0.18}$ & $2.16_{-0.14}^{+0.22}$ & $1.84_{-0.12}^{+0.12}$ & $4.18_{-0.24}^{+0.12}$ \\
 & Mixed Protocluster & 579 & & $1.92_{-0.07}^{+0.12}$ & $2.79_{-0.13}^{+0.21}$ & & $6.45_{-0.12}^{+0.38}$ & $0.66_{-0.10}^{+0.40}$ & $2.12_{-0.12}^{+0.24}$ & $2.98_{-0.24}^{+0.36}$ \\
 & $M \geq 2000\,M_{_\sun}$ & 216 & & $2.23_{-0.06}^{+0.03}$ & $3.45_{-0.04}^{+0.02}$ & & $8.38_{-0.02}^{+0.05}$ & $0.82_{-0.12}^{+0.06}$ & $1.80_{-0.04}^{+0.04}$ & $4.30_{-0.12}^{+0.30}$ 
  \enddata
  \tablenotetext{a}{See Appendix \ref{res:types} for definitions of these subsets.}
  \tablenotetext{b}{The lognormal fitting routine tends to find mass bounds very near the initial guess ($M_\mathrm{min} = 10^2$~\msun, $M_\mathrm{max} = 10^4$~\msun), but parameter correlations show that the exact bounds have little impact on the [$\hat{\mu},\hat{\sigma}$] returned.}
\end{deluxetable*}

\subsubsection{Mass Functions of Astrophysically Motivated Subsets}\label{res:cut_stub}

Theories of star formation hold that a strong power-law behavior should be evident for dense molecular cloud structures, particularly leading toward the stellar IMF \citep[][]{Krumholz:2005,Hennebelle:2011,Hennebelle:2013b,Padoan:2011,Federrath:2012}.  The lack of a clear single power law fit to the complete Distance Catalog prompted us to fit functional forms to astrophysically motivated subsets of the data to identify homogenous sets of objects predicted by theory.  While the fits to many of the subsets, did not reveal any strongly compelling behavior (see Appendix~\ref{res:types}), there are two whose power-law fits are fairly well constrained.  The set of nearby objects (\dsun~$\leq 6.5$\,kpc) may be described by an $\hat{\alpha} = 2.0\pm0.1$ power law, and set of criteria aimed at isolating molecular cloud clumps (2\,kpc~$\leq$ \dsun~$\leq$ 10\,kpc and $M \geq 300$\,\msun) is fit by a $\hat{\alpha} = 1.9\pm0.1$ power-law.  The heliocentric distance cut removes cloud-scale objects that are more strongly affected by the angular filtering function of the BGPS data pipeline, leaving principally molecular cloud clumps, similar to the mixed criteria.

The overall lack of clean fits to either the complete Distance Catalog or most of the subsets indicates that the sample is largely homogenous and the interstellar medium is not cleanly, tightly described by simple relations.   These effects are discussed in depth in \S\ref{disc:mfn}.  The results of the studied astrophysical subsets are presented in Appendix~\ref{res:types}, along with the accompanying figures similar to Figures~\ref{fig:pl_whole} and~\ref{fig:ln_whole}.  The fit results are all shown in Table~\ref{table:mfn_fit_values}.

\subsection{The Distribution of Dense Molecular Gas in the Galactic Plane}\label{res:gal}

\subsubsection{Face-On View of the Milky Way}

\begin{figure*}[!t]
  \centering
  \includegraphics[width=2.3in]{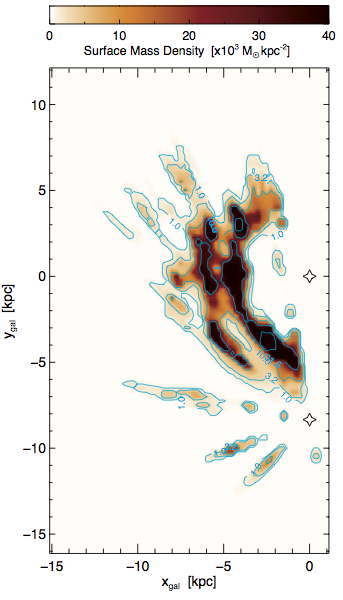}
  \includegraphics[width=4.0in]{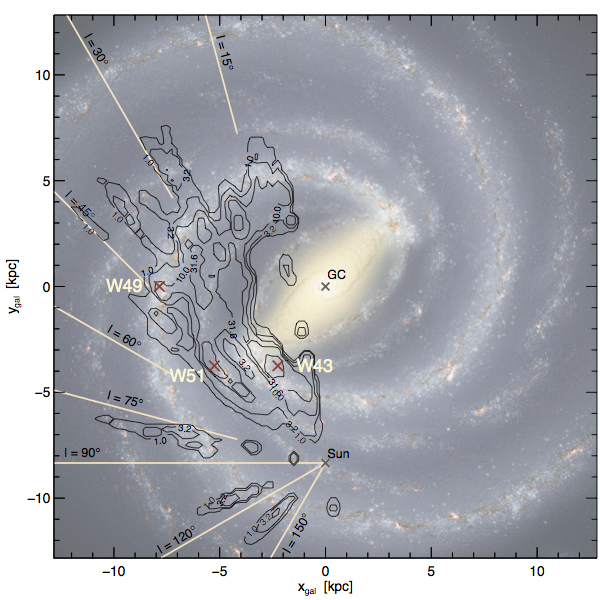}
  \caption[Dense gas mass surface density over the Galactic disk for Distance Catalog sources.]{Dense gas mass surface density over the Galactic disk for Distance Catalog sources.  \emph{Left}: Color scale and contours indicate the dense gas surface mass density ($\times 10^3\,M_{_\sun}$\,kpc$^{-2}$) in each 0.25\,kpc~$\times$ 0.25\,kpc pixel.  The Sun and Galactic Center are identified by stars at $(x_\mathrm{gal},y_\mathrm{gal}) = (0,-8.34)$\,kpc and $(0,0)$\,kpc, respectively.  \emph{Right}: The same mass surface density contours plotted atop an artist's rendering of the Milky Way based on \spitzer\ data (R. Hurt: NASA/JPL-Caltech/SSC).  Several of the mass concentrations correspond to prominent star-formation complexes originally discovered by \citet[][marked with red `$\mathsf{X}$'s]{Westerhout:1958}.  As discussed in EB15, we have excluded from consideration of much of \rgal~$< 4$\,kpc for kinematic reasons; sources appearing inside this circle have robust distance estimates through association with a trigonometric parallax measurement.}
  \label{fig:gal_mass}
\end{figure*}

\citet[][their Figure~15]{EB14a} presented a map of the locations of dense molecular cloud structures in the Galactic disk with respect to a rendering of Galactic structure based on \spitzer\ data.  Since these objects span several orders of magnitude in mass (Figure~\ref{fig:phys_cat}), their locations alone do not necessarily indicate the mass distribution of dense molecular gas.  To estimate the Galactic disk mass surface density of the objects in the BGPS Distance Catalog, we constructed a grid consisting of 0.25\,kpc~$\times$ 0.25\,kpc bins into which masses computed for individual sources were placed.  Since typical distance uncertainties are $\sim0.5$\,kpc, using 0.25\,kpc bins approximately Nyquist samples the data.  Taking advantage of the information contained in source DPDFs, we produced a map (Figure~\ref{fig:gal_mass}, \emph{left}) by randomly drawing $10^3$ distances for each object, and adding the mass derived using Equation~(\ref{eqn:bgps}), divided by $10^3$ to reflect the probabilistic contribution of that mass to the appropriate grid cell.  This use of the full DPDF leads to some of the smearing effects along the line of sight towards isolated regions in the outer Galaxy (\eg $\ell \approx 110\degr, 130\degr$), where peculiar motions have a larger effect on kinematic distances.  Furthermore, given that BGPS coverage for $\ell > 90\degr$ is neither blind nor contiguous, lack of mass in Figure~\ref{fig:gal_mass} in the outer Galaxy should not be construed to imply a void in the Galactic plane.

Mass surface density contours are plotted atop an image of the Galaxy produced using \spitzer\ near-infrared data (R. Hurt: NASA/JPL-Caltech/SSC) in the right panel of Figure~\ref{fig:gal_mass}.  As with the corresponding map from EB15, there are many caveats in the interpretation of this image.  First, there are virtually no BGPS sources with well-constrained distances at \rgal~$\lesssim 4$\,kpc owing to non-circular motions induced by the Galactic bar,\footnote{The DPDF formalism relies primarily upon kinematic distances whereby velocity measurements of dense molecular gas tracers are mapped onto Galactic position using a rotation curve assuming circular motion about the Galactic center.  The Galactic bar induces strong radial streaming motions, rendering regions under its influence unusable for kinematic distance measurement.} with the exception of objects associated with trigonometric parallaxes measurement of maser transitions in regions of high-mass star formation \citep[\eg][]{Reid:2014}.  Second, given that the typical uncertainty for well-constrained distance estimates from EB15 is $\sim\,0.5$\,kpc, the map and contours have a $\sim\,1$\,kpc effective resolution, significantly coarser than the underlying image, although commensurate with \emph{Herschel} maps of nearby galaxies \citep[mean distance $\approx~10$~Mpc;][]{Kennicutt:2011}.

Third, there is a significant concentration of mass near $(x_\mathrm{gal},y_\mathrm{gal}) = (-4,3)$\,kpc that appears to lie in a void in the underlying image.  As discussed in EB15, this could represent either molecular gas at that location not identified in the \spitzer\ data (which is a ``best guess'' picture of the Galaxy), or sources in the W43 star-forming complex at the end of the Galactic bar erroneously placed at the far kinematic distance.  Comparison of Figure~\ref{fig:gal_mass} with the Galactocentric locations of GMCs in the \thco\ Galactic Ring Survey \citep[GRS;][]{Jackson:2006} reveals a similar large collection of gas at this location \citep[][their Figure~5]{RomanDuval:2010}.  Most of the BGPS sources in this region are associated with one or more \HII\ regions with robust KDA resolution from the \HII\ Region Discovery Surveys \citep[HRDS;][]{Bania:2010,Bania:2012} and the distances to GRS clouds were primarily resolved using the same \HI\ absorption techniques as the HRDS (EB15).

The BGPS-GRS correlation implies either the existence of a large collection of mass in an apparently empty region of the \spitzer\ image or a fundamental limitation in the application of \HI\ absorption techniques for distance resolution.  Evidence in support of the former comes from both \citet[][]{Russeil:2011}, who find interarm clumps for this region in the Hi-GAL $\ell = 30\degr$ SDP field, and a study of the M51 spiral galaxy that reveals a population of GMCs downstream of a major spiral arm \citep[][]{Egusa:2011}.  Furthermore, the appearance of interarm gas could also be the effect of a spiral feature possessing a systematic noncircular velocity, biasing kinematic distances.

Identified in the right panel of Figure~\ref{fig:gal_mass} are the high-mass star-forming regions W43, W51 and W49, which stand out in the mass diagram as high surface density areas.  Despite the rough tracing of some spiral structure, this map is not a definitive distribution of dense molecular gas in the Milky Way.  First of all, objects used to produce this map represent only 20\% of the full BGPS catalog (by number, although 33\% by flux density).  Second, given the coarse resolution of the map, any conclusions about detected spiral structure should be used cautiously as the map serves to show where the detected mass lies, and not a complete census of mass in the Galaxy.  The gap observed between $60\degr \leq \ell \leq 75\degr$, however, is real and corresponds to a similar minimum in \twco\ column seen in the \citet[][]{Dame:2001} maps.  The expansion of robust distance determination to other dust-continuum surveys, however, will have high utility for tracing the Galactic dense molecular gas distribution across the entire disk.

\subsubsection{Vertical and Radial Distributions of Star-Forming Mass}

\begin{figure}[!t]
  \centering
  \includegraphics[width=3.1in]{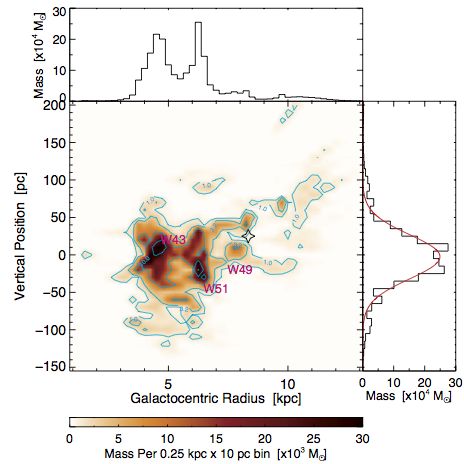}
  \caption[Vertical and radial mass distributions.]{Vertical and radial mass distributions.  The color-scale image in the center represents the azimuthal integration of mass, computed in analogous fashion to Figure~\ref{fig:gal_mass}, \emph{left}.  Since bins at larger Galactocentric radius encompass larger volumes, the absolute value of the color-scale is not physically meaningful, but rather represents the relative concentrations of mass.  At top is the mass distribution versus \rgal, marginalizing over vertical position, and at right is the distribution versus $z$, marginalizing over Galactocentric radius.  The fit to the mass-weighted $z$ histogram is consistent with the number-count histogram from EB15.}
  \label{fig:gal_vert}
\end{figure}

The azimuthally integrated vertical and radial distributions of the mass of star-forming clouds provide an alternate view of Galactic star-formation distribution, and are shown in Figure~\ref{fig:gal_vert}.  The Galactocentric radius and vertical position were computed using the $(\ell,b,$\dsun$) \rightarrow (R_\mathrm{gal},\phi,z)$ conversion matrix from Appendix~C of \citet{EllsworthBowers:2013}, which accounts for the 25\,pc vertical solar offset above the Galactic midplane.   As in Figure~\ref{fig:gal_mass}, masses and Galactocentric positions were derived from $10^3$ independent samples from the DPDF for each object.  The color image at the center represents the amount of mass in each 0.25\,kpc (radial) $\times$ 10\,pc (vertical) bin.  Since this is an azimuthal integration, the color scale intensity of the image is physically meaningless other than to identify overdense regions; pixels at larger \rgal\ represent a significantly larger volume than those nearer the Galactic center.  Identified in the image are the W43, W51, and W49 regions that stand out in the face-on image of the Galaxy.  These three regions dominate the mass distribution, but there are still clear collections of mass in the 4-5\,kpc range, and around 6\,kpc.  The warping of the disk to large \rgal\ is visible in the mass distribution, but we caution that BGPS observations in the outer Galaxy ($\ell > 90\degr$) were targeted towards regions of known star formation, and do not represent the same blind sample as the inner Galaxy coverage. 

To the right of the image is the marginalization over \rgal, or the the vertical histogram of the mass distribution, shown as total mass per bin.  A Gaussian fit to the this mass-weighted histogram ($\mu = -2.4\pm1.8$\,pc, FWHM~= $63.7\pm3.6$\,pc) is consistent with the distribution derived from the number-count histogram from EB15.  This implies that the mass is roughly evenly distributed amongst catalog sources regardless of vertical position, and that there is no particular bias in the mass of dense molecular cloud structures near the Galactic midplane.


\section{DISCUSSION}\label{ch4:discuss}


\subsection{What is a BGPS source?}\label{disc:whatis}

Based on a subset of BGPS objects throughout the Galactic plane observed in the lowest inversion transitions of \nhhh\ around 24~GHz, \citet{Dunham:2010,Dunham:2011c} began a discussion about the nature of BGPS sources.  Using the larger sample of objects with well-constrained distance estimates from EB15, we continue this discussion.

\begin{figure}[!t]
  \centering
  \includegraphics[width=3.1in]{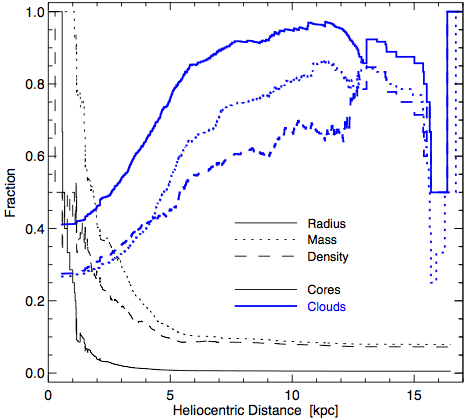}
  \caption[Fraction of source type as a function of heliocentric distance.]{Fraction of source type as a function of heliocentric distance.  Shown in thin black lines is the fraction of sources in the BGPS Distance Catalog nearer than the indicated heliocentric distance that are cores, and thick blue lines indicate the fraction of sources farther that are clouds.  Line styles indicate core and cloud divisions as determined by physical radius (solid), mass (dotted), and number density (dashed).  Both ends of the distance range suffer from small number statistics.  (This figure is styled after D11, their Figure 21.)}
  \label{fig:dunham}
\end{figure}

Observationally, dense molecular cloud structures are typically defined in terms of cloud-, clump-, and core-scale objects.  The BGPS is generally sensitive to clump-scale objects, but we investigate to what extent clouds and cores are present in the data.  Figure~\ref{fig:dunham} illustrates the relative concentrations of non-clump objects at various heliocentric distances, and is modeled after D11 (their Figure~21).  Comparison between the two figures yields insight into the effects of using a larger sample.  The thin black lines in Figure~\ref{fig:dunham} mark the fraction of sources closer than a given distance with properties consistent with molecular cloud cores based on physical radius (solid; $R \leq 0.125$\,pc), mass (dotted; $M \leq 27.5\,M_{_\sun}$), and number density (dashed; $n \geq 10^4$\,\cc), as defined in \S\ref{res:ensemble}.  Many ($>70\%$) of the objects at \dsun~$\leq 1$\,kpc fall under the molecular cloud core definition, and conversely, the majority of core-scale objects are seen at \dsun~$\leq 2-3$\,kpc.  This is commensurate with the completeness and angular transfer functions illustrated in Figure~\ref{fig:phys_relate}.

Similarly, thick blue lines in Figure~\ref{fig:dunham} mark the fraction of sources located farther than a given distance that are characterized as clouds ($R\geq1.25$\,pc, $M\geq750\,M_{_\sun}$, and $n\leq750$\,\cc).  Of particular note is the disparity in the distributions between different definitions for GMC-scale objects.  Beyond 6\,kpc, $\geq90\%$ of objects in the Distance Catalog have large physical radius.  As found by D11 for the \nhhh-observed subset, the density curve never reaches 90\% because there exists a substantial fraction of high-density sources in the distance range 6\,kpc~$\leq$ \dsun~$\leq 13$\,kpc (see Figure~\ref{fig:phys_relate}, \emph{bottom left}).  Furthermore, the geometrical uncertainties in both the cataloging process and derivation of physical properties decrease the reliability of this quantity for object classification.  The combination of improved cataloging routines with higher fidelity to the underlying structure and future high angular resolution studies of the Galactic plane (\eg with CCAT) may alleviate this disparity.  One principal deviation from the D11 results, however, is the mass fraction of cloud-scale objects as a function of distance.  Whereas for the \nhhh\ subset the cloud fraction by mass closely tracked the radius fraction, in Figure~\ref{fig:dunham} the cloud fraction by mass is significantly smaller.  This may be due either to the targeted nature of the D11 observations or differences between the BGPS version~1 and version~2 catalogs (see G13).

With the enlarged set of physical properties presented here, we confirm the previous conclusions of studies of BGPS objects that the majority of sources detected are clump-scale objects in the distance range 2\,kpc~$\lesssim$ \dsun~$\lesssim 10$\,kpc, cores nearer by, and cloud-scale objects beyond.  Uncertainties in the geometry of dense molecular cloud structures remains one of the most significant issues in terms of understanding the number density distribution and its role in the evolution of dense regions.

\subsection{Larson's Laws \& The Continuum of Structure}\label{disc:larson}

\subsubsection{The Mass-Radius Relationship}\label{res:larson}

The physical properties of molecular clouds and their substructures appear to follow a series of scaling relationships known as Larson's Laws, following their introduction in a literature study of molecular clouds by \citet{Larson:1981}.  The most frequently used of these scaling relationships is the size-linewidth relationship, which holds that virial and/or turbulent motions across a molecular cloud structure increase with the square root of the size of that structure.  This relationship is valid for single molecular line tracers \citep[\eg][]{Solomon:1987,Heyer:2009}, but the linewidths of different molecular transitions observed in the same dense molecular cloud structure may vary by an order of magnitude.  The heterogenous nature of the spectroscopic data gathered for BGPS distance estimation therefore prevents their use in deriving such a relationship.

\begin{figure}[!t]
  \centering
  \includegraphics[width=3.1in]{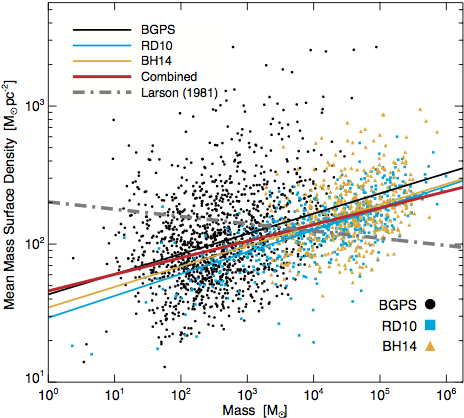}
  \caption{The mass surface density - mass ($\Sigma - M$) relationship for molecular cloud structures.  Black dots identify BGPS Distance Catalog objects from Table~\ref{table:physics}, while cyan squares mark the values for GRS GMCs from \citet[][labeled RD10]{RomanDuval:2010} and orange triangles the GRS structures discussed in \citet[][labeled BH14]{Battisti:2014}.  The black, cyan, and orange lines plot the respective power-law fits to the individual data sets, and the red line corresponds to the power-law fit for the combined data.  The original result from \citet{Larson:1981} is shown as a thick gray dot-dashed line.  Fit parameters for each line are listed in Table~\ref{table:larson_fit}.}
  \label{fig:larson}
\end{figure}


\begin{deluxetable}{cccc}
  \tablecolumns{4}
  \tablewidth{0pt}
  \tabletypesize{\small}
  \tablecaption{Power Law Fit Values for Larson-like Surface Density - Mass Relationship\label{table:larson_fit}}
  \tablehead{
    \colhead{} & \multicolumn{2}{c}{$\Sigma = a\,M^b$} & \colhead{} \\
    \cline {2-3}
    \colhead{Data Set} & \colhead{$a$} & \colhead{$b$} & \colhead{$\rho$\tablenotemark{a}}
  }
  \startdata
BGPS & $43^{+12}_{-9}$ & 0.15 (0.04) & -0.96\\
RD10 & $29^{+14}_{-9}$ & 0.16 (0.04) & -0.98\\
BH14 & $35^{+50}_{-20}$ & 0.15 (0.08) & -0.99\\
Combined & $46^{+7}_{-6}$ & 0.12 (0.02) & -0.95\\
  \citet{Larson:1981} & 202 & $-0.053$ & \nodata
  \enddata
  \tablenotetext{a}{Parameter correlation coefficient $\rho = \sigma_\mathrm{ab} / \sigma_\mathrm{a}\sigma_\mathrm{b}$.}
\end{deluxetable}

The first two \citeauthor{Larson:1981} relationships (the size-linewidth and mass-linewidth relationships) may be combined to form a mass-size relationship, sometimes called Larson's 3\srd Law.\footnote{This name is also given to the number density-size relationship similarly derived from the first two \citet{Larson:1981} relations.}  Using the physical radius $R = L/2$ rather than the overall length of the molecular feature for consistency with the remainder of our analysis, we derive
\beqn\label{eqn:lar_mr}
\frac{M}{M_{_\sun}} = 460\,\left( \frac{R}{\mathrm{pc}}\right)^{1.90}~
\eeqn
as the form of the \citeauthor{Larson:1981} mass-radius relationship.  This power-law relationship has been reproduced by simulations of isothermal supersonic turbulence \citep[][]{Kritsuk:2013}.

While Equation~(\ref{eqn:lar_mr}) appears to have theoretical backing, it is important to note the effect of observational systematics on the derived relationship.  Combining Equations~(\ref{eqn:mass}) and (\ref{eqn:prad}) leads to a systematic $M \propto R^2$ relationship, which has been advanced as the source of Larson's 3\srd Law \citep[\eg][and references therein]{BallesterosParedes:2012}.  Owing to the similarity between Equation~(\ref{eqn:lar_mr}) and the systematic relationship, we investigate the \emph{deviation} from $M \propto R^2$ for both BGPS sources and measurements of GMCs from the \thco(1-0) data of the GRS.

To remove the effect of heliocentric distance, we compare the mean mass surface density ($\Sigma$) with the mass of each object.  Figure~\ref{fig:larson} plots the 1,369 BGPS Distance Catalog objects with real $\theta_R$ as black circles, where the vertical axis is computed via $\Sigma =$ \mml$ / \pi R^2$.  The wide scatter in points is likely due in part to the non-uniform geometry of objects (as discussed earlier).  To evaluate the correspondence of the relationship for the BGPS objects (\ie clump-scale structures) with larger structures, we also plot in Figure~\ref{fig:larson} results of two studies of \thco\ clouds from the GRS.  GMCs studied by \citet[][hereafter RD10]{RomanDuval:2010} are shown as cyan squares, and the clouds containing BGPS sources from \citet[][hereafter BH14]{Battisti:2014} are depicted with orange triangles.  While both of these studies used the GRS \thco\ data, they used different source identification algorithms, resulting in the differences shown in Figure~\ref{fig:larson}.  For each data set, we fitted a linear function in logarithmic space that is robust against outliers\footnote{Rather than a ``least squares'' fit, we utilized a ``least absolute deviation'' method to minimize the effects of outliers on the resulting fit \citep[see][\S 15.7.3]{Press:2007}.} to the data as $\log \Sigma = \log a + b \log M$, and the resulting power-law fit values are listed in Table~\ref{table:larson_fit}.  The individual fits for each data set are shown in the same color as the data points.  Furthermore, we also fit a power law through the entire collection of points (red) to span a further order of magnitude in each dimension than any one data set alone.

Casting Equation~(\ref{eqn:lar_mr}) in the same form as the data of Figure~\ref{fig:larson} yields
\beqn\label{eqn:sigma_m}
\frac{\Sigma}{M_{_\sun}\,\mathrm{pc}^{-2}} = 202\,\left( \frac{M}{M_{_\sun}}\right)^{-0.053}~,
\eeqn
indicating a nearly-flat mean mass surface density as a function of mass, and is shown as the dot-dashed gray line in Figure~\ref{fig:larson}.  The slope is so shallow as to cause only a factor of two difference in $\Sigma$ over some six orders of magnitude of mass.

Strikingly, the data from both continuum (BGPS) and \thco\ (RD10, BH14) observations consistently show a trend of \emph{increasing} $\Sigma$ with object mass, with the fits to the three individual sets returning nearly identical power-law indices.  The similarity of these fit values indicates no significant difference between physical properties computed from dust continuum and molecular transition line observations of dense molecular cloud structures.  These fit values significantly differ from the value in Equation~(\ref{eqn:sigma_m}).  Although contrasting with the \citeauthor{Larson:1981} slope, these fits still represent less than an order of magnitude change in $\Sigma$ over six orders of magnitude in mass.

Caution should be taken when drawing conclusions from this result.  While the intrinsic scatter within each data set is large, none of the data sets could be consistent with the predicted \citeauthor{Larson:1981} relationship ($b = -0.053$) within the fit uncertainties.  However, both the Larson relationship and the fits shown in Table~\ref{table:larson_fit} deviate only slightly from the systematic $M\sim R^2$.  The fact that the properties of dense molecular cloud structures from three separate studies all reveal nearly the same, slight departure from the systematic $b = 0$ indicates a real, though slight, trend in mass surface density as a function of mass.  In its simplest interpretation, this relation seems to imply that higher mass GMCs are also denser and therefore could possibly form stars more efficiently.  This interpretation must be tempered, however, by the inability of current survey data-reduction techniques and cataloging algorithms to accurately represent large-scale low-surface-brightness structures in the interstellar medium.

\subsubsection{The Continuum of Structure}

The continuum of structure within molecular cloud complexes is a fundamental property whose connection with the intimate processes of star formation requires further study.  Much of the literature discusses objects as clouds, clumps, or cores, and assigns specific properties and physical significance to each \citep[\eg][and references therein]{McKee:2007,Bergin:2007,Kennicutt:2012}.  As seen in Figure~\ref{fig:larson}, however, observed dense molecular cloud structures exhibit a continuum of properties across many orders of magnitude in scale; here we discuss the implications of this continuum.

The near uniformity of mean mass surface density of dense molecular cloud structures provides a crucial postulate for the construction of the Galactic distribution of dense molecular gas.  Whether an artifact of observational systematics or a consequence of supersonic turbulence \citep[\eg][]{Hopkins:2012a}, the observed uniform relationship indicates that the fractal quality of the molecular interstellar medium\footnote{The interstellar medium, and molecular clouds in particular, display fractal dimensions \citep[\eg][]{Beech:1987}, wherein a self-similarity relationship exists over many orders of magnitude and no preferred scale exists between the limits of Galactic disk scale heights and runaway gravitational collapse of molecular cloud cores.} is perhaps more fundamental and a stronger driver of molecular cloud complex evolution than the individual classes of objects (\ie clouds, clumps, cores) working independently.  There exist real physical divisions as the fractal nature of molecular clouds is driven by the the dominance of turbulence \citep[][]{Kritsuk:2013}; regimes dominated by other physical processes (\ie gravitational collapse) obey other scaling relationships.\footnote{Following a Jeans collapse scenario, $M\sim R$, given uniform composition and temperature.}

Aside from questions of the evolution of the ISM, what are the implications of these scaling relations on observational data?  As a basic reference point, we start from the \citeauthor{Larson:1981} result of Equation~(\ref{eqn:lar_mr}) and convert the physical properties to observable quantities via Equations~(\ref{eqn:bgps}) and (\ref{eqn:prad}).  The resulting relationship between observed flux density, deconvolved angular radius, and heliocentric distance is given by
\beqn\label{eqn:flux_rad}
\left( \frac{S_{_{1.1}}}{\mathrm{Jy}} \right) = 3.34~\left(\frac{\theta_R}{\mathrm{arcmin}} \right)^{1.90} \left(\frac{d_{_\sun}}{\mathrm{kpc}} \right)^{-0.10}~.
\eeqn
This equation implies that for dense molecular cloud structures, the observed flux density is nearly independent of heliocentric distance, and depends only on the angular radius on the sky, which is the same as saying the mass surface density is independent of mass (as seen in Figure~\ref{fig:larson}).  This relationship is plotted in Figure~\ref{fig:flux_rad} (\emph{top}) as green lines for \dsun~= 1\,kpc (solid), 5\,kpc (dotted), and 20\,kpc (dashed), and matches the slope of data quite well.  Therefore, in maps from (sub-)millimeter continuum surveys, the physical scales of detected objects are indistinguishable, meaning that a core looks like a clump looks like a cloud.  It is only through robust distance estimation or application of the size-linewidth relationship that the scales of molecular cloud structures may be interpreted.  This is an important point because the use of ancillary data is essentially the \emph{only} means for deriving physically meaningful results from these surveys.

In EB15, it was noted that the surface brightness ($SB$) distributions for the full BGPS catalog and various subsets (including the Distance Catalog) were largely independent of cataloged flux density.  Indeed, by combining the surface brightness equation from EB15 (their Equation~3) with Equation~(\ref{eqn:flux_rad}), we obtain
\beqn
SB = 13.3\ \mathrm{MJy\ sr}^{-1}~\left(\frac{S_{_{1.1}}}{\mathrm{Jy}} \right)^{-0.05} \left(\frac{d_{_\sun}}{\mathrm{kpc}} \right)^{-0.105}~,
\eeqn
which indicates that the surface brightness of dense molecular cloud structures, regardless of size scale and heliocentric distance, should be nearly identical.  Indeed, \citet{Solomon:1987} comment that for the standard $\sigma_v \propto R^{0.5}$ size-linewidth relationship, virial equilibrium translates into constant average surface density, and hence constant average surface brightness.  The observed surface brightness along any line of sight is then proportional to the number of structures, or total mass, along that line of sight.  If one assumed an axisymmetric model and sampled a large fraction of the plane, the radial dependence (and normalization) of the molecular gas along the line of sight could be solved for.

\subsection{Mass Distributions and Mass Functions}\label{disc:mfn}

\subsubsection{Aggregate Mass Distribution versus Subsets}\label{disc:whole_part}

The aggregate mass distribution shown in the left panels of Figures~\ref{fig:pl_whole} and \ref{fig:ln_whole} represents an inhomogenous set of dense molecular cloud structures across the Galactic plane.  The entire collection of sources appears to be fitted well by a lognormal function over more than two orders of magnitude, and displays one or two power laws at the high-mass end.  Prompted by theoretical predictions for a power-law form of the so-called clump mass function \citep[\eg][]{Donkov:2012,Veltchev:2013}, we made a series of astrophysically motivated cuts (see Appendix~\ref{res:types}) in an attempt to isolate a more-homogenous population of sources.

While the classifications of molecular cloud versus clump versus core are intuitively appealing, the mass distributions of BGPS sources divided into core/clump and cloud groups (Table~\ref{table:mfn_fit_values}) return functional fits very similar to the aggregate set.  Three possible explanations exist, likely working in concert: (1) the dividing lines between object classes are somewhat arbitrary, (2) uncertainties in the heliocentric distances mix objects across classification boundaries, and (3) the distinctions themselves are artificial in light of the continuum of structure in molecular cloud complexes.  It is likely that a survey with improved angular resolution may be able to disentangle mass function parameters as a function of source classification or physical properties.  Furthermore, use of multi-band Hi-GAL data will obviate the need to assume a dust temperature (or to marginalize over the dust temperature, as was done with the Monte Carlo trials), eliminating a source of uncertainty in derived properties.

The best power-law fits were for subsets primarily limited in heliocentric distance, and are characterized by nearly-symmetric [$\hat{\alpha},M_\mathrm{min}$] joint confidence intervals (similar to the right panel of Figure~\ref{fig:pl_whole}).  These subsets were the ``nearby'' heliocentric distance and ``mixed protocluster'' cuts, which exclude the most distant objects, with the latter additionally excluding nearby, low-mass objects.  While both are strongly power-law the ``nearby'' subset has a steeper power-law index ($\hat{\alpha} = 2.04\pm0.09$ versus $1.92^{+0.13}_{-0.06}$), although the two are marginally consistent with each other.  

The lack of a strong power-law fit for the ``distant'' sample or the two cloud-scale subsets (in $R$ and $n$) may be partially caused by the apparent steepening in the mass distribution at high-mass as discussed below.  The subsets that \emph{do} exhibit power-law structure, however, indicate that the presence of a true underlying power law (if present) is not obscured and converted to a lognormal by sample inhomogeneity (see \S\ref{res:inevit}).

\subsubsection{Steepness of the High-Mass Power-Law Fit}\label{disc:steep}

One curious aspect of the mass function fits is the propensity for the high-mass end to display a fairly steep power-law tail ($\alpha \approx 2.2-2.3$).  While this power-law index is similar to the high-mass end of the \citet{Salpeter:1955} IMF, objects detected by the BGPS in this mass range are generally cloud-scale in physical radius.  Since Galactic GMCs tend to follow a power-law mass function with $\alpha \approx 1.6 - 1.8$ \citep[][]{Blitz:1993}, the present result is somewhat contradictory.

A steep power-law function (\ie $\alpha > 2$) implies that most of the total mass is in low-mass objects and that high-mass objects are rarer than more shallow functions (\ie there are fewer per mass bin; Equation~\ref{eqn:mfn}).  Could the apparent steepening be caused by ``missing'' high-mass sources in the present Distance Catalog?  Simple simulations show that removal of sources from a shallow power-law distribution does have the effect of imitating a steeper power law, given that a progressively larger fraction of sources are removed with increasing mass. 

Between the atmospheric subtraction in the BGPS data-reduction pipeline removing emission on large angular scales ($\gtrsim 300\arcsec$; G13) and the tendency of the cataloging routine to break up large-scale emission into smaller chunks based on intermediate peaks in the flux-density distribution, the net effect is to remove the largest scale objects and convert single large structures into multiple smaller ones.  What effect do these have on the derived mass distribution?  Converting the mass-size scaling relationship of Equation~(\ref{eqn:lar_mr}) into observational quantities via Equation~(\ref{eqn:prad}) yields
\beqn\label{eqn:mass_arad}
M = 43.8\,M_{_\sun}~\left(\frac{\theta_R}{\mathrm{arcmin}} \right)^{1.90} \left(\frac{d_{_\sun}}{\mathrm{kpc}} \right)^{1.90}~.
\eeqn
The effective $\theta_R \lesssim 2\farcm5$ limit of the BGPS means that clouds at \dsun~= 5\,kpc (near the peak of the heliocentric distance distribution) with $M \gtrsim 5000$~\msun\ should be either truncated or broken into multiple smaller sources.  At a distance of 10\,kpc, the filtering affects sources with $M \gtrsim 2\times10^4$~\msun.  Indeed, in the Physics Catalog of Table~\ref{table:physics}, there are only 7 objects with \mml\ above this value.  In terms of the mass distribution, the breaking up of large structures into smaller ones adds objects to the low-mass end of the mass distribution, steepening the power law.  The power-law indices presented in Table~\ref{table:mfn_fit_values} should, therefore, be viewed as upper limits.  Accurate measurement of the high-mass power-law distribution of dense molecular cloud structures, therefore, requires retaining some amount of large-scale structure in survey images and catalogs. 

\subsubsection{The Inevitability of a Lognormal + Power-Law Mass Function?}\label{res:inevit}

\begin{figure*}[!t]
  \centering
  \includegraphics[width=3.1in]{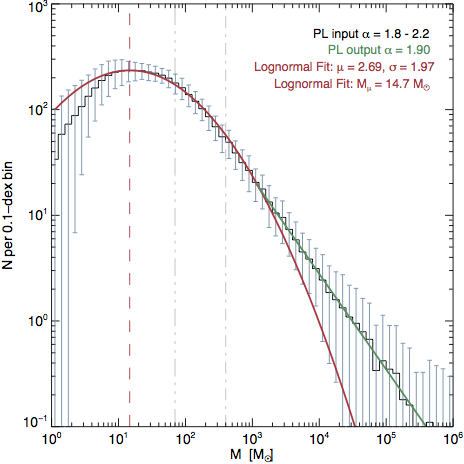}
  \includegraphics[width=3.1in]{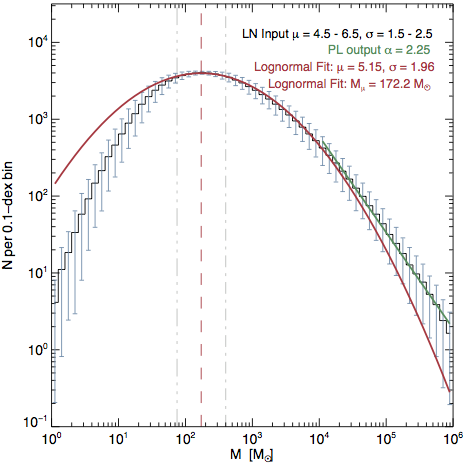}
  \caption[Simulated aggregate mass distributions of star-forming regions at a variety of distances.]{Simulated aggregate mass distributions of star-forming regions (SRFs) at a variety of distances.  Only sources in those regions that would be detected by the BGPS (\ie $S \geq 3\sigma$) are included.  \emph{Left}: Composite of 100 SFRs, each with a power-law mass distribution with index $1.8 \leq \alpha \leq 2.2$.  The error bars represent the spread among 100 Monte Carlo realizations.  The green line represents the power-law fit to the high-mass end of the distribution, and the red curve is the lognormal fit to the peak.  Vertical dot-dashed lines represent the 50\% ($M=70\,M_{_\sun}$) and 90\% ($M=400\,M_{_\sun}$) mass completeness levels, and the vertical dashed red line marks the $M_\mu$ of the lognormal fit.  \emph{Right}: Composite of 100 SFRs, each with a lognormal mass distribution with mean $4.5 \leq \mu \leq 6.5$ and width $1.5 \leq \sigma \leq 2.5$.  Other features as at left.}
  \label{fig:inevit}
\end{figure*}

Observationally, the aggregate mass distribution, as shown in Figure~\ref{fig:pl_whole} (\emph{left}), should be expected to be well fitted by a lognormal function with a power-law tail regardless of the actual underlying distribution of source mass.  Given the finite angular resolution and flux-limited nature of the BGPS (and similar surveys), it is possible to simulate the observed mass distribution of an inhomogenous sample spread across the Galactic plane.

We conducted a series of Monte Carlo trials, whereby we ``observed'' 100 star-forming regions (SFRs) whose distances were drawn from the \dsun\ distribution of the Distance Catalog.  Each SFR consisted of $10^3$ objects drawn purely from either a power-law (Trial 1) or lognormal (Trial 2) mass distribution.  A physical radius was assigned using Equation~(\ref{eqn:lar_mr}) to estimate whether any given object would be resolved by the BGPS, and the extent to which its flux density (computed from the mass) would be diluted in the beam.  Using an average BGPS noise level, sources were culled from the list that would not be detected ($S_\nu < 3\sigma$).  The aggregate distribution of ``detected'' sources was recorded, and the process repeated for $N=100$ Monte Carlo realizations.  

The results of Trial 1, where the underlying power-law mass distribution was allowed to range over $1.8 \leq \alpha \leq 2.2$ with $M_\mathrm{min} = 1$~\msun are shown in the left panel of Figure~\ref{fig:inevit}.   The aggregate shows a strong power-law tail with $\alpha = 1.9$, and a lognormal distribution at lower mass, peaking at $M_\mu \approx 15\,M_{_\sun}$.  While the location of the lognormal peak is well below the completeness levels of the BGPS data set, these simple simulations do not account for real features in the data, such as source confusion and hierarchical (fractal) structure, and likely overestimate the number of low-mass sources visible.  Trial 2 (right panel) constructed individual lognormal mass distributions using the ranges $4.5 \leq \mu \leq 6.5$ and $1.5 \leq \sigma \leq 2.5$.  The resulting aggregate mass distribution shows a steeper ($\alpha \approx 2.3$) power-law tail at high mass, but has a lognormal fit much closer to that of the full BGPS set in Figure~\ref{fig:ln_whole}.  These simulations are not meant to definitively show the form of the underlying mass distributions, but rather to demonstrate the inevitability of a lognormal distribution with a power-law tail at high mass for inhomogenous data sets such as the BGPS, regardless of the underlying source distribution.  The inhomogeneity should be reduced in the astrophysically motivated subsets (Appendix~\ref{res:types}), hence we do not necessarily expect them to approach lognormal distributions with muted high-mass power laws.

\subsubsection{Comparisons with the Literature}\label{disc:lit}

There has been much discussion in the literature about dense molecular cores and their relationship with the stellar IMF \citep[see][]{Shirley:2003}, but it is only with the recent advent of large-scale Galactic plane surveys that large, statistically robust samples of clump-scale objects have become available.  For comparison with the present results, we discuss the recent studies of clump-scale objects by \citet{Olmi:2013} from the science demonstration fields from the \herschel\ Hi-GAL survey \citep[][]{Molinari:2010a} and by \citet{Wienen:2015} from the $\lambda = 870$\,\micron\ ATLASGAL survey \citep[][]{Schuller:2009}.

Turning first to the \citet{Olmi:2013} analysis, direct comparison with the BGPS results is not fully straightforward, as the source-finding algorithm used by those authors is based on a Mexican Hat Wavelet \citep[\eg][]{Barnard:2004} and typically finds structures on scales smaller than those identified in the BGPS catalog (which used a seeded-watershed algorithm).  Furthermore, Hi-GAL sources were identified as well-separated sources visible in each of the $\lambda = 70$\,\micron, 160\,\micron, and 250\,\micron\ bands; the beam size at 250\,\micron\ is 17\arcsec, approximately half the width of the BGPS 33\arcsec\ beam.  These two features imply that the \citeauthor{Olmi:2013} data set will find more, smaller sources than the BGPS.

Comparing source lists in the $\ell = 30\degr$ field, the BGPS~V2 catalog identifies 492 sources, whereas \citeauthor{Olmi:2013} report identifying 1,950 sources n the same longitude range.  Although the Hi-GAL latitude range extends twice as far as the nominal BGPS coverage, the bulk of the \herschel-measured emission lies within the boundaries of the BGPS mosaics.  For the $\ell = 59\degr$ field, the differences are much more astounding; the BGPS catalog contains 17 sources while \citeauthor{Olmi:2013} find 3,402.  A visual comparison of the data sets reveals that the Hi-GAL tile contains significant filamentary structure at modest flux density whereas the BGPS mosaic exhibits nearly uniform noise, indicating that atmospheric removal has obliterated much of this low-level signal.  The vast majority of the Hi-GAL sources in this tile lie along these filaments \citep[see][their Figure~3]{Molinari:2010b}.  Careful comparison of Hi-GAL and BGPS maps, however, shows that much of what appears to be atmospheric 1/f noise actually corresponds to the cirrus-like filamentary structure made choppy by the filtering.

With these differences between the two data sets in mind, we compare the power-law fits to the mass distribution of sources.  Since \citet{Olmi:2013} use a uniform distance for each of the two fields ($\ell=30\degr$ is assigned a distance of 7.6\,kpc and $\ell=59\degr$ is assigned to 3.6\,kpc), it is possible to compare their mass function fits to our subsets in heliocentric distance.  Taking first the $\ell=30\degr$ field, the large distance assigned to the Hi-GAL sources corresponds to the set of distant BGPS objects (\dsun$~> 6.5$\,kpc; see Appendix~\ref{mfn:dsun}).  While our sample is not fitted well by a power-law, the maximum-likelihood parameters are $\hat{\alpha} = 1.93^{+0.19}_{-0.05}$ for $M \gtrsim 2000$\,\msun.  This is in comparison to the steeper \citeauthor{Olmi:2013} fit values ($\hat{\alpha},\,M_{\min}) = (2.15,\,212\,M_{_\sun}$).  The lognormal fit presented by those authors has a peak near $M_\mu = e^{4.58} = 100\,M_{_\sun}$, somewhat smaller than the fit to the entire BGPS set (Table~\ref{table:mfn_fit_values}), following the trend of their power-law fit.

The $\ell=59\degr$ field is placed at a distance corresponding to our nearby sample in heliocentric distance (\dsun$~< 6.5$\,kpc).  The nearby sample is fitted well by a power-law with $\hat{\alpha} =  2.01^{+0.11}_{-0.09}$ for $M \gtrsim 620$\,\msun.  As with the $\ell=30\degr$ field, \citet{Olmi:2013} find a steeper power-law that extends to lower mass ($\hat{\alpha},\,M_{\min}) = (2.20,\,7.3\,M_{_\sun}$).  Because this power-law fit extends to such low mass, it is tempting to conclude that this Hi-GAL tile is imaging core-scale rather than clump-scale objects.  Since sources must be identified in the $\lambda = 250$\,\micron\ images whose angular resolution is $\approx 25\arcsec$, however, this corresponds to a minimum observable size of $\approx0.4$\,pc at the assigned 3.6\,kpc distance.

The more recent \citet{Wienen:2015} analysis of clump-scale dense molecular cloud structures in the ATLASGAL data provides a more direct comparison set with the BGPS Distance Catalog.  While the \citeauthor{Olmi:2013} results derive from two specific regions, ATLASGAL sources span the entire inner Galactic plane ($-60\degr \leq \ell \leq 60\degr$).  As such, it suffers from the same spread in heliocentric distance (and hence resolving power) as the BGPS and sweeps up objects across the spectrum of molecular cloud structures.  We compare the BGPS Distance Catalog distribution with the ATLASGAL sources from the first Galactic quadrant, a largely overlapping population.  \citet{Wienen:2015} note that only a 70\% distance matching rate (within 2\,kpc) exists between their distance assignments and those from EB15.  This relatively large disagreement seems to arise primarily from the fraction of objects placed at the far kinematic distance (44\% of ATLASGAL sources versus 22\% of BGPS sources).  Those authors dismiss the significance of the distance mismatch by claiming that distance errors average out by virtue of the large ATLASGAL sample size.  However, if the population of true distances is similar for the two surveys, the larger fraction of ATLASGAL sources placed at the far kinematic distance would systematically shift the mass distribution and affect the mass function fits.  While a detailed distance catalog comparison is beyond the scope of this paper, the significant overlap between the ATLASGAL and BGPS survey data merits a careful comparison of source identification schemes, derived distances, and implications for mass distributions.

Notwithstanding the discrepancies between the two distance catalogs, differences in source properties arise from the angular transfer function of the ATLASGAL data, which ranges from the beam-scale 19\farcs2 to a maximum recoverable scale of $\approx 150\arcsec$ \citep[][]{Schuller:2009}.  This beam size is $\approx 60\%$ that of the BGPS, meaning that smaller-scale structure is resolved in the ATLASGAL mosaics.  The corresponding physical size at the 4\,kpc heliocentric distance peak in the source distribution is $\approx 0.4$\,pc, meaning that the source catalog is still dominated by clump-scale objects \citep[][]{Bergin:2007}.  On the large-scale end, the ATLASGAL data reduction pipeline effectively removes uniform emission on scales larger than $\sim2\farcm5$, whereas the BGPS angular transfer function is $\gtrsim 0.8$ for scales smaller than $\sim5\arcmin$ (G13), leading to the recovery of larger objects.

The above-mentioned differences may play a role in the comparison between the mass function fits described in \S\ref{res:mfns} and those presented by \citet[][]{Wienen:2015}.  While those authors experimented with fitting power laws to binned data, we compare only the results obtained using the maximum-likelihood method of \citet[][]{Clauset:2009}.  A power-law function fit to the entire ATLASGAL source collection yields $\hat{\alpha} = 1.82 \pm 0.02$, somewhat flatter than our value of $\hat{\alpha} = 1.94^{+0.34}_{-0.10}$.  The source of the difference is unclear, unless there exist sources of significant mass in the ATLASGAL catalog within the \rgal\ = 4\,kpc kinematic avoidance zone (EB15) for which kinematic distances are not computed for BGPS sources.  Furthermore, the very small error bars reported by those authors conflict with the wider spread in power-law index resulting from the Monte Carlo realizations of the BGPS mass distribution shown in Figure~\ref{fig:pl_whole}, and likely do not take into account the effects of distance uncertainty.  The only subset of the data to which \citeauthor{Wienen:2015} fit mass functions is the cut 2\,kpc $\leq$ \dsun $\leq$ 5\,kpc, encompassing the peak of the heliocentric distance distribution.  For this subset, they find a power-law index of $\hat{\alpha} = 2.26\pm0.05$, somewhat steeper than either our set of nearby sources (\dsun $\leq 6.5$\,kpc; $\hat{\alpha} = 2.0\pm0.1$) or the mixed protocluster criterion (2\,kpc $\leq$ \dsun $\leq$ 10\,kpc, $M \geq 300$\,\msun; $\hat{\alpha} = 1.9\pm0.1$).  One possible explanation for the steeper ATLASGAL fits is the filtering of large-scale emission into smaller sources due to the angular transfer function (\S\ref{disc:steep}).  Also included in the \citet{Wienen:2015} analysis is a lognormal fit to this distance subset, yielding a peak near 700\,\msun, considerably larger than the lognormal fits to BGPS Distance Catalog sources (\S\ref{res:lnwhole}).

For both these comparison studies, a better understanding of the discrepancies between the mass function fits will come when all data sets are analyzed with the same cataloging method to remove systematic effects introduced by source identification.  Furthermore, a more detailed study of the filtering effects of data pipeline angular transfer functions on source identification within large-scale structure may provide a more solid basis of comparison between results from different (sub-)millimeter Galactic plane surveys.

\subsection{Estimating the Dense Gas Mass Fraction}\label{disc:dgmf}

\begin{figure}[!t]
  \centering
  \includegraphics[width=3.1in]{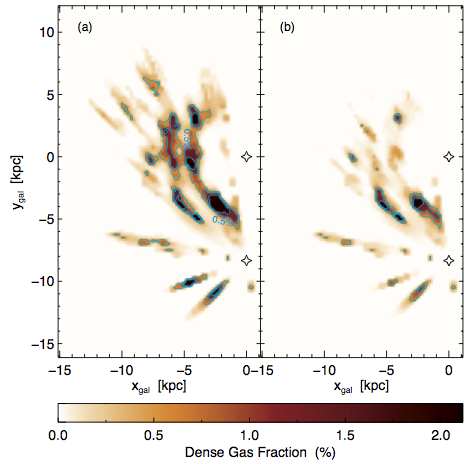}
  \caption{Two estimates of the dense gas mass fraction estimate as a function of Galactocentric position.  In both panels, contours indicate linear increments of 0.5\%, and the stars mark the locations of the Galactic center ($x_\mathrm{gal},y_\mathrm{gal}) = (0,0)$\,kpc, and the Sun ($x_\mathrm{gal},y_\mathrm{gal}) = (0,-8.34)$\,kpc.  (\emph{a}) The face-on map was computed by dividing the mass surface density of Figure~\ref{fig:gal_mass} by the azimuthally averaged \htwo\ model of \citet{Wolfire:2003}.  (\emph{b})  Same, except the face-on mass surface density map used consists only of sources with $n \geq 10^3$\,\cc.}
  \label{fig:dgmf}
\end{figure}

The Galactic mass distribution in Figure~\ref{fig:gal_mass} represents the surface density of the dense molecular cloud structures observed by the BGPS across the Galactic plane.  Although the mass in that figure is diluted into $0.25$~kpc~$\times\ 0.25$~kpc pixels, it still represents a large-scale measurement of the dense gas mass on scales approximating the spatial resolution of external galaxies in the \herschel\ KINGFISH survey \citep[][]{Kennicutt:2011}.  Placing the Milky Way in the context of fully visible galactic disks is an important step toward linking locally observable small-scale processes with the global pictures gleaned from external galaxies.  In addition to studying the mass surface density of dense molecular cloud structures, we can ask what the dense gas mass fraction is.

Because our estimate of the dense gas mass surface density is measured on kiloparsec scales, it is important to select a model or measurement of the total molecular gas content with similar resolution.  For this purpose, we utilized the azimuthally averaged \htwo\ distribution modeled by \citet{Wolfire:2003} from the \twco\ data presented in \citet{Bronfman:1988}.  This model does not take into account azimuthal asymmetries such as spiral arms or the Galactic bar, but it provides a baseline value to estimate the fraction of the surface density in the Galactic disk measured by the BGPS Distance Catalog.  

Two estimates of the dense gas mass fraction map are shown in Figure~\ref{fig:dgmf}.  The first (panel \emph{a}) was computed by directly dividing the surface density in Figure~\ref{fig:gal_mass} by the \citet{Wolfire:2003} model.  Since the BGPS is sensitive to large-scale diffuse objects at large heliocentric distance (see Figure~\ref{fig:phys_relate}), this first estimate includes many objects whose bulk density is not significantly enhanced over the base \citeauthor{Wolfire:2003} model.  The second estimate of the dense gas mass fraction (panel \emph{b}) was computed in the same way as the first estimate, but includes only objects with $n \geq 10^3$\,\cc.  While this value is not generally considered ``dense'' gas, it represents a view of the distribution of the denser molecular cloud structures detected by the BGPS.

The vast majority of pixels in the maps have $< 1\%$ dense gas mass fraction, but some have $>5\%$ (note that the color scale saturates at $\approx 2.1\%$).  Recall that the surface mass density in Figure~\ref{fig:gal_mass} represents the mean of BGPS sources over 0.25\,kpc~$\times$ 0.25\,kpc pixels, whereas the typical size of GMCs is $\sim10$\,pc.  Therefore, the dense gas mass fraction shown in Figure~\ref{fig:dgmf} should be considered as a lower limit, averaging over a large area, especially in light of the total mass shown in the left panel coming from only 20\% of the BGPS catalog (by number, but 33\% by flux density).  This point should be further emphasized for the right panel, with the number density threshold required for inclusion.

For context, two recent studies have analyzed the dense gas fraction of molecular cloud structures hosting BGPS dust-continuum sources.  \citet{Battisti:2014} compared the masses from BGPS sources with the masses of the parent GRS GMCs, and found a mean value of $0.11^{+0.12}_{-0.06}$.  This study did not apply a specific number density threshold, but rather investigated the fraction of the \thco\ GMC mass visible in dust continuum emission.  In contrast, \citet[][]{Ginsburg:2015} studied the high-mass W51 star forming complex, specifically analyzing the dense gas fraction.  Those authors found that $\sim 10-15\%$ of the W51 GMC mass consists of gas with $n \geq 10^4$\,\cc.  Given the caveats about the nature of Figure~\ref{fig:dgmf}, the present measurement is consistent with both of these values.

In interpreting the dense gas fraction maps in Figure~\ref{fig:dgmf}, two issues bear discussing.  The first is the ``broad-brush'' nature of the \citet{Wolfire:2003} model of \htwo\ gas.  Not only are there sizable uncertainties in both disentangling of (optically thick) \twco\ emission into the near and far kinematic distances and the variable nature of the $X_\mathrm{CO}$ conversion factor as a function of environment \citep[][]{Bolatto:2013}, but enhancements in the surface density of \htwo\ in spiral arm features (coincident with the measured position of BGPS sources) would tend to decrease the measured dense gas mass fraction.

The second issue is the finite resolution of current (sub-)millimeter continuum surveys and the nature of detected objects as a function of heliocentric distance.  Beyond \dsun~$\approx 10$\,kpc, the majority of detected objects are cloud-scale with lower mean density (see Figure~\ref{fig:phys_relate}).  What is the threshold for objects to be included in the dense gas mass fraction?  One illustration of this dilemma is the previously discussed concentration of dense gas at ($x_\mathrm{gal},y_\mathrm{gal}) \approx (-4,3)$\,kpc in  Figure~\ref{fig:dgmf}(\emph{a}).  While this may correspond to gas in the Sagittarius arm, is the BGPS detecting just the dense portion, or whole clouds (\dsun~$\approx 12$\,kpc)?  Despite these caveats, however, this type of measurement offers a key insight into the structure of the Milky Way within the context of galaxies in the local universe.


\section{SUMMARY}\label{ch4:summary}

A major payoff of large-scale blind continuum surveys of the Galactic plane at (sub-)millimeter wavelengths is deriving the physical properties of regions hosting high-mass star formation.  A detailed census of these dense molecular cloud structures can help constrain star-formation and galactic-evolution theories.  In this work, we used the DPDF distance-estimation formalism of EB15 to compute the physical properties for the set of 1,710 BGPS V2 sources in the Distance Catalog.  The significant aspects of this paper are:

\begin{enumerate}

\item We utilized the full information available in the Bayesian DPDF framework introduced in \citet{EllsworthBowers:2013} and EB15 to compute physical property probability density functions.  Although the DPDFs themselves are computed directly on a linear distance scale, the application of the DPDF, in concert with probability density functions for source flux density with uncertainty and an estimate of the temperature distribution of sources, to the computation of physical properties requires Monte Carlo methods.  We present a catalog of the masses, physical radii and number densities for the objects in the BGPS Distance Catalog, with uncertainties derived from the individual probability density functions.

\item Unlike studies of isolated regions at a uniform distance, where flux-density completeness limits map directly onto mass completeness limits, large-scale blind dust-continuum surveys of the Galactic plane suffer from Malmquist bias and care must be taken to estimate the mass completeness levels for these heterogenous samples.  We demonstrate a means for estimating the mass completeness \emph{function} using the rms noise at the locations of sources in the survey.  We estimate that the BGPS Distance Catalog is 90\% complete for $M\geq 400\,M_{_\sun}$ and 50\% complete for $M\geq 75\,M_{_\sun}$.

\item To understand the role of clump-scale features in molecular cloud complexes, we compute both power-law and lognormal fits to the mass distribution.  Fits to the entire collection of sources yield a power-law index $\alpha = 1.94^{+0.34}_{-0.10}$ for objects with masses $10^3\,M_{_\sun} \lesssim M \lesssim 10^5\,M_{_\sun}$, and a lognormal ($\mu,\,\sigma) = (5.3,\,2.0$), where the peak of the distribution is at $M_\mu = e^{5.3} = 200\,M_{_\sun}$.  We also fit both functional forms to a variety of astrophysically motivated subsets in an attempt to identify a collection of BGPS sources that represent a ``clump mass function'' or ``protocluster mass function''.  The subset of nearby sources (\dsun $\leq 6.5$\,kpc) is fitted well by a power-law with index $\hat{\alpha} = 2.01^{+0.11}_{-0.09}$, and the set of clump-mass objects at moderate distances ($M \geq 300$\,\msun\ and 2\,kpc $\leq$ \dsun$ \leq$ 10\,kpc) is fitted well by a power-law with index $\hat{\alpha} = 1.92^{+0.12}_{-0.07}$.  For nearly all subsets, however, we find power-law indices $1.85 \leq \alpha \leq 2.05$.  The exception to this rule is the set of sources at $M \geq 2000\,M_{_\sun}$, which has $\alpha \approx 2.2$; this steeper than expected power-law index may be caused by observation biases in the BGPS against large angular-extent sources.  Lognormal fits to the subsets reveal that the less-homogenous collections of sources are better fit by this form.  Collections of sources at similar distances show stronger power-law behavior.  Furthermore, simulations of sources at a variety of heliocentric distances reveal that these inhomogenous samples produce an observed lognormal mass distribution with a power-law tail regardless of the underlying mass distributions.

\item Combining the heliocentric distances from the DPDFs with the masses derived in this work, we compute the distribution of dense molecular gas in the Galactic plane.  Both the face-on and azimuthally integrated \rgal$-z$ projections of the Galactic mass distributions reveal the prominent high-mass star-forming regions W43, W51, and W49 as significant concentrations of mass.  The mass-weighted vertical mass distribution has a scale height ($\approx 30$\,pc) consistent with the number-count distribution from EB15, implying that mass is uniformly distributed among sources regardless of vertical position.  Dividing the disk's mass surface density (face-on view) by a smooth model for Galactic \htwo\ distribution yields an estimate of the dense gas mass fraction, which peaks around 5\% for the highest mass complexes.  This is consistent with the measured 11\% value from the \citet{Battisti:2014} study of GRS clouds, given that BGPS source mass spreads into pixels 250\,pc on a side commensurate with the typical uncertainty in the DPDFs), whereas typical sizes of GRS clouds are $\lesssim 10$\,pc.  Our measurement is a lower limit due to the inclusion of only sources with well-constrained distances (20\% of the BGPS V2 catalog by number, 33\% by flux density).

\end{enumerate}

The physical properties of BGPS-detected dense molecular cloud structures span the range from core to cloud, and illustrate the fractal nature of molecular cloud complexes, embodied in the \citet{Larson:1981} scaling relationships.  While there are theoretical motivations for these designations, such as gravitational collapse for protoclusters (clumps) or individual stellar systems (cores), the systematic observational effects in current surveys makes it difficult to accurately categorize individual objects, especially since physical processes, such as gravitational binding, are environment specific.  What data such as those presented here can do, however, is begin to provide observational constraints on the evolution of molecular cloud complexes and the context for high-mass star formation.

\begin{figure*}[!t]
  \centering
  \includegraphics[width=6.5in]{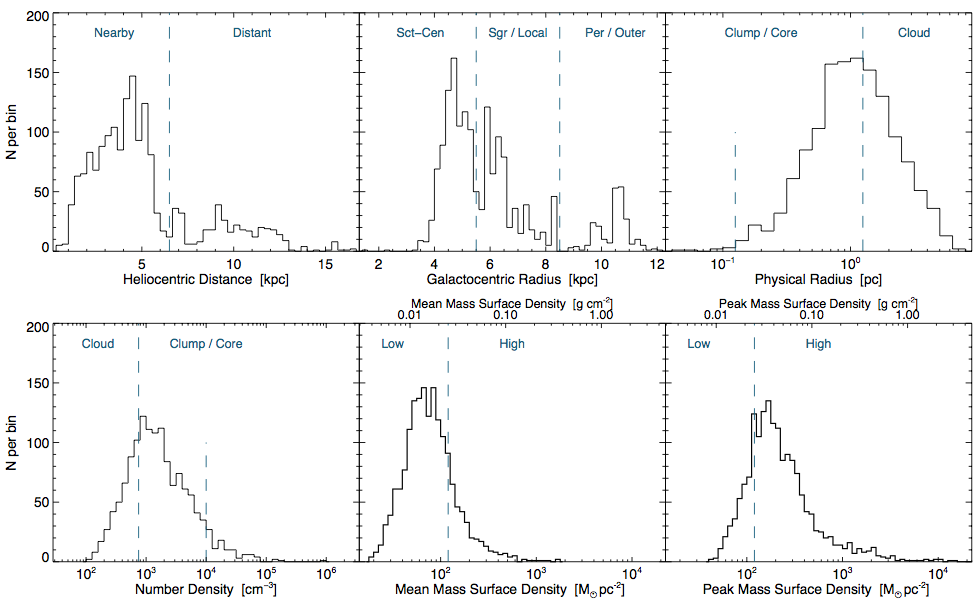}
  \caption{Physical property distributions for BGPS Distance Catalog sources used for division into astrophysically meaningful subsets.  \emph{Top Row:}  From left to right are shown the heliocentric distance, Galactocentric radius, and physical radius distributions of objects.  \emph{Bottom Row:}  From left to right are shown the number density, mean mass surface density, and peak mass surface density distributions of objects.  In each panel the vertical dashed line(s) indicates the dividing line between subsets, with each labeled at top, as listed in Table~\ref{table:mfn_fit_values}.  With the exception of the \dsun\ and \rgal\ panels, only sources with real $\theta_R$ are shown (see \S\ref{res:catalog}).}
  \label{fig:cuts}
\end{figure*}

\acknowledgments

The authors thank A. Clauset for providing implementations of the power-law fitting methods used.  This work was supported by the National Science Foundation through NSF grant AST-1008577.  The BGPS project was supported in part by NSF grant AST-0708403, and was performed at the Caltech Submillimeter Observatory (CSO), supported by NSF grants AST-0540882 and AST-0838261.  The CSO was operated by Caltech under contract from the NSF.  ER is supported by a Discovery Grant from NSERC of Canada.  NJE is supported by NSF grant AST-1109116.

\appendix
\section{Mass Functions Based on Astrophysical Cuts}\label{res:types}

BGPS Distance Catalog objects have wide-ranging distributions of heliocentric distance, Galactocentric radius, physical radius, number density, and mass surface density.  These wide distributions suggest that the present sample is a rather heterogenous set of objects; conclusions drawn from their aggregate mass distribution may be affected by global changes in physical conditions and properties and not necessarily representative of any one type of physical object.  Indeed, we know the BGPS generally identifies cores to a distance of 1\,kpc, clumps from 1\,kpc to 7\,kpc, and clouds beyond 7\,kpc (D11).  As such, it is instructive to divide the total collection of sources into astrophysically meaningful subsets and analyze the resulting mass function fits individually.

The physical quantities chosen to discriminate objects for further analysis are shown in Figure~\ref{fig:cuts}.  Each panel of the figure shows the distribution of sources with dashed vertical lines marking the boundaries between the subsets.  The text in each panel identifies the physically significant subset that corresponds to the entries in Table~\ref{table:mfn_fit_values}.  Note that in the physical radius and number density panels, there is a third region for molecular cloud cores, but they are considered with the clump-scale objects to avoid biases from small-number statistics.  In addition, we examined the effect of an object possessing a dense gas versus \thco\ primary \vlsr\ assignment (see EB15) and attempted several heterogenous cuts in search of a ``protocluster'' mass function (\S\ref{mfn:other}).  The implications of these subsets and their functional fits are discussed in \S\ref{disc:mfn}.

\begin{figure*}[!t]
  \centering
  \includegraphics[width=6.5in]{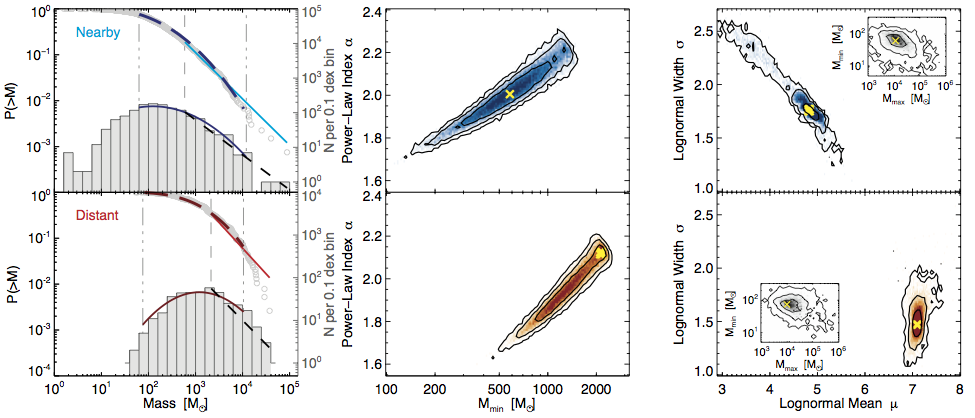}
  \caption[Mass functions of subsets in heliocentric distance.]{Mass functions of subsets in heliocentric distance.  Top row illustrates the nearby subset (\dsun~$\leq 6.5$\,kpc; lines in blue) and the bottom row more distant objects (lines in red).  \emph{Left}: The mass distribution shown as the CCDF (gray circles) and differential mass distribution (gray histogram).  The power-law fits are shown as solid light-colored (CCDF) and dashed black (histogram) lines, and the lognormal fits as long-dashed (CCDF) and solid (histogram) dark-colored lines.  \emph{Middle}: Joint confidence intervals for the power-law fit, as in Figure~\ref{fig:pl_whole}.  \emph{Right}: Joint confidence intervals for the lognormal fit, as in Figure~\ref{fig:ln_whole}.}
  \label{fig:cut_dsun}
\end{figure*}

The functional fits to the mass functions for the following subsets were computed in the same manner as for the entire Distance Catalog, as discussed in \S\ref{res:mfns}.  For compactness, Figures~\ref{fig:cut_dsun} - \ref{fig:cut_other} show the mass distributions and fit parameter joint-confidence intervals in three panels horizontally, with subsets of the same criteria shown as rows.  The left column shows the mean of $2\times10^4$ Monte Carlo mass distributions for the given subset, computed as described in \S\ref{res:mass_dist}.  The middle column displays the [$\hat{\alpha},M_\mathrm{min}$] power-law parameter joint-confidence intervals, as in the right panel of Figure~\ref{fig:pl_whole}, while the right column plots the [$\hat{\mu},\hat{\sigma}$] and [$M_\mathrm{min},M_\mathrm{max}$] lognormal parameter joint confidence intervals (as in Figure~\ref{fig:ln_whole}).  For both the middle and right columns, the rows have the same plotting ranges for easy visualization of the differences between the fits to the subsets.

\subsection{Heliocentric Distance}\label{mfn:dsun}

The heliocentric distance histogram for Distance Catalog objects (Figure~\ref{fig:cuts}, \emph{top left}) exhibits a minimum at \dsun~$\approx6.5$\,kpc.  Beam-scale BGPS objects at this distance have $R \approx 1$\,pc, indicating that most of the sources beyond this point are likely GMC-scale.  Dividing the sample here creates a ``nearby'' group containing 78\% of the objects, with the remaining 22\% in the ``distant'' group.  The mass distributions and functional fit parameters are shown in Figure~\ref{fig:cut_dsun}.

The nearby sample (top row) is not polluted with more distant objects, and the distribution strongly resembles a power law with $\hat{\alpha} = 2.01^{+0.11}_{-0.09}$ for $M \gtrsim 620$\,\msun, as evidenced by the quasi-symmetric joint confidence intervals in the middle panel.  The lognormal fit (right panel) shows a strong preference for values near the joint maximum-likelihood ($\hat{\mu},\,\hat{\sigma}) = (4.85,\,1.72$).  The low-probability tail towards the top-left corner indicates the need for a wider gaussian to fit the data if the mean is moved to lower mass.

In contrast, the distant sample (bottom row) does not exhibit a strong power-law shape, as the middle panel suggests a continuum of [$\hat{\alpha},M_\mathrm{min}$] across the 20,000 Monte Carlo trials, with a marginalized maximum-likelihood power law index $\hat{\alpha} = 1.93_{-0.05}^{+0.19}$.  The small sample size of the distant group also means that each independent realization of the mass distribution will demonstrate more variability in the fit parameters than for a large-$N$ sample.  Despite the volatility of the power-law fits, however, the lognormal fits have a tightly constrained mean around $M_\mu = e^{7.10} \approx 1200\,M_{_\sun}$.  The significant difference between the $\hat{\mu}$ for the nearby and distant subsets clearly illustrates the systematic $M \sim d_{_\sun}^2$.

\subsection{Galactocentric Radius}\label{mfn:rgal}

\begin{figure*}[!t]
  \centering
  \includegraphics[width=6.5in]{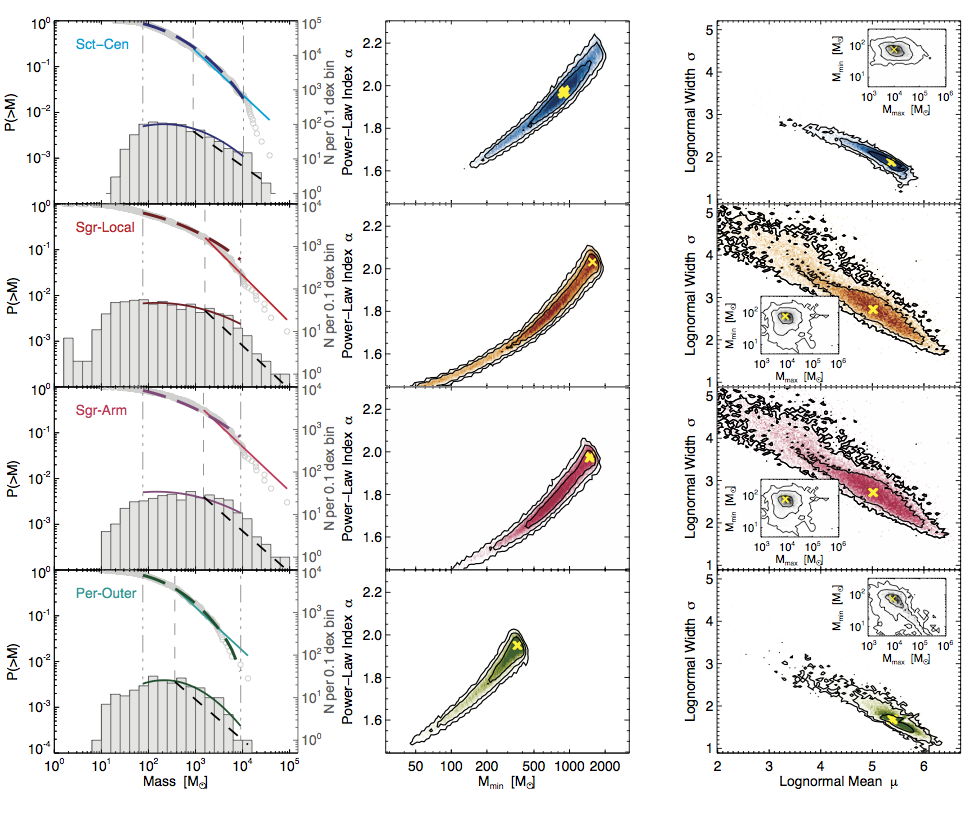}
  \caption[Mass functions of subsets in Galactocentric radius.]{Mass functions of subsets in Galactocentric radius.  Top row illustrates objects in the Scutum-Centarus arm (\rgal~$\leq 5.5$\,kpc; lines in blue), and the second row objects in the Sagittarius and Local arms (5.5\,kpc~$\leq$ \rgal~$\leq$ 8.5\,kpc; lines in red).  The third row are the subset of sources from the second row at heliocentric distances greater than 2.8\,kpc to remove contamination from local sources, meant to illustrate the Sagittarius arm only (lines in pink).  The bottom row objects in the Perseus and Outer arms (lines in green).  Columns are as in Figure~\ref{fig:cut_dsun}.}
  \label{fig:cut_rgal}
\end{figure*}

Galactic environment can influence the properties and evolution of star-forming regions \citep[][]{Kennicutt:1998}, and we use Galactocentric radius as a proxy for Galactic environment to study the differences in the mass distribution from the Molecular Ring out to toward the truncation of the stellar disk \citep[\rgal~$\approx 15$\,kpc;][]{Ruphy:1996}.  We note that disambiguating the effects of heliocentric distance from Galactocentric radius is somewhat difficult due to the Sun's location within the Galactic disk (\ie objects with small \rgal\ have larger \dsun; see Figure~\ref{fig:phys_relate} for the effects on the physical properties).  The histogram of source \rgal\ (Figure~\ref{fig:cuts}, \emph{top center}) shows several concentrations of sources for further investigation.  The molecular ring / Scutum-Centarus Arm feature contains objects at \rgal~$\leq 5.5$\,kpc (50\%), sources at 5.5\,kpc~$\leq$ \rgal~$\leq$ 8.5\,kpc are likely part of either the Sagittarius or Local Arms (36\%), and anything beyond the solar circle (\rgal $\geq 8.5$\,kpc) is part of either the Perseus or Outer Arms (14\%).  This assignment is not one-to-one, as seen in the object-location face-on view of the Milky Way from EB15 (their Figure~15) where some objects apparently identified with the Perseus arm on the far side of the tangent point near $30\degr \lesssim \ell \lesssim 50\degr$ lie within the middle \rgal\ group.  For the purposes of this analysis, distance from the Galactic center is the factor of interest; arm names are simply used as descriptive identifiers.

\begin{figure*}[!t]
  \centering
  \includegraphics[width=6.5in]{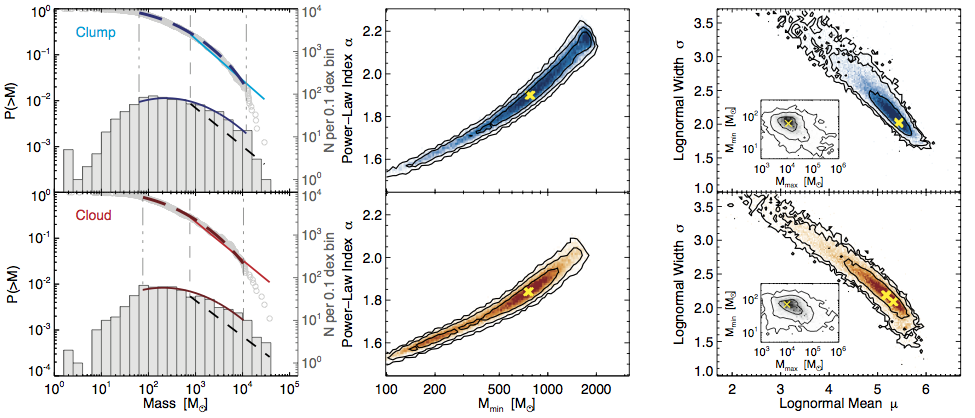}
  \caption[Mass functions of subsets in physical radius.]{Mass functions of subsets in physical radius.  Top row illustrates clump-scale objects ($R \leq 1.25$\,pc; lines in blue), and the bottom row cloud-scale objects (lines in red).  Columns are as in Figure~\ref{fig:cut_dsun}.}
  \label{fig:cut_prad}
\end{figure*}

\begin{figure*}[!t]
  \centering
  \includegraphics[width=6.5in]{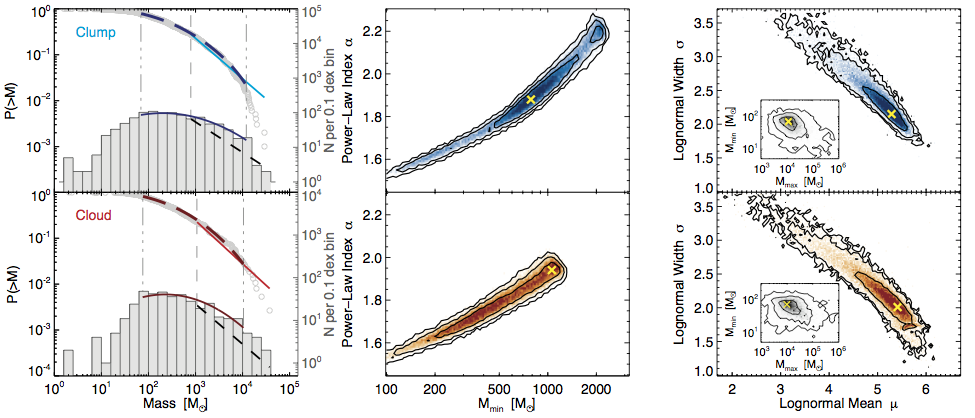}
  \caption[Mass functions of subsets in number density.]{Mass functions of subsets in number density.  Top row illustrates clump-scale objects ($n \geq 750$\,\cc; lines in blue), and the bottom row cloud-scale objects (lines in red).  Columns are as in Figure~\ref{fig:cut_dsun}.}
  \label{fig:cut_nden}
\end{figure*}

The individual mass distributions and functional fits to each subset are shown in Figure~\ref{fig:cut_rgal}.  Objects in the Scutum-Centarus arm (top row) illustrate a somewhat-constrained power-law fit with $\hat{\alpha} = 1.98_{-0.11}^{+0.09}$ for $M\gtrsim 10^3\,M_{_\sun}$, although the same high-mass steepening at $M\gtrsim10^4\,M_{_\sun}$ found for the complete sample is suggested in the CCDF (gray circles).  The lognormal fit for this subset is characterized by ($\hat{\mu},\,\hat{\sigma}) = (5.4,\,1.7)$, with a tail extending towards lower $\mu$, higher $\sigma$, as is a general feature of these fits.  Neither functional form yields tight parameter constraints in this case, suggesting the existence of substantial remaining inhomogeneities.

The second row of Figure~\ref{fig:cut_rgal} illustrates intermediate \rgal\ sources, mostly associated with the Sagittarius and Local arms.  The wide spread in the mass distribution is a consequence of the large range of \dsun\ corresponding to this Galactocentric annulus.  Although the maximum-likelihood power-law parameter values somewhat favor the high-mass end of the distribution ($\hat{\alpha} = 2.02_{-0.28}^{+0.02}$ for $M\gtrsim 1600$\,\msun), the middle panel demonstrates a long tail of Monte Carlo trials to low-$\alpha$.  While this spread in the power-law fit parameters seems to indicate a continuously curved mass distribution, the right panel shows no particular localization of the lognormal parameters.  The maximum-likelihood parameters, ($\hat{\mu},\,\hat{\sigma}$) = (4.95,\,2.74), reasonably fits the wide plateau of the distribution, but the extremely heterogenous nature of this sample suggest further cuts are necessary.  As a test, objects likely to be in the Local Arm (\dsun~$\lesssim 2.8$\,kpc, see EB15) were excluded from the fits shown in the third row of Figure~\ref{fig:cut_rgal}.  While the power-law fit is marginally more constrained ($\hat{\alpha} = 1.97_{-0.24}^{+0.01}$ for $M\gtrsim 1600$\,\msun), the lognormal fit is indistinguishable from the entire intermediate \rgal\ subset.

Objects in the Perseus and Outer arms beyond the Solar circle are depicted in the bottom row of Figure~\ref{fig:cut_rgal}, where the small size of this subset ($N=234$) compared to the others leads to the sharp truncation in the power-law parameters around $M_\mathrm{min} = 400\,M_{_\sun}$, with a maximum-likelihood $\hat{\alpha} = 1.94_{-0.10}^{+0.03}$.  The lognormal fits (right panel) are constrained approximately as well as for the complete sample, but with a slightly larger mean.  This subset is generally drawn from the outer Galaxy targeted fields in the BGPS, and likely is not completely representative of the outermost regions of the Galactic disk.

\subsection{Clumps vs. Clouds}\label{mfn:clump}

\begin{figure*}[!t]
  \centering
  \includegraphics[width=6.5in]{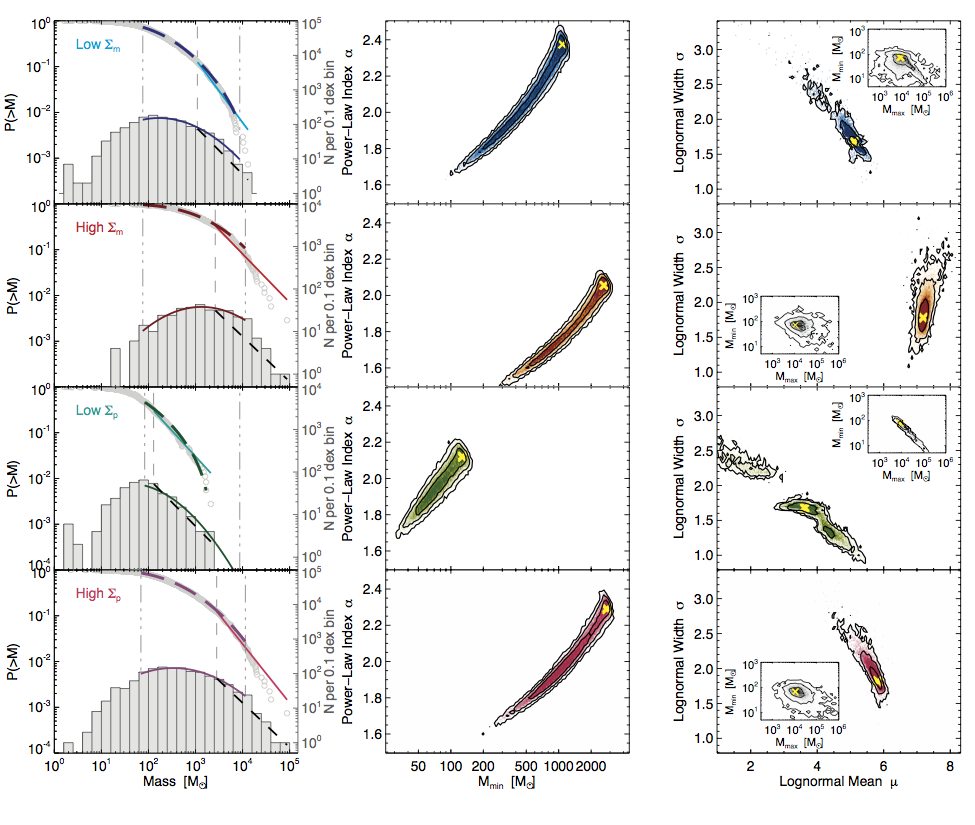}
  \caption{Mass functions of subsets in mass surface density.  The top two rows depict objects divided by mean surface density, either below $\Sigma_\mathrm{m}~= 120\,\mpc$ (top row, blue lines) or above (second row, red lines).  The bottom two rows show sources divided by peak surface density, either below $\Sigma_\mathrm{p}~= 120\,\mpc$ (third row, green lines) or above (bottom row, pink lines).  Columns are as in Figure~\ref{fig:cut_dsun}.}
  \label{fig:cut_sden}
\end{figure*}

Being shallower functions heliocentric distance, the derived physical radii and number densities of BGPS-detected dense molecular cloud structures can possibly divide the catalog into subsets more closely matched to physical distinctions in the interstellar medium (Figure~\ref{fig:cuts}, \emph{top right} and \emph{bottom left}).  There are relatively few objects falling in the ``core'' category ($R \leq 0.125$\,pc or $n \geq 10^4$\,\cc), and we combine these with the ``clump'' subset here.  We therefore divide sources into ``core/clump'' (59\% of the physical radius distribution and 73\% of the number density distribution) and ``cloud'' subsets.  The lack of a one-to-one mapping between these two physical tracers is likely due to variations in source geometry away from the cylindrical model used to compute $n$.  Furthermore, there are likely observational biases present that skew this relationship, including the blending of substructure into larger-radius complexes when convolved with the telescope beam \citep[see][]{Merello:2015}.

Functional fits to the mass distributions are shown in Figures~\ref{fig:cut_prad} (physical radius), and \ref{fig:cut_nden} (number density).  Given the similarities between the mass distributions for the two methods of classifying dense molecular cloud structures, we discuss them together.  The clump-scale objects are shown in the top rows of each figure.  For both physical property criteria, the power-law fits stretch over wide range of parameter space with indices that span a range $1.6 \lesssim \hat{\alpha} \lesssim 2.2$, although both have a maximum-likelihood value near $\hat{\alpha} = 1.9$.  The lognormal fits also return similar values for the two clump criteria, with peaks at slightly lower mass and widths slightly larger than the lognormal fit to the complete data set.

The two cloud criteria (bottom row of Figures~\ref{fig:cut_prad} and \ref{fig:cut_nden}) also show great similarity in their functional fits.  The primary difference in the power-law fits comes from the smaller sample size for number density group ($N=376$ for clouds with $n < 750$\,\cc\ versus $N=562$ for the clouds with $R > 1.25$\,pc) leading to the truncation of the joint confidence intervals at large $M_\mathrm{min}$ in the bottom middle panel of Figure~\ref{fig:cut_nden}.  The lognormal fits are very similar, both to each other and to the clump fits in the top rows of these figures.  These sets of fits suggest that dividing the dense molecular cloud structures into categories based on physical radius or number density alone do not provide homogenous or distinct subsets for which meaningful mass functions may be fit that help constrain the theories of molecular cloud evolution.

\subsection{Mass Surface Density}\label{mfn:sden}

\begin{figure*}[!t]
  \centering
  \includegraphics[width=6.5in]{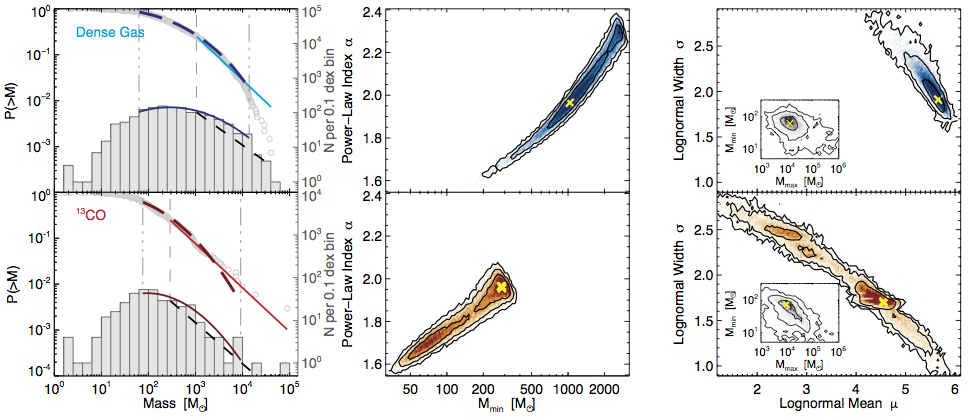}
  \caption{Mass functions of subsets in velocity type.  Top row includes objects whose \vlsr\ was derived from spectroscopic observation of a dense gas tracer (\eg \hcop(3-2); lines in blue), and the bottom row are objects whose kinematic information comes only from the \thco\ isolation technique of EB15 (lines in red).  Columns are as in Figure~\ref{fig:cut_dsun}.}
  \label{fig:cut_vtype}
\end{figure*}

Investigation of Larson's 3\srd Law in terms of deviations away from a systematic $M \sim R^2$ curve showed that while the mean mass surface density of Distance Catalog objects spans less than two orders of magnitude around $\Sigma = 100\,\mpc$, there appears to be a slight upward trend as a function of mass (\S\ref{res:larson}).  In studies of star forming regions, the value $A_V = 8$~mag is often cited as a threshold for ongoing star formation due to self-shielding from ionizing radiation \citep[][]{McKee:1989}.  Depending on the dust physics assumptions used, this translates into a mass surface density of $\Sigma = 120 - 160\,\mpc$ \citep[][]{Lada:2013}.  The histogram of mean mass surface density for Distance Catalog sources is shown in Figure~\ref{fig:cuts}, \emph{bottom center}, and the vertical dashed line marks the $A_V = 8$~mag value.  The top axis of this panel lists the $\Sigma$ values in g~cm$^{-2}$, as this unit is discussed in terms of thresholds for star formation.  In similar fashion, we considered the \emph{peak} mass surface density, computed from the peak flux density reported in the BGPS catalog and the 33\arcsec\ size of the BGPS beam, and the resulting source histogram is shown in the bottom right panel of Figure~\ref{fig:cuts} with similar markings as the mean mass surface density.

In Figure~\ref{fig:cut_sden}, we show the mass distributions and functional fits for the four subsets based on mass surface density.  The top row depicts the sources with mean $\Sigma$ below $120\,\mpc$, and comprises some 80\% of the Distance catalog.  This set is not well-fit by a power law ($\hat{\alpha} = 2.38^{+0.03}_{-0.36}$), but does have a moderately constrained lognormal fit ($\hat{\mu},\hat{\sigma})~= (5.1,1.7$).  The positive correlation between mass and mean $\Sigma$ can be seen in the differences between the lognormal fits of these two subsets.

The fits for the subsets divided by peak mass surface density are shown in the bottom two rows of Figure~\ref{fig:cut_sden}.  In this case, the bulk of the sources (78\%) have peak $\Sigma$ \emph{above} the $120\,\mpc$ threshold.  The lack of high-mass sources in the third row of the figure indicates that there are no large diffuse sources whose total mass is large despite having low $\Sigma$; this is commensurate with the BGPS angular transfer function (G13).  

\subsection{Type of Radial Velocity Measurement}\label{mfn:vtype}

Kinematic distances for BGPS sources rely primarily upon \vlsr\ measurements from dense gas tracers (\eg \hcop(3-2), \nhhh(1,1)) in concert with a Galactic rotation curve.  \citet{Shirley:2013} showed for BGPS sources that detection in one of the dense gas lines is a strong function of millimeter flux density.  To expand the available catalog of objects for which kinematic distances may be computed, EB15 demonstrated a means for combining continuum data with diffuse-gas tracing \thco(1-0) to extract \vlsr\ in agreement with the dense-gas values at a rate of 95\%.  This method was applied to dense molecular cloud structures not detected in one of the dense-gas surveys to expand the number of objects in the BGPS Distance Catalog.  The type of \vlsr\ assignment, therefore, offers an additional cut for investigating mass function fits.

The top row of Figure~\ref{fig:cut_vtype} shows the mass distribution and functional fit parameters for the set of objects detected in one of the dense gas tracers (see EB15 for details).  Since this set comprises 76\% of the Distance Catalog, it is unsurprising that the functional fits are nearly identical to those of the entire Catalog shown in Figures~\ref{fig:pl_whole} and \ref{fig:ln_whole}.  The set of sources identified only with \thco(1-0) is generally lower mass, and has less-constrained fits.

\subsection{Attempts to Find a ``Protocluster'' Mass Function}\label{mfn:other}

\begin{figure*}[!t]
  \centering
  \includegraphics[width=6.5in]{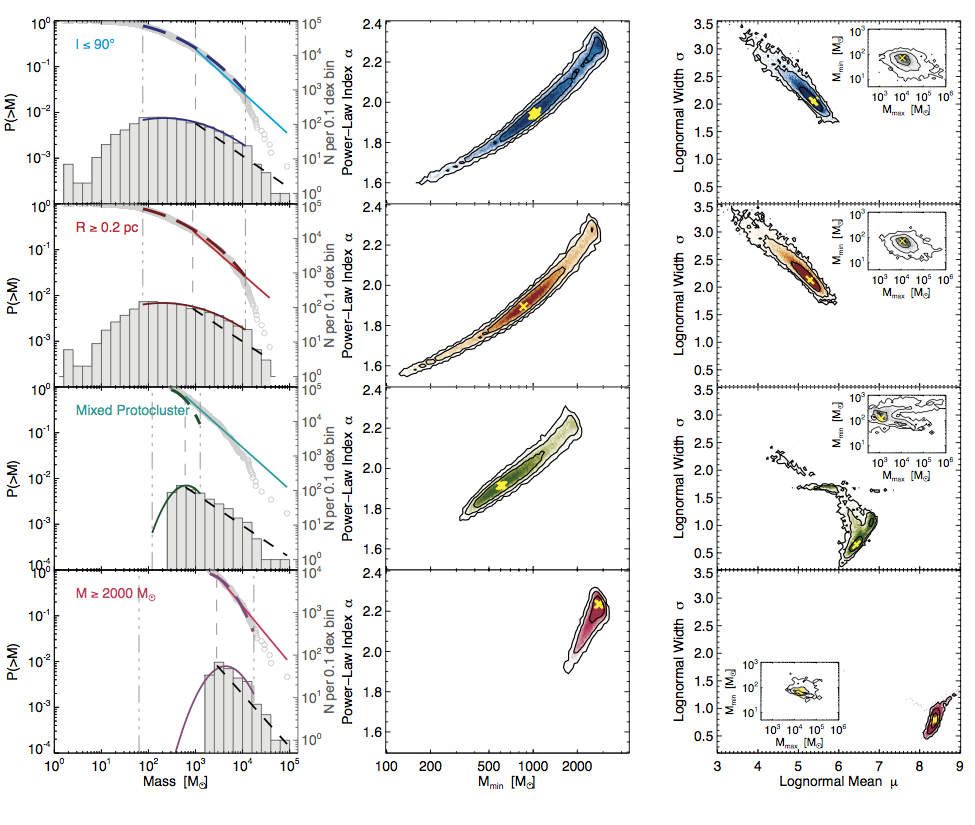}
  \caption{Mass functions of protocluster-type subsets.  Top row only includes objects in the blind portion of the survey ($\ell \leq 90\degr$; lines in blue), the second row objects of probable protocluster physical radius ($R \geq 0.2$\,pc; lines in red), the third row objects with a mixed set of criteria (2\,kpc~$\leq$ \dsun~$\leq$ 10\,kpc and $M\geq300\,M_{_\sun}$; lines in green), and the bottom row the most high-mass objects ($M\geq2000\,M_{_\sun}$; lines in pink).  Columns are as in Figure~\ref{fig:cut_dsun}.}
  \label{fig:cut_other}
\end{figure*}

In addition to the strict cuts along the physical quantities described in Figure~\ref{fig:cuts}, additional subsets of the data were constructed in an attempt to define a ``protocluster'' mass function.  The subsets are shown in Figure~\ref{fig:cut_other} along with their respective functional fits, and are discussed below.

The first two subsets are the blindly surveyed portion of the BGPS ($\ell \leq 90\degr$) and all objects with physical radius $R \geq 0.2$\,pc.  The outer Galaxy portions of the BGPS were targeted specifically based on known regions of star formation or concentrations of emission in the \twco(1-0) maps of \citet{Dame:2001}, meaning they are typically nearby clouds and suffer significant selection bias.  Utilizing sources only from the inner Galaxy, where survey coverage was not (generally) predicated on prior knowledge of the structure of emission,\footnote{The ``generally'' arises because of the several regions along the inner Galactic plane where the BGPS latitude coverage was flared out from the nominal $|b| \leq 0\fdg5$ to $|b| \leq 1\fdg5$ in areas with known large vertical extent of star-formation activity \citep[see][]{Aguirre:2011}.} allows for a more unbiased sampling of sources (and therefore mass distribution) for fitting a mass function. The second subset removes all core-scale objects and those with non-real solutions for $\theta_R$ (which have low flux-density and therefore likely low mass, see \S\ref{res:catalog}).  The removal of low-mass objects does not appreciably affect the high-mass end of the mass distribution, and both subsets have power-law fit parameters $\hat{\alpha} \approx 1.9,\ M_\mathrm{min} \approx 900\,M_{_\sun}$, consistent with the fit parameters for the complete set.  The loci of points in middle panels of Figure~\ref{fig:cut_other} bear a strong resemblance to that in Figure~\ref{fig:pl_whole}.  The lognormal fits similarly reflect the entire sample, and do not add additional insight.

The other two subsets shown in Figure~\ref{fig:cut_other} begin to more strongly resemble power-law distributions.  The third row is the subset of sources with 2\,kpc~$\leq$ \dsun~$\leq 10$\,kpc and $M \geq 300$\,\msun\ chosen to reflect a mixed criterion in search of protoclusters.  These conditions remove the most nearby objects (likely cores) and most distant objects (likely full clouds), and restricts consideration to a mass range consistent with star clusters \citep[][]{Kennicutt:2012}.  The power-law fit joint confidence regions produce a narrow range of parameters with maximum-likelihood values at $\hat{\alpha} = 1.92_{-0.07}^{+0.12},\ M_\mathrm{min} = 620_{-160}^{+380}$\,\msun.  The power-law index is consistent with many of the fits, and extends to near the 90\% mass completeness level.  The lognormal fit displays an unusual set of contours, indicating that this functional form is not an appropriate description of the mass distribution.  Finally, the bottom row of Figure~\ref{fig:cut_other} illustrates the fit to just the high-mass end of the BGPS distribution ($M \geq 2000$\,\msun).  The power-law fit parameters line up with the uppermost collection of points in the fit to the full sample, confirming the steep distribution ($\hat{\alpha} = 2.23_{-0.06}^{+0.03}$) for these sources.  While these largest objects should conform to the observed GMC power-law index $\alpha \approx 1.6-1.8$ \citep[][]{Blitz:1993}, we discuss in \S\ref{disc:mfn} some observational effects that might lead to the apparent steepening.  The lognormal fit is tightly constrained, but since the minimum mass of this subset is substantially larger than the completeness levels, no real turnover in the data is expected.

~
~


\begin{thebibliography}{}
\expandafter\ifx\csname natexlab\endcsname\relax\def\natexlab#1{#1}\fi

\bibitem[{{Aguirre} {et~al.}(2011){Aguirre}, {Ginsburg}, {Dunham}, {Drosback},
  {Bally}, {Battersby}, {Bradley}, {Cyganowski}, {Dowell}, {Evans}, {Glenn},
  {Harvey}, {Rosolowsky}, {Stringfellow}, {Walawender}, \&
  {Williams}}]{Aguirre:2011}
{Aguirre}, J.~E., {Ginsburg}, A.~G., {Dunham}, M.~K., {et~al.} 2011, \apjs,
  192, 4

\bibitem[{{Ballesteros-Paredes} {et~al.}(2012){Ballesteros-Paredes},
  {D'Alessio}, \& {Hartmann}}]{BallesterosParedes:2012}
{Ballesteros-Paredes}, J., {D'Alessio}, P., \& {Hartmann}, L. 2012, \mnras,
  427, 2562

\bibitem[{{Bania} {et~al.}(2012){Bania}, {Anderson}, \& {Balser}}]{Bania:2012}
{Bania}, T.~M., {Anderson}, L.~D., \& {Balser}, D.~S. 2012, \apj, 759, 96

\bibitem[{{Bania} {et~al.}(2010){Bania}, {Anderson}, {Balser}, \&
  {Rood}}]{Bania:2010}
{Bania}, T.~M., {Anderson}, L.~D., {Balser}, D.~S., \& {Rood}, R.~T. 2010,
  \apjl, 718, L106

\bibitem[{{Barnard} {et~al.}(2004){Barnard}, {Vielva}, {Pierce-Price}, {Blain},
  {Barreiro}, {Richer}, \& {Qualtrough}}]{Barnard:2004}
{Barnard}, V.~E., {Vielva}, P., {Pierce-Price}, D.~P.~I., {et~al.} 2004,
  \mnras, 352, 961

\bibitem[{{Bastian} {et~al.}(2010){Bastian}, {Covey}, \&
  {Meyer}}]{Bastian:2010}
{Bastian}, N., {Covey}, K.~R., \& {Meyer}, M.~R. 2010, \araa, 48, 339

\bibitem[{{Battersby} {et~al.}(2011){Battersby}, {Bally}, {Ginsburg},
  {Bernard}, {Brunt}, {Fuller}, {Martin}, {Molinari}, {Mottram}, {Peretto},
  {Testi}, \& {Thompson}}]{Battersby:2011}
{Battersby}, C., {Bally}, J., {Ginsburg}, A., {et~al.} 2011, \aap, 535, A128

\bibitem[{{Battisti} \& {Heyer}(2014)}]{Battisti:2014}
{Battisti}, A.~J., \& {Heyer}, M.~H. 2014, \apj, 780, 173, (BH14)

\bibitem[{{Beech}(1987)}]{Beech:1987}
{Beech}, M. 1987, \apss, 133, 193

\bibitem[{{Bergin} \& {Tafalla}(2007)}]{Bergin:2007}
{Bergin}, E.~A., \& {Tafalla}, M. 2007, \araa, 45, 339

\bibitem[{{Blitz}(1993)}]{Blitz:1993}
{Blitz}, L. 1993, in Protostars and Planets III, ed. E.~H. {Levy} \& J.~I.
  {Lunine}, 125--161

\bibitem[{{Bolatto} {et~al.}(2013){Bolatto}, {Wolfire}, \&
  {Leroy}}]{Bolatto:2013}
{Bolatto}, A.~D., {Wolfire}, M., \& {Leroy}, A.~K. 2013, \araa, 51, 207

\bibitem[{{Bronfman} {et~al.}(1988){Bronfman}, {Cohen}, {Alvarez}, {May}, \&
  {Thaddeus}}]{Bronfman:1988}
{Bronfman}, L., {Cohen}, R.~S., {Alvarez}, H., {May}, J., \& {Thaddeus}, P.
  1988, \apj, 324, 248

\bibitem[{{Chabrier}(2003)}]{Chabrier:2003}
{Chabrier}, G. 2003, \pasp, 115, 763

\bibitem[{{Clauset} {et~al.}(2009){Clauset}, {Shalizi}, \&
  {Newman}}]{Clauset:2009}
{Clauset}, A., {Shalizi}, C.~R., \& {Newman}, M.~E.~J. 2009, SIAM Review, 51,
  661

\bibitem[{{Cohen} {et~al.}(1980){Cohen}, {Cong}, {Dame}, \&
  {Thaddeus}}]{Cohen:1980}
{Cohen}, R.~S., {Cong}, H., {Dame}, T.~M., \& {Thaddeus}, P. 1980, \apjl, 239,
  L53

\bibitem[{{Dame} {et~al.}(2001){Dame}, {Hartmann}, \& {Thaddeus}}]{Dame:2001}
{Dame}, T.~M., {Hartmann}, D., \& {Thaddeus}, P. 2001, \apj, 547, 792

\bibitem[{{Dame} {et~al.}(1987){Dame}, {Ungerechts}, {Cohen}, {de Geus},
  {Grenier}, {May}, {Murphy}, {Nyman}, \& {Thaddeus}}]{Dame:1987}
{Dame}, T.~M., {Ungerechts}, H., {Cohen}, R.~S., {et~al.} 1987, \apj, 322, 706

\bibitem[{{Donkov} {et~al.}(2012){Donkov}, {Veltchev}, \&
  {Klessen}}]{Donkov:2012}
{Donkov}, S., {Veltchev}, T.~V., \& {Klessen}, R.~S. 2012, \mnras, 423, 889

\bibitem[{{Dunham} {et~al.}(2011){Dunham}, {Rosolowsky}, {Evans}, {Cyganowski},
  \& {Urquhart}}]{Dunham:2011c}
{Dunham}, M.~K., {Rosolowsky}, E., {Evans}, II, N.~J., {Cyganowski}, C., \&
  {Urquhart}, J.~S. 2011, \apj, 741, 110, (D11)

\bibitem[{{Dunham} {et~al.}(2010){Dunham}, {Rosolowsky}, {Evans}, {Cyganowski},
  {Aguirre}, {Bally}, {Battersby}, {Bradley}, {Dowell}, {Drosback}, {Ginsburg},
  {Glenn}, {Harvey}, {Merello}, {Schlingman}, {Shirley}, {Stringfellow},
  {Walawender}, \& {Williams}}]{Dunham:2010}
{Dunham}, M.~K., {Rosolowsky}, E., {Evans}, II, N.~J., {et~al.} 2010, \apj,
  717, 1157

\bibitem[{{Egusa} {et~al.}(2011){Egusa}, {Koda}, \& {Scoville}}]{Egusa:2011}
{Egusa}, F., {Koda}, J., \& {Scoville}, N. 2011, \apj, 726, 85

\bibitem[{{Ellsworth-Bowers} {et~al.}(2015){Ellsworth-Bowers}, {Rosolowsky},
  {Glenn}, {Ginsburg}, {Evans}, {Battersby}, {Shirley}, \& {Svoboda}}]{EB14a}
{Ellsworth-Bowers}, T.~P., {Rosolowsky}, E., {Glenn}, J., {et~al.} 2015, \apj,
  799, 29, (EB15)

\bibitem[{{Ellsworth-Bowers} {et~al.}(2013){Ellsworth-Bowers}, {Glenn},
  {Rosolowsky}, {Mairs}, {Evans}, {Battersby}, {Ginsburg}, {Shirley}, \&
  {Bally}}]{EllsworthBowers:2013}
{Ellsworth-Bowers}, T.~P., {Glenn}, J., {Rosolowsky}, E., {et~al.} 2013, \apj,
  770, 39

\bibitem[{{Elmegreen}(1985)}]{Elmegreen:1985}
{Elmegreen}, B.~G. 1985, in Birth and Infancy of Stars, ed. R.~{Lucas},
  A.~{Omont}, \& R.~{Stora}, 257--277

\bibitem[{{Enoch} {et~al.}(2007){Enoch}, {Glenn}, {Evans}, {Sargent}, {Young},
  \& {Huard}}]{Enoch:2007}
{Enoch}, M.~L., {Glenn}, J., {Evans}, II, N.~J., {et~al.} 2007, \apj, 666, 982

\bibitem[{{Enoch} {et~al.}(2006){Enoch}, {Young}, {Glenn}, {Evans}, {Golwala},
  {Sargent}, {Harvey}, {Aguirre}, {Goldin}, {Haig}, {Huard}, {Lange},
  {Laurent}, {Maloney}, {Mauskopf}, {Rossinot}, \& {Sayers}}]{Enoch:2006}
{Enoch}, M.~L., {Young}, K.~E., {Glenn}, J., {et~al.} 2006, \apj, 638, 293

\bibitem[{{Federrath} \& {Klessen}(2012)}]{Federrath:2012}
{Federrath}, C., \& {Klessen}, R.~S. 2012, \apj, 761, 156

\bibitem[{{Giannetti} {et~al.}(2013){Giannetti}, {Brand}, {S{\'a}nchez-Monge},
  {Fontani}, {Cesaroni}, {Beltr{\'a}n}, {Molinari}, {Dodson}, \&
  {Rioja}}]{Giannetti:2013}
{Giannetti}, A., {Brand}, J., {S{\'a}nchez-Monge}, {\'A}., {et~al.} 2013, \aap,
  556, A16

\bibitem[{{Ginsburg} {et~al.}(2015){Ginsburg}, {Bally}, {Battersby},
  {Youngblood}, {Darling}, {Rosolowsky}, {Arce}, \& {Lebr{\'o}n
  Santos}}]{Ginsburg:2015}
{Ginsburg}, A., {Bally}, J., {Battersby}, C., {et~al.} 2015, \aap, 573, A106

\bibitem[{{Ginsburg} {et~al.}(2013){Ginsburg}, {Glenn}, {Rosolowsky},
  {Ellsworth-Bowers}, {Battersby}, {Dunham}, {Merello}, {Shirley}, {Bally},
  {Evans}, {Stringfellow}, \& {Aguirre}}]{Ginsburg:2013}
{Ginsburg}, A., {Glenn}, J., {Rosolowsky}, E., {et~al.} 2013, \apjs, 208, 14,
  (G13)

\bibitem[{{G{\'o}mez} {et~al.}(2014){G{\'o}mez}, {Wyrowski}, {Schuller},
  {Menten}, \& {Ballesteros-Paredes}}]{Gomez:2014}
{G{\'o}mez}, L., {Wyrowski}, F., {Schuller}, F., {Menten}, K.~M., \&
  {Ballesteros-Paredes}, J. 2014, \aap, 561, A148

\bibitem[{{Hennebelle} \& {Chabrier}(2011)}]{Hennebelle:2011}
{Hennebelle}, P., \& {Chabrier}, G. 2011, \apjl, 743, L29

\bibitem[{{Hennebelle} \& {Chabrier}(2013)}]{Hennebelle:2013b}
---. 2013, \apj, 770, 150

\bibitem[{{Heyer} {et~al.}(2009){Heyer}, {Krawczyk}, {Duval}, \&
  {Jackson}}]{Heyer:2009}
{Heyer}, M., {Krawczyk}, C., {Duval}, J., \& {Jackson}, J.~M. 2009, \apj, 699,
  1092

\bibitem[{{Hildebrand}(1983)}]{Hildebrand:1983}
{Hildebrand}, R.~H. 1983, \qjras, 24, 267

\bibitem[{{Hopkins}(2012)}]{Hopkins:2012a}
{Hopkins}, P.~F. 2012, \mnras, 423, 2016

\bibitem[{{Hopkins}(2013)}]{Hopkins:2013d}
---. 2013, \mnras, 433, 170

\bibitem[{{Hopkins} {et~al.}(2014){Hopkins}, {Kere{\v s}}, {O{\~n}orbe},
  {Faucher-Gigu{\`e}re}, {Quataert}, {Murray}, \& {Bullock}}]{Hopkins:2014e}
{Hopkins}, P.~F., {Kere{\v s}}, D., {O{\~n}orbe}, J., {et~al.} 2014, \mnras,
  445, 581

\bibitem[{{Jackson} {et~al.}(2006){Jackson}, {Rathborne}, {Shah}, {Simon},
  {Bania}, {Clemens}, {Chambers}, {Johnson}, {Dormody}, {Lavoie}, \&
  {Heyer}}]{Jackson:2006}
{Jackson}, J.~M., {Rathborne}, J.~M., {Shah}, R.~Y., {et~al.} 2006, \apjs, 163,
  145

\bibitem[{{Juvela} {et~al.}(2015){Juvela}, {Ristorcelli}, {Marshall},
  {Montillaud}, {Pelkonen}, {Ysard}, {McGehee}, {Paladini}, {Pagani},
  {Malinen}, {A.}, {Rivera-Ingraham}, {Lefevre}, {Toth}, {Montier}, {Bernard},
  \& {Martin}}]{Juvela:2015}
{Juvela}, M., {Ristorcelli}, I., {Marshall}, D.~J., {et~al.} 2015, ArXiv
  e-prints, arXiv:1501.07092, (A\&A, in press)

\bibitem[{{Kamenetzky} {et~al.}(2014){Kamenetzky}, {Rangwala}, {Glenn},
  {Maloney}, \& {Conley}}]{Kamenetzky:2014}
{Kamenetzky}, J., {Rangwala}, N., {Glenn}, J., {Maloney}, P.~R., \& {Conley},
  A. 2014, \apj, 795, 174

\bibitem[{{Katz} {et~al.}(1996){Katz}, {Weinberg}, \& {Hernquist}}]{Katz:1996}
{Katz}, N., {Weinberg}, D.~H., \& {Hernquist}, L. 1996, \apjs, 105, 19

\bibitem[{{Kennicutt} \& {Evans}(2012)}]{Kennicutt:2012}
{Kennicutt}, R.~C., \& {Evans}, N.~J. 2012, \araa, 50, 531

\bibitem[{{Kennicutt} {et~al.}(2011){Kennicutt}, {Calzetti}, {Aniano},
  {Appleton}, {Armus}, {Beir{\~a}o}, {Bolatto}, {Brandl}, {Crocker}, {Croxall},
  {Dale}, {Meyer}, {Draine}, {Engelbracht}, {Galametz}, {Gordon}, {Groves},
  {Hao}, {Helou}, {Hinz}, {Hunt}, {Johnson}, {Koda}, {Krause}, {Leroy}, {Li},
  {Meidt}, {Montiel}, {Murphy}, {Rahman}, {Rix}, {Roussel}, {Sandstrom},
  {Sauvage}, {Schinnerer}, {Skibba}, {Smith}, {Srinivasan}, {Vigroux},
  {Walter}, {Wilson}, {Wolfire}, \& {Zibetti}}]{Kennicutt:2011}
{Kennicutt}, R.~C., {Calzetti}, D., {Aniano}, G., {et~al.} 2011, \pasp, 123,
  1347

\bibitem[{{Kennicutt}(1998)}]{Kennicutt:1998}
{Kennicutt}, Jr., R.~C. 1998, \apj, 498, 541

\bibitem[{{Kere{\v s}} {et~al.}(2009){Kere{\v s}}, {Katz}, {Dav{\'e}},
  {Fardal}, \& {Weinberg}}]{Keres:2009}
{Kere{\v s}}, D., {Katz}, N., {Dav{\'e}}, R., {Fardal}, M., \& {Weinberg},
  D.~H. 2009, \mnras, 396, 2332

\bibitem[{{Kim} {et~al.}(2014){Kim}, {Abel}, {Agertz}, {Bryan}, {Ceverino},
  {Christensen}, {Conroy}, {Dekel}, {Gnedin}, {Goldbaum}, {Guedes}, {Hahn},
  {Hobbs}, {Hopkins}, {Hummels}, {Iannuzzi}, {Keres}, {Klypin}, {Kravtsov},
  {Krumholz}, {Kuhlen}, {Leitner}, {Madau}, {Mayer}, {Moody}, {Nagamine},
  {Norman}, {Onorbe}, {O'Shea}, {Pillepich}, {Primack}, {Quinn}, {Read},
  {Robertson}, {Rocha}, {Rudd}, {Shen}, {Smith}, {Szalay}, {Teyssier},
  {Thompson}, {Todoroki}, {Turk}, {Wadsley}, {Wise}, {Zolotov}, \& {AGORA
  Collaboration29}}]{Kim:2014}
{Kim}, J.-h., {Abel}, T., {Agertz}, O., {et~al.} 2014, \apjs, 210, 14

\bibitem[{{Kritsuk} {et~al.}(2013){Kritsuk}, {Lee}, \& {Norman}}]{Kritsuk:2013}
{Kritsuk}, A.~G., {Lee}, C.~T., \& {Norman}, M.~L. 2013, \mnras, 436, 3247

\bibitem[{{Kroupa}(2001)}]{Kroupa:2001}
{Kroupa}, P. 2001, \mnras, 322, 231

\bibitem[{{Krumholz} {et~al.}(2005){Krumholz}, {McKee}, \&
  {Klein}}]{Krumholz:2005}
{Krumholz}, M.~R., {McKee}, C.~F., \& {Klein}, R.~I. 2005, \nat, 438, 332

\bibitem[{{Lada} {et~al.}(2013){Lada}, {Lombardi}, {Roman-Zuniga}, {Forbrich},
  \& {Alves}}]{Lada:2013}
{Lada}, C.~J., {Lombardi}, M., {Roman-Zuniga}, C., {Forbrich}, J., \& {Alves},
  J.~F. 2013, \apj, 778, 133

\bibitem[{{Larson}(1981)}]{Larson:1981}
{Larson}, R.~B. 1981, \mnras, 194, 809

\bibitem[{{Martin} {et~al.}(2012){Martin}, {Roy}, {Bontemps},
  {Miville-Desch{\^e}nes}, {Ade}, {Bock}, {Chapin}, {Devlin}, {Dicker},
  {Griffin}, {Gundersen}, {Halpern}, {Hargrave}, {Hughes}, {Klein}, {Marsden},
  {Mauskopf}, {Netterfield}, {Olmi}, {Patanchon}, {Rex}, {Scott}, {Semisch},
  {Truch}, {Tucker}, {Tucker}, {Viero}, \& {Wiebe}}]{Martin:2012}
{Martin}, P.~G., {Roy}, A., {Bontemps}, S., {et~al.} 2012, \apj, 751, 28

\bibitem[{{McKee}(1989)}]{McKee:1989}
{McKee}, C.~F. 1989, \apj, 345, 782

\bibitem[{{McKee} \& {Ostriker}(2007)}]{McKee:2007}
{McKee}, C.~F., \& {Ostriker}, E.~C. 2007, \araa, 45, 565

\bibitem[{{Merello} {et~al.}(2015){Merello}, {Evans}, {Shirley}, {Rosolowsky},
  {Ginsburg}, {Bally}, {Battersby}, \& {Dunham}}]{Merello:2015}
{Merello}, M., {Evans}, II, N.~J., {Shirley}, Y.~L., {et~al.} 2015, ArXiv
  e-prints, arXiv:1501.05965, (ApJS, accepted)

\bibitem[{{Molinari} {et~al.}(2010{\natexlab{a}}){Molinari}, {Swinyard},
  {Bally}, {Barlow}, {Bernard}, {Martin}, {Moore}, {Noriega-Crespo}, {Plume},
  {Testi}, {Zavagno}, {Abergel}, {Ali}, {Anderson}, {Andr{\'e}}, {Baluteau},
  {Battersby}, {Beltr{\'a}n}, {Benedettini}, {Billot}, {Blommaert}, {Bontemps},
  {Boulanger}, {Brand}, {Brunt}, {Burton}, {Calzoletti}, {Carey}, {Caselli},
  {Cesaroni}, {Cernicharo}, {Chakrabarti}, {Chrysostomou}, {Cohen},
  {Compiegne}, {de Bernardis}, {de Gasperis}, {di Giorgio}, {Elia}, {Faustini},
  {Flagey}, {Fukui}, {Fuller}, {Ganga}, {Garcia-Lario}, {Glenn}, {Goldsmith},
  {Griffin}, {Hoare}, {Huang}, {Ikhenaode}, {Joblin}, {Joncas}, {Juvela},
  {Kirk}, {Lagache}, {Li}, {Lim}, {Lord}, {Marengo}, {Marshall}, {Masi},
  {Massi}, {Matsuura}, {Minier}, {Miville-Desch{\^e}nes}, {Montier}, {Morgan},
  {Motte}, {Mottram}, {M{\"u}ller}, {Natoli}, {Neves}, {Olmi}, {Paladini},
  {Paradis}, {Parsons}, {Peretto}, {Pestalozzi}, {Pezzuto}, {Piacentini},
  {Piazzo}, {Polychroni}, {Pomar{\`e}s}, {Popescu}, {Reach}, {Ristorcelli},
  {Robitaille}, {Robitaille}, {Rod{\'o}n}, {Roy}, {Royer}, {Russeil},
  {Saraceno}, {Sauvage}, {Schilke}, {Schisano}, {Schneider}, {Schuller},
  {Schulz}, {Sibthorpe}, {Smith}, {Smith}, {Spinoglio}, {Stamatellos},
  {Strafella}, {Stringfellow}, {Sturm}, {Taylor}, {Thompson}, {Traficante},
  {Tuffs}, {Umana}, {Valenziano}, {Vavrek}, {Veneziani}, {Viti}, {Waelkens},
  {Ward-Thompson}, {White}, {Wilcock}, {Wyrowski}, {Yorke}, \&
  {Zhang}}]{Molinari:2010b}
{Molinari}, S., {Swinyard}, B., {Bally}, J., {et~al.} 2010{\natexlab{a}}, \aap,
  518, L100

\bibitem[{{Molinari} {et~al.}(2010{\natexlab{b}}){Molinari}, {Swinyard},
  {Bally}, {Barlow}, {Bernard}, {Martin}, {Moore}, {Noriega-Crespo}, {Plume},
  {Testi}, {Zavagno}, {Abergel}, {Ali}, {Andr{\'e}}, {Baluteau}, {Benedettini},
  {Bern{\'e}}, {Billot}, {Blommaert}, {Bontemps}, {Boulanger}, {Brand},
  {Brunt}, {Burton}, {Campeggio}, {Carey}, {Caselli}, {Cesaroni}, {Cernicharo},
  {Chakrabarti}, {Chrysostomou}, {Codella}, {Cohen}, {Compiegne}, {Davis}, {de
  Bernardis}, {de Gasperis}, {Di Francesco}, {di Giorgio}, {Elia}, {Faustini},
  {Fischera}, {Fukui}, {Fuller}, {Ganga}, {Garcia-Lario}, {Giard}, {Giardino},
  {Glenn}, {Goldsmith}, {Griffin}, {Hoare}, {Huang}, {Jiang}, {Joblin},
  {Joncas}, {Juvela}, {Kirk}, {Lagache}, {Li}, {Lim}, {Lord}, {Lucas},
  {Maiolo}, {Marengo}, {Marshall}, {Masi}, {Massi}, {Matsuura}, {Meny},
  {Minier}, {Miville-Desch{\^e}nes}, {Montier}, {Motte}, {M{\"u}ller},
  {Natoli}, {Neves}, {Olmi}, {Paladini}, {Paradis}, {Pestalozzi}, {Pezzuto},
  {Piacentini}, {Pomar{\`e}s}, {Popescu}, {Reach}, {Richer}, {Ristorcelli},
  {Roy}, {Royer}, {Russeil}, {Saraceno}, {Sauvage}, {Schilke},
  {Schneider-Bontemps}, {Schuller}, {Schultz}, {Shepherd}, {Sibthorpe},
  {Smith}, {Smith}, {Spinoglio}, {Stamatellos}, {Strafella}, {Stringfellow},
  {Sturm}, {Taylor}, {Thompson}, {Tuffs}, {Umana}, {Valenziano}, {Vavrek},
  {Viti}, {Waelkens}, {Ward-Thompson}, {White}, {Wyrowski}, {Yorke}, \&
  {Zhang}}]{Molinari:2010a}
---. 2010{\natexlab{b}}, \pasp, 122, 314

\bibitem[{{Motte} {et~al.}(2007){Motte}, {Bontemps}, {Schilke}, {Schneider},
  {Menten}, \& {Brogui{\`e}re}}]{Motte:2007}
{Motte}, F., {Bontemps}, S., {Schilke}, P., {et~al.} 2007, \aap, 476, 1243

\bibitem[{{Netterfield} {et~al.}(2009){Netterfield}, {Ade}, {Bock}, {Chapin},
  {Devlin}, {Griffin}, {Gundersen}, {Halpern}, {Hargrave}, {Hughes}, {Klein},
  {Marsden}, {Martin}, {Mauskopf}, {Olmi}, {Pascale}, {Patanchon}, {Rex},
  {Roy}, {Scott}, {Semisch}, {Thomas}, {Truch}, {Tucker}, {Tucker}, {Viero}, \&
  {Wiebe}}]{Netterfield:2009a}
{Netterfield}, C.~B., {Ade}, P.~A.~R., {Bock}, J.~J., {et~al.} 2009, \apj, 707,
  1824

\bibitem[{{Offner} {et~al.}(2014){Offner}, {Clark}, {Hennebelle}, {Bastian},
  {Bate}, {Hopkins}, {Moraux}, \& {Whitworth}}]{Offner:2013}
{Offner}, S.~S.~R., {Clark}, P.~C., {Hennebelle}, P., {et~al.} 2014, Protostars
  and Planets VI, 53

\bibitem[{{Olmi} {et~al.}(2013){Olmi}, {Angl{\'e}s-Alc{\'a}zar}, {Elia},
  {Molinari}, {Montier}, {Pestalozzi}, {Pezzuto}, {Polychroni}, {Ristorcelli},
  {Rodon}, {Schisano}, {Smith}, {Testi}, \& {Thompson}}]{Olmi:2013}
{Olmi}, L., {Angl{\'e}s-Alc{\'a}zar}, D., {Elia}, D., {et~al.} 2013, \aap, 551,
  A111

\bibitem[{{Ossenkopf} \& {Henning}(1994)}]{Ossenkopf:1994}
{Ossenkopf}, V., \& {Henning}, T. 1994, \aap, 291, 943

\bibitem[{{Padoan} \& {Nordlund}(2002)}]{Padoan:2002}
{Padoan}, P., \& {Nordlund}, {\AA}. 2002, \apj, 576, 870

\bibitem[{{Padoan} \& {Nordlund}(2011)}]{Padoan:2011}
---. 2011, \apj, 730, 40

\bibitem[{{Padoan} {et~al.}(1997){Padoan}, {Nordlund}, \&
  {Jones}}]{Padoan:1997}
{Padoan}, P., {Nordlund}, A., \& {Jones}, B.~J.~T. 1997, \mnras, 288, 145

\bibitem[{{Peretto} \& {Fuller}(2010)}]{Peretto:2010b}
{Peretto}, N., \& {Fuller}, G.~A. 2010, \apj, 723, 555

\bibitem[{Press {et~al.}(2007)Press, Teukolsky, Vetterling, \&
  Flannery}]{Press:2007}
Press, W.~H., Teukolsky, S.~A., Vetterling, W.~T., \& Flannery, B.~P. 2007,
  Numerical Recipes 3rd Edition: The Art of Scientific Computing, 3rd edn. (New
  York, NY, USA: Cambridge University Press)

\bibitem[{{Reid} {et~al.}(2014){Reid}, {Menten}, {Brunthaler}, {Zheng}, {Dame},
  {Xu}, {Wu}, {Zhang}, {Sanna}, {Sato}, {Hachisuka}, {Choi}, {Immer},
  {Moscadelli}, {Rygl}, \& {Bartkiewicz}}]{Reid:2014}
{Reid}, M.~J., {Menten}, K.~M., {Brunthaler}, A., {et~al.} 2014, \apj, 783, 130

\bibitem[{{Ridge} {et~al.}(2006){Ridge}, {Di Francesco}, {Kirk}, {Li},
  {Goodman}, {Alves}, {Arce}, {Borkin}, {Caselli}, {Foster}, {Heyer},
  {Johnstone}, {Kosslyn}, {Lombardi}, {Pineda}, {Schnee}, \&
  {Tafalla}}]{Ridge:2006a}
{Ridge}, N.~A., {Di Francesco}, J., {Kirk}, H., {et~al.} 2006, \aj, 131, 2921

\bibitem[{{Roman-Duval} {et~al.}(2010){Roman-Duval}, {Jackson}, {Heyer},
  {Rathborne}, \& {Simon}}]{RomanDuval:2010}
{Roman-Duval}, J., {Jackson}, J.~M., {Heyer}, M., {Rathborne}, J., \& {Simon},
  R. 2010, \apj, 723, 492, (RD10)

\bibitem[{{Rosolowsky} {et~al.}(2010){Rosolowsky}, {Dunham}, {Ginsburg},
  {Bradley}, {Aguirre}, {Bally}, {Battersby}, {Cyganowski}, {Dowell},
  {Drosback}, {Evans}, {Glenn}, {Harvey}, {Stringfellow}, {Walawender}, \&
  {Williams}}]{Rosolowsky:2010}
{Rosolowsky}, E., {Dunham}, M.~K., {Ginsburg}, A., {et~al.} 2010, \apjs, 188,
  123

\bibitem[{{Rosolowsky} {et~al.}(2008){Rosolowsky}, {Pineda}, {Foster},
  {Borkin}, {Kauffmann}, {Caselli}, {Myers}, \& {Goodman}}]{Rosolowsky:2008}
{Rosolowsky}, E.~W., {Pineda}, J.~E., {Foster}, J.~B., {et~al.} 2008, \apjs,
  175, 509

\bibitem[{{Ruphy} {et~al.}(1996){Ruphy}, {Robin}, {Epchtein}, {Copet},
  {Bertin}, {Fouque}, \& {Guglielmo}}]{Ruphy:1996}
{Ruphy}, S., {Robin}, A.~C., {Epchtein}, N., {et~al.} 1996, \aap, 313, L21

\bibitem[{{Russeil} {et~al.}(2011){Russeil}, {Pestalozzi}, {Mottram},
  {Bontemps}, {Anderson}, {Zavagno}, {Beltr{\'a}n}, {Bally}, {Brand}, {Brunt},
  {Cesaroni}, {Joncas}, {Marshall}, {Martin}, {Massi}, {Molinari}, {Moore},
  {Noriega-Crespo}, {Olmi}, {Thompson}, {Wienen}, \& {Wyrowski}}]{Russeil:2011}
{Russeil}, D., {Pestalozzi}, M., {Mottram}, J.~C., {et~al.} 2011, \aap, 526,
  A151

\bibitem[{{Salpeter}(1955)}]{Salpeter:1955}
{Salpeter}, E.~E. 1955, \apj, 121, 161

\bibitem[{{Schlingman} {et~al.}(2011){Schlingman}, {Shirley}, {Schenk},
  {Rosolowsky}, {Bally}, {Battersby}, {Dunham}, {Ellsworth-Bowers}, {Evans},
  {Ginsburg}, \& {Stringfellow}}]{Schlingman:2011}
{Schlingman}, W.~M., {Shirley}, Y.~L., {Schenk}, D.~E., {et~al.} 2011, \apjs,
  195, 14

\bibitem[{{Schuller} {et~al.}(2009){Schuller}, {Menten}, {Contreras},
  {Wyrowski}, {Schilke}, {Bronfman}, {Henning}, {Walmsley}, {Beuther},
  {Bontemps}, {Cesaroni}, {Deharveng}, {Garay}, {Herpin}, {Lefloch}, {Linz},
  {Mardones}, {Minier}, {Molinari}, {Motte}, {Nyman}, {Reveret}, {Risacher},
  {Russeil}, {Schneider}, {Testi}, {Troost}, {Vasyunina}, {Wienen}, {Zavagno},
  {Kovacs}, {Kreysa}, {Siringo}, \& {Wei{\ss}}}]{Schuller:2009}
{Schuller}, F., {Menten}, K.~M., {Contreras}, Y., {et~al.} 2009, \aap, 504, 415

\bibitem[{{Scoville} \& {Solomon}(1975)}]{Scoville:1975}
{Scoville}, N.~Z., \& {Solomon}, P.~M. 1975, \apjl, 199, L105

\bibitem[{{Shirley} {et~al.}(2003){Shirley}, {Evans}, {Young}, {Knez}, \&
  {Jaffe}}]{Shirley:2003}
{Shirley}, Y.~L., {Evans}, II, N.~J., {Young}, K.~E., {Knez}, C., \& {Jaffe},
  D.~T. 2003, \apjs, 149, 375

\bibitem[{{Shirley} {et~al.}(2013){Shirley}, {Ellsworth-Bowers}, {Svoboda},
  {Schlingman}, {Ginsburg}, {Rosolowsky}, {Gerner}, {Mairs}, {Battersby},
  {Stringfellow}, {Dunham}, {Glenn}, \& {Bally}}]{Shirley:2013}
{Shirley}, Y.~L., {Ellsworth-Bowers}, T.~P., {Svoboda}, B., {et~al.} 2013,
  \apjs, 209, 2

\bibitem[{{Solomon} {et~al.}(1987){Solomon}, {Rivolo}, {Barrett}, \&
  {Yahil}}]{Solomon:1987}
{Solomon}, P.~M., {Rivolo}, A.~R., {Barrett}, J., \& {Yahil}, A. 1987, \apj,
  319, 730

\bibitem[{{Stinson} {et~al.}(2013){Stinson}, {Brook}, {Macci{\`o}}, {Wadsley},
  {Quinn}, \& {Couchman}}]{Stinson:2013}
{Stinson}, G.~S., {Brook}, C., {Macci{\`o}}, A.~V., {et~al.} 2013, \mnras, 428,
  129

\bibitem[{{Swift} \& {Beaumont}(2010)}]{Swift:2010}
{Swift}, J.~J., \& {Beaumont}, C.~N. 2010, \pasp, 122, 224

\bibitem[{{Tassis} {et~al.}(2010){Tassis}, {Christie}, {Urban}, {Pineda},
  {Mouschovias}, {Yorke}, \& {Martel}}]{Tassis:2010}
{Tassis}, K., {Christie}, D.~A., {Urban}, A., {et~al.} 2010, \mnras, 408, 1089

\bibitem[{{Veltchev} {et~al.}(2013){Veltchev}, {Donkov}, \&
  {Klessen}}]{Veltchev:2013}
{Veltchev}, T.~V., {Donkov}, S., \& {Klessen}, R.~S. 2013, \mnras, 432, 3495

\bibitem[{{Westerhout}(1958)}]{Westerhout:1958}
{Westerhout}, G. 1958, \bain, 14, 215

\bibitem[{{Wienen} {et~al.}(2015){Wienen}, {Wyrowski}, {Menten}, {Urquhart},
  {Csengeri}, {Walmsley}, {Bontemps}, {Russeil}, {Bronfman}, {Koribalski}, \&
  {Schuller}}]{Wienen:2015}
{Wienen}, M., {Wyrowski}, F., {Menten}, K.~M., {et~al.} 2015, ArXiv e-prints,
  arXiv:1503.00007, (A\&A, accepted)

\bibitem[{{Wolfire} {et~al.}(2003){Wolfire}, {McKee}, {Hollenbach}, \&
  {Tielens}}]{Wolfire:2003}
{Wolfire}, M.~G., {McKee}, C.~F., {Hollenbach}, D., \& {Tielens}, A.~G.~G.~M.
  2003, \apj, 587, 278

\end{thebibliography}
\end{document}